\documentstyle[amscd,amssymb,verbatim,12pt]{amsart}
\theoremstyle{plain}

\theoremstyle{definition}

\numberwithin{equation}{section}

\def\dspace{\baselineskip=0.3 in}
\begin{document}
\dspace
\title[ DUAL NATURE OF THE RICCI SCALAR.................]{ DUAL NATURE OF THE
  RICCI  SCALAR AND ITS CERTAIN CONSEQUENCES}

\author[S.K.Srivastava  ]%
        {    }

\maketitle

\centerline{\bf S.K.Srivastava }

\centerline{ Department of Mathematics, North Eastern Hill University,}

\centerline{ NEHU Campus,Shillong - 793022 ( INDIA ) }

\centerline{e-mail:srivastava@@nehu.ac.in ; sushil@@iucaa.ernet.in }

\vspace{1cm}

\centerline{\bf  Abstract}  

\smallskip

 Ricci scalar is the key ingredient of non-Newtonian theory of gravity,
where space-time geometry has a crucial role. Normally, it is supposed to be a
geometrical field, but interestingly it also behaves like a physical
field. Thus it plays dual role in the arena of gravitation. This article is an
$overview$ of the work related to dual roles of the Ricci scalar. A scalar is a mathematical
concept representing a spinless particle. Here, particle concept,
manifesting physical aspect of the Ricci scalar, is termed as $riccion$. It is a scalar
particle with (mass)$^2$ inversely proportinal to the gravitational constant. Many
interesting consequences of dual role of the Ricci scalar are discussed here. It
causes inflationary scenario in the early universe without taking an another
scalar like ``inflaton'' also. It is found that a $riccion$ behaves like an
$instanton$ also. This feature inspires ``primordial inflation''. It is interesting
to see that a $riccion$, obtained from higher-dimensional space-time, decouples
into fermion and anti-fermion  pair if parity is violated. One-loop
renormalization of $riccion$ indicates fractal geometry at high
energy. Homogeneous and inhomogeneous models of the early universe are derived
using dual roles of the Ricci scalar. Production of spinless and spin-1/2
particles, due to riccion coupling, is discussed here. Contribution of $riccion$ to the cosmic dark energy is obtained here
through one-loop renormalization and it is shown that dark enargy decays to
dark matter during expansion of the universe. It inspires a new cosmological
scenario consistent with observational evidences.  

\vspace{2cm}

\centerline{\bf 1. Introduction}

 Einstein's Equivalence Principle (EEP)and  General Covariance (GC)  are two basic principles of the General Relativity (GR). According to EEP, in the sufficiently small region, laws of motion can be described in the same form as these are in non-accelerated systems free from gravitational field. It means that, in the sufficiently small region, laws of the Nature can be understood through Special Relativity(SR) in the framework of Minkowskian space-time. This concept is analogous to the concept of manifolds which are locally euclidean  geometrical structures. This analogy provided an insight to understand gravity through the geometry of the space-time. As a result, application of the riemannian geometry appeared in the arena of the physical phenomenon like gravitation. Thus validity of EEP implies manifestation of gravitation through metrizable and differentiable manifolds of the curved space-time. So, using geometry of the curved space-time, Einstein could generalize his theory of SR to GR as he found that even a free particle gets accelerated there. It means that the geometry induces a physical force which  can not be eliminated by any means, as it is due to intrinsic property of the space-time. Moreover, this force vanishes locally, but it is realized at large scales. Gravity also has  similar properties. So, this geometrical force was identified with gravity. On this basis, Einstein claimed that geometry manifests gravitation. As a result, contrary to Newtonian theory , Einstein theory of gravity emerged as a non-linear theory different from other forces in Nature.

The action for the Einstein's theory of gravitation is constructed using EEP and GC , where lagrangian density is $R/{16{\pi}G}$ with $R$ being the Ricci scalar and $G$ the gravitational constant. $R$ depends on components of the metric tensor as well as its first and second order derivatives with respect to space-time coordinates. Thus the metric tensor is the key ingredient of Einstein's theory of gravity.

At the classical level, this theory is in good accord with the experimental results. But, at quantum level, it is not problem - free. The main problem is non - renormalzability of this theory. So the need to construct a  theory, different from GR, was realized. During the last three decades, different attempts were made to construct renormalizable quantum gravity. Among these attempts, higher-derivative gravity is an important candidate.  Its action  is obtained by adding higher-order terms of the curvature tensor like $R^2, R_{\mu\nu}R^{\mu\nu}, R_{\mu\nu\rho\sigma}R^{\mu\nu\rho\sigma}, {\Box}R, R^3$ etc. to the Einstein-Hilbert lagrangian $R/16{\pi}G$ ( where $R_{\mu\nu}$ and $R_{\mu\nu\rho\sigma}$ are components of Ricci and Riemann curvature tensor respectively). Here $\Box$ is the  d'Alembertian operator in curved space-time defined as

$$\Box = \frac{1}{\sqrt{- g}} \frac{\partial}{\partial x^{\mu}}\Big(\sqrt{- g} g^{\mu\nu} \frac{\partial}{\partial x^{\nu}} \Big) \eqno(1.1)$$

\noindent with $g_{\mu\nu}$ being components of the metric tensor defined
through the distance function, in curved space-time as $dS^2 =
g_{\mu\nu}dx^{\mu}dx^{\nu} (\mu, \nu =0,1,2,3).$ Here $g$ is the determinant
of $g_{\mu\nu}$. Higher-derivative terms are also relevant in string inspired
theories e.g. brane-gravity and Gauss-Bonnet theory \cite{rm}.

 The modified action contains higher-order derivatives of the metric tensor which is consistent with  EEP and GC . At the quantum level, $g_{\mu\nu}$ are supposed to be spin-2 gravitational field, called gravitons. A multiplicatively renormalizable and asymptotically free theory of gravity was obtained from the modified action \cite{ks,es,et}, but it is not unitary. On this ground , the theory of higher- derivative gravity is rejected.

 Now the question arises whether it is the end of the story.Definitely, it is not so. Investigations, from time to time, have shown that there is an another aspect of the higher-derivative gravity which is very interesting and problem- free.

 The main ingredient in the action of higher-derivative gravity is the Ricci scalar $R$. It is been found that $R$ behaves like a spinless physical field also in addition to its usual role as a geometrical field. In what follows, we review important works of physicists showing physical aspect of the Ricci scalar manifested by higher-derivative gravity.

In 1967,B.S.DeWitt \cite{bs} and in 1968, Zel'dovich $\&$ Novikov \cite{yab68}
conjectured that inclusion of higher-derivative terms, in the gravitational
action, could avoid singularity in cosmological solutions. This idea was
followed by Zel'dovich and Pitaevski and , in 1971, they discussed
possibility of avoiding singularity in the early universe model using
higher-derivative gravity \cite{yab71}. In 1979, Gurovich and Starobinski
pointed out that if an arbitrary term $f(R)$, subject to the condition $lim_{R
  \to 0} f(R)/R = 0,$ is added to the Einstein-Hilbert lagrangian, ghost terms
do not appear \cite{vts}. Earlier to this work, Stelle \cite{ks} had also mentioned that,contrary to the square of Weyl tensor
$$C_{\mu\nu\rho\sigma}C^{\mu\nu\rho\sigma}= R_{\mu\nu\rho\sigma}
R^{\mu\nu\rho\sigma}- 2 R_{\mu\nu}R^{\mu\nu} +(1/3)R^2,$$

\noindent introduction of $R^2$ or $f(R)$ does not cause the ghost problem . In 1980, Starobinsky proposed independently that if the sign of the $R^2$ term in the higher-derivative gravitational action is chosen properly, one could obtain only one scalar particle with positive energy and positive squared mass. He called it as ``scalaron''\cite{aas,aak}. In 1984, using the lagrangian density, $L=R + {\epsilon}R^2$, B.Whitt demonstrated that Einstein's gravity could be described by modified metric tensor components ${\tilde g}_{\mu\nu}= (1+ 2 {\epsilon}R) g_{\mu\nu}$ and the scalar curvature $R$ with the minimal coupling to gravity, given by equations \cite{bw},

$$ {\tilde R}_{\mu\nu} - \frac{1}{2}{\tilde g}_{\mu\nu} {\tilde R} = 8 \pi G {\tilde T}_{\mu\nu}(R),$$

where

$${\tilde T}_{\mu\nu}(R) = \frac{6 \epsilon^2}{8 \pi G (1+ 2 {\epsilon}R)^2} \Big[{\partial}_{\mu}R {\partial}_{\nu}R -{\tilde g}_{\mu\nu}\{(1/2){\partial}^{\sigma}R {\partial}_{\sigma}R  + \frac{R^2}{12 \epsilon} \} \Big].$$

In this case, action for the scalar field $R$ is 

$$S(R) = \int {d^4x} 6 \epsilon^2 \sqrt{- {\tilde g}}(1+ 2 {\epsilon}R)^{-2} \Big[{\partial}^{\sigma}R {\partial}_{\sigma}R  + \frac{R^2}{6 \epsilon} \Big].$$

In these equations, the scalar  $R$ behaves like a scalar physical field
with $ \frac{1}{6 \epsilon}$ as squared mass. Here $\epsilon$ is the coupling
constant with squared length dimension. The same year, Hawking and Luttrell
also found the similar result \cite{swh,nhb}. Two years later, using the
lagrangian density like Whitt in the conformally invariant flat space-time,
M.B.Mijic et al \cite{mbmks} had shown that the scalar curvature $R$ behaved
like a damped harmonic oscillator with the restoring force given by  $
\frac{1}{6 \epsilon}$. Further, they have discussed that $R$ also drives
inflation when $\epsilon$ is large. In the same model, it is demonstrated
that, after inflationary phase, the universe passes through the period when
$R$ oscillates rapidly causing particle - production. These particles reheat
the universe on annihilation \cite{wms}.

In higher - derivative gravitational models, discussed above, either the gravitational constant $G$ is taken equal to 1 or lagrangian density is taken as

 $$\frac{1}{8 \pi G}(R + \epsilon R^2).$$

As a result, (mass)$^2$ for $R$ does not contain the term $16 \pi G.$ But, in
a gravitational theory, role of $G$ is crucial. So, in \cite{skp93}, lagrangian for higher-derivative gravity  was taken as 

 $$ \frac{R}{16 \pi G} +\quad {\rm higher - derivative \quad terms}$$

\noindent giving proper role to $G$ and the physical aspect of $R$. It is
found that trace of field equations from this action shows that the Ricci
scalar behaves like a physical field with squared mass dependent on $G$. Here,
it is important to mention that the physical aspect of $R$ is manifested only
when higher - derivative terms have significant presence in the gravitational
action,  compared to $ \frac{R}{16 \pi G}$. It is possible at sufficiently
high energy. It means that physical aspect of the Ricci scalar can be realized at high energy. Thus , it is reasonable to infer that, in the high energy regime, the Ricci scalar shows dual role (i) like a spinless physical field as well as (ii) like a geometrical field.

In a field theory, field is regarded as a mathematical concept associated with
a physical particle. So, if the Ricci scalar behaves as a physical field at
high energy, there should be a particle manifesting this behaviour. Here
onwards, this particle is called as $riccion$, which is different from scalar
mode of $graviton$. The difference between $riccion$ and $graviton$ is
highlighted in Appendix A.

Natural units, defined as $\kappa_B = \hbar = c = 1$ ( where $\kappa_B$ is Boltzman's constant, $\hbar$ is Planck's constant divided by $2\pi$ and $c$ is the speed of light) are used here with  $\rm GeV$ is used as a fundamental unit such that $1{\rm GeV} = 1.16 \times 10^{13}K = 1.78 \times 10^{-24} gm$ , $ 1{\rm GeV}^{-1} = 1.97 \times 10^{-14}cm = 6.58 \times 10^{-25} sec$.

\bigskip

\centerline{\bf 2. A toy model to demonstrate the Dual nature}

\centerline{\bf of the Ricci scalar}

\smallskip

As mentioned above, dual nature of the Ricci scalar is manifested by
higher-derivative gravitational action. In this section, for simplicity, we
take the following action for $R^2$ - gravity to demonstrate the same. Here,
$R^2$ - gravitational action is taken as
\cite{skp93,skps93,sk94,skp96,skp97,sk98,sk99,sks99,sk00, sk03}
$$ S_g = \int{d^4x}{\sqrt{-g}}\Big[ 
{\frac{R}{16{\pi}G}}  + {\tilde\alpha} R_{mu\nu} R^{\mu\nu} +
{\tilde\beta}R^2 ],  \eqno(2.1)$$
where   ${\tilde\alpha}$ and ${\tilde\beta}$  are
 dimensionless
coupling constants. The Newtonian gravitational constant $ G = M_P^{-2}$ ($M_P$ is the Planck mass).

The invariance of $S_g$ , given by eq.(2.1) , under  
transformations
$g_{\mu\nu}\to g_{\mu\nu} 
+{\delta}g_{\mu\nu}$ yields the gravitational field equations  
$${\frac{1}{16{\pi}G}}(R_{\mu\nu} -{\frac{1}{2}}g_{\mu\nu}R) + 
{\tilde\alpha}( 2R^{\rho}_{\mu;\nu\rho} -{\Box}R_{\mu\nu} $$
 $$- {\frac{1}{2}}
g_{\mu\nu}{\Box}R + 2 R^{\rho}_{\mu}R_{\rho\nu} - {\frac{1}{2}}
g_{\mu\nu}R^{\rho\sigma}R_{\rho\sigma}) + {\tilde\beta}
( 2 R_{;\mu\nu} -
2 g_{\mu\nu}{\Box}R  $$ $$- {\frac{1}{2}} g_{\mu\nu} R^2 + 
2 R R_{\mu\nu})
 = 0 , \eqno(2.2)$$
where $\{;\}$ denotes covariant derivative in curved space - time and  ${\Box}$ is defined by eq.(1.1).

Taking trace of equations (2.2), one obtains

$${\Box} R +  m^2 R = 0 , \eqno(2.3a)$$ where
$$m^2 = [32{\pi}({\tilde\alpha} + 3{\tilde\beta})G]^{-1}.
  \eqno(2.3b)$$

This equation shows that $R$ can be treated as a mathematical concept representing a spinless particle with squared mass , given by eq.(2.3b). To understand the physical meaning of the equation (2.3), it is useful to take an analogy of the Klein - Gordon equation for a scalar field $\phi$ 

$${\Box}\phi +  m^2_{\phi} \phi = 0  .\eqno(2.4)$$

Equation (2.3) seems to be analogous to the equation (2.4), but there are two problems : 

\noindent (i)Mass dimension of $\phi$ is 1 as eq.(2.4) is derived from the action   
$$S_{\phi} ={\frac{1}{2}} \int {d^4x} {\sqrt{-g}}
[{g^{\mu\nu}}{\partial}_{\mu}\phi{\partial}_{\nu}\phi -  m^2_{\phi} \phi^2] .\eqno(2.5)$$
But mass dimension of $R$ is 2, being combination of square of first derivatives and second derivative of components of the metric tensor, $g_{\mu\nu}$ ($g_{\mu\nu}$ are dimensionless as $dS^2 = g_{\mu\nu} d x^{\mu} d x^{\nu}.)$ with respect to space - time co-ordinates. 

\noindent (ii) $\Box$ and $R$, both depend on $g_{\mu\nu}$, but $\phi$ does not depend on it.

In what follows, these problems are resolved. First problem is 
  solved multiplying the equation (2.3a) by $\eta$ ( where , in natural units, 
$\vert{\eta}\vert = 1 $ with $(mass)^{-1}$) and
recognizing ${\eta}R$ as ${\tilde R} .$  As a result,the equation (2.3a) 
is rewritten as 
$${\Box}{\tilde R} +  m^2 {\tilde R} = 0 \eqno(2.6)$$ 
It is easy to find that mass dimension of ${\tilde R}$ is 1 as
mass dimension of $R$ is 2 (mentioned above).

 The second  problem is resolved through the following
discussion. In the small
vicinity of a space - time point, $g_{\mu\nu}(x)$ can be expanded as 
\begin{eqnarray*}
g_{\mu\nu}(x) & = & g_{\mu\nu}(0) + {\frac{1}{3}}
R_{\mu\alpha\nu\beta}(0) x^{\alpha}x^{\beta} 
- {\frac{1}{6}}{{\partial}_{\gamma}}R_{\mu\alpha\nu\beta}(0)
x^{\alpha}x^{\beta}x^{\gamma} + \\ &&  [
{\frac{1}{20}}R_{\mu\alpha\nu\beta;\gamma\delta} +  {\frac{2}{45}} 
R_{\alpha\mu\beta\lambda}R^{\lambda}_{\gamma\nu\delta}](0)
x^{\alpha}x^{\beta}x^{\gamma}x^{\delta} + ....... ,
\end{eqnarray*}
$$ \eqno(2.7)$$ where
$g_{\mu\nu}(0) = {\eta}_{\mu\nu}$.

Using this expansion, one can easily see that the Ricci scalar $R$ is
non-zero in a locally inertial co-ordinate system, unless the entire
space-time manifold is Minkowskian. Also, in the locally inertial
co-ordinate system, ${\Box}$ operator (defined by eq.(2.2b)) looks like  

$${\Box} = \eta^{\mu\nu}\frac{\partial^2}{\partial x^{\mu}\partial x^{\nu}},$$
showing that, in a locally inertial co-ordinate system, ${\Box}$ is a
similar operator for $\tilde R$ as it is for other scalar field $\phi.$
According to the principle of equivalence and principle of covariance, this
characteristic feature of ${\Box}$ and $\tilde R$ (these being scalars) will be maintained at
global scales also \cite{sk03}.

On the basis of these analyses, it is obtained that  the Ricci scalar not only behaves as a geometrical field,
 but also
as a spinless matter field analogous to $\phi.$

To exploit the matter aspect of the Ricci scalar, $\tilde R$ is
 treated as
a basic field. So there should be an action $S_{\tilde R}$
 yielding
eq.(2.6)as a result of invariance of $S_{\tilde R}$ under transformation
${\tilde R} \to {\tilde R} + \delta{\tilde R}$ in the same manner
 as $S_{\phi}$ yields 
eq.(2.4)using invariance under transformation $\phi \to \phi +
\delta\phi.$

If such an action $S_{\tilde R}$ exists,
$$\delta S_{\tilde R} = -  \int{d^4x}{\sqrt{-g}}\quad \delta{\tilde R}
[{\Box} +  m^2]{\tilde R}, \eqno(2.8)$$
implying eq.(2.7) subject to  the condition $\delta S_{\tilde R} = 0.$
From eq.(2.8),
\begin{eqnarray*}
\delta S_{\tilde R} &=&   \int{d^4x}{\sqrt{-g}}\quad
\Big[ g^{\mu\nu}{\partial_{\mu}{\tilde R}}{\partial_{\nu}(\delta{\tilde
R})} -  m^2{\tilde R} \delta{\tilde R}\Big] \\ &=&  \int{d^4x}{\sqrt{-g}}\quad
\delta\Big\{\Big [{1 \over2}( g^{\mu\nu}{\partial_{\mu}{\tilde
R}}{\partial_{\nu}{\tilde R}} - m^2{\tilde R}^2)\Big]\Big\} ,  
\end{eqnarray*}
$$ \eqno(2.9)$$

In a locally inertial co-ordinate system,

$$\int{d^4x}{\sqrt{-g}}\quad
\delta\Big\{\Big [{1 \over2} {\partial^{\mu}{\tilde
R}}{\partial_{\mu}{\tilde R}} - m^2{\tilde R}^2 \Big]\Big\}$$

$$=\int{d^4X}\quad
\delta\Big\{\Big [{1 \over2} {\partial^{\mu}{\tilde
R}}{\partial_{\mu}{\tilde R}} - m^2{\tilde R}^2 \Big]\Big\}$$

$$=\delta\int{d^4X}\quad
\Big\{\Big [{1 \over2} {\partial^{\mu}{\tilde
R}}{\partial_{\mu}{\tilde R}} - m^2{\tilde R}^2 \Big]\Big\}$$
as ${\tilde R}$,being a scalar, is equivalent at local as well as global scales. Here $X^i (i=0,1,2,3)$ are local and  $x^i (i=0,1,2,3)$ are global coordinates.

The action 

$$\int{d^4X}\quad
\Big\{\Big [{1 \over2} {\partial^{\mu}{\tilde
R}}{\partial_{\mu}{\tilde R}} - m^2{\tilde R}^2 \Big]\Big\},$$
being covariant, is invariant under transformations $X^i \to x^i$ (i.e. from local to global coordinates), so

$$\int{d^4X}\quad
\Big\{\Big [{1 \over2} {\partial^{\mu}{\tilde
R}}{\partial_{\mu}{\tilde R}} - m^2{\tilde R}^2 \Big]\Big\}=\int{d^4x}{\sqrt{-g}}\quad
\Big\{\Big [{1 \over2} {\partial^{\mu}{\tilde
R}}{\partial_{\mu}{\tilde R}} - m^2{\tilde R}^2 \Big]\Big\}$$

 Thus,  eq.(2.9) reduces to
$$\delta S_{\tilde R} = \delta \int{d^4x}
\Big\{{\sqrt{-g}}\Big [{1 \over2} g^{\mu\nu}{\partial_{\mu}{\tilde
R}}{\partial_{\nu}{\tilde R}} - m^2{\tilde R}^2 \Big]\Big\}$$
yielding
$$S_{\tilde R} =  \int{d^4x}
{\sqrt{-g}}\quad\Big [{1 \over2} g^{\mu\nu}{\partial_{\mu}{\tilde
R}}{\partial_{\nu}{\tilde R}} - m^2{\tilde R}^2 \Big], \eqno(2.10)$$

 Thus eq.(2.10) gives the required action for
the spinless physical field ${\tilde R}.$ It is different from $scalaron$, proosed by Starobinsky
\cite{aas, aak} in two ways (i) mass dimension of $riccion$ is 1, whereas the same
for a $scalaron$ is 2 and (ii) (mass)$^2$ for a $riccion$ is inversely
proportional to $G$ (the gravitational constant), but a $scalaron$ is
independent of $G$.

\vspace{0.5cm}

\noindent \centerline{\bf 3. Riccion as an instanton  }

\bigskip

Here gravitational action is taken as \cite{sk99}

$$ S_g = \int{d^4x}{\sqrt{-g}}\Big[ 
{\frac{R}{16{\pi}G}}  - {\tilde\alpha} R_{mu\nu} R^{\mu\nu} -
{\tilde\beta}R^2 + \lambda \eta^2 ( R^3 + 9 {\Box}R^2 ) ],  \eqno(3.1)$$
where   ${\tilde\alpha}$, ${\tilde\beta}$ and $\lambda$  are
 dimensionless
coupling constants. $G$ and $\eta$ are defined in the previous section.
Though term containing ${\Box}R^2$ is a surface term, still it is needed here. The 
motivation
for this type of action has been clearified below.

The action (3.1) yields gravitational field equations as
\begin{eqnarray*}  
&&{\frac{1}{16{\pi}G}}(R_{\mu\nu} -{\frac{1}{2}}g_{\mu\nu}R) - 
{\tilde\alpha}( 2R^{\rho}_{\mu;\nu\rho} -{\Box}R_{\mu\nu}- {\frac{1}{2}}
g_{\mu\nu}{\Box}R + 2 R^{\rho}_{\mu}R_{\rho\nu}\\ &&- {\frac{1}{2}}
g_{\mu\nu}R^{\rho\sigma}R_{\rho\sigma}) - {\tilde\beta}
( 2 R_{;\mu\nu} -
2 g_{\mu\nu}{\Box}R - {\frac{1}{2}} g_{\mu\nu} R^2 + 
2 R R_{\mu\nu}) \\&&+ \lambda \eta^2 (12 R^2_{;\mu\nu} -
\frac{15}{2}g_{\mu\nu}{\Box}R^2 + 18{\Box}R^2 - {\frac{1}{2}} g_{\mu\nu} R^3 +
3 R^2 R_{\mu\nu})  = 0 .
\end{eqnarray*}
$$ \eqno(3.2)$$

Trace of equations (3.2) is obtained as
$${\Box} R -  m^2 R + \lambda \eta^2 ({\tilde\alpha} + 3{\tilde\beta})^{-1} R^3 = 0 , \eqno(3.3 a)$$ where
$$m^2 = [32{\pi}({\tilde\alpha} + 3{\tilde\beta})G]^{-1}.
  \eqno(3.3b)$$ 

 Multiplying eq.(3.3) by $\eta$ and recognizing $\eta R$ as $\tilde R$,eq.(3.3) is re-written as

$${\Box}{\tilde R} = - \frac{\lambda}{2({\tilde\alpha} + 3{\tilde\beta})} {\tilde R}\Big({\tilde R}^2 - \frac{M_P^2}{32{\pi}{\lambda}}\Big)
 \eqno(3.4)$$

Proceeding on the line of the previous section, action for riccion is obtained as
 
$$S_{\tilde R} =  \int{d^4x}
{\sqrt{-g}}\quad\Big [{1 \over2} g^{\mu\nu}{\partial_{\mu}{\tilde
R}}{\partial_{\nu}{\tilde R}} - V({\tilde R})\Big] \eqno(3.5a)$$
with

$$ V({\tilde R}) = \frac{\lambda}{8({\tilde\alpha} + 3{\tilde\beta})} \Big({\tilde R}^2 -\frac{M_P^2}{32{\pi}{\lambda}}\Big)^2
\eqno(3.5b)$$

Lagrangian, obtained from eq.(3.5), is 
$$ L(t) =
 \int{d^3x}
{\sqrt{-g}}\quad\Big [{1 \over2} g^{\mu\nu}{\partial_{\mu}{\tilde
R}}{\partial_{\nu}{\tilde R}} - V({\tilde R})\Big] $$
with $ V({\tilde R})$ given by eq.(3.5b). The corresponding hamiltonian \cite{sk99} is
$$ H(t) = E(t) =
 \int{d^3x}
{\sqrt{-g}}\quad\Big [{1 \over2} g^{\mu\nu}{\partial_{\mu}{\tilde
R}}{\partial_{\nu}{\tilde R}} + V({\tilde R})\Big], \eqno(3.6) $$
where integration is performed over $ t = constant$ hypersurface.
As the potential $V({\tilde R})$ does not contain the velocity like term $\dot {\tilde R}$, the hamiltonian gives the total energy $E$ of riccion ${\tilde R}$ in this case.

Now, one can find the condition for the finite euclidean action corresponding
to the lagrangian density for ${\tilde R}$, which gives the classical path for
imaginary time called the instanton \cite{ tdl,tpc}.

If the early universe is spatially flat, homogeneous and isotropic, its geometry is given by the line-element

$$ dS^2 = dt^2 - a^2(t)[ dx^2 + dy^2 + dz^2 ] \eqno(3.7) $$

In the background geometry of this model, the riccion energy (given by eq.(3.6)) can be written as

$$ E = a^3(t) v \Big\{\frac{1}{2} \Big(\frac{d{\tilde R}}{dt}\Big)^2 + V({\tilde R}) \Big\},  \eqno(3.8) $$
where $v$ is the volume of the 3-dimensional space and $V({\tilde R})$ is the potential,given by eq.(3.5b).

Now, using the transformation $t \to - i \tau$, one obtains that the Ricci scalar $R$ transforms to $-R$. As a result,

$$\tilde R \to - \tilde R \eqno(3.9) $$

and
$$ E = a^3(t) v \Big\{ -\frac{1}{2} \Big(\frac{d{\tilde R}}{dt}\Big)^2 + V({\tilde R}) \Big\}.  \eqno(3.10) $$
Thus, in imaginary time, $\tilde R$ can go from the state $\tilde R = - \frac{M_P}{\sqrt{32{\pi}{\lambda}({\tilde\alpha} + 3{\tilde\beta})}}$ to the state  $\tilde R = + \frac{M_P}{\sqrt{32{\pi}{\lambda}({\tilde\alpha} + 3{\tilde\beta})}}$ with $E=0$. So, setting $E=0$ in eq.(3.8), one obtains the classical trajectory in the $\tilde R - V({\tilde R})$ with imaginary time as

$$\tilde R = \frac{M_P}{4 \sqrt{2{\pi}{\lambda}}} tanh \Big(\frac{M_P \tau}{8\sqrt{\frac{\lambda}{2 \pi}}}\Big), \eqno(3.11)$$
which is an instanton solution. If coupling constants $\lambda$, $\tilde\alpha$ and $\tilde\beta$ satisfy the equation

$$\lambda ({\tilde\alpha} + 3{\tilde\beta})= 1, \eqno(3.12)$$

the action for this trajectory, in the imaginary time system, can be calculated as

\begin{eqnarray*}
S_E &=& v \int^{\infty}_{\infty} {d \tau} a^3(\tau)\Big[ \frac{1}{2} \Big(\frac{d{\tilde R}}{dt}\Big)^2 +  \frac{\lambda^2}{8} \Big({\tilde R}^2 -\frac{M_P^2}{32{\pi \lambda}}\Big)^2 \Big]\\ &=& \frac{v \lambda^2}{4} \int^{\infty}_{\infty} {d \tau} a^3(\tau)\Big({\tilde R}^2 -\frac{M_P^2}{32{\pi \lambda}}\Big)^2 ,
\end{eqnarray*}

$$\eqno(3.13) $$
using eq.(3.10) with $E = 0.$ In what follows, it will be shown that $S_E$ is finite, which is essential for riccion to behave like an instanton. 

The basic equation showing the physical aspect of $\tilde R$, in this case, is eq.(3.4). So,$\tilde R$,given by eq.(3.11), should satisfy it for the sake of consistency. In the background geometry of the early universe,given by eq.(3.7), eq.(3.4) is re-written as

$${\ddot {\tilde R}} + \frac{3 \dot a}{a} {\dot {\tilde R}} = -\frac{\lambda^2}{2} {\tilde R} \Big({\tilde R}^2 - \frac{M_P^2}{32{\pi}{\lambda}}\Big)   \eqno(3.14)$$

Connecting eqs.(3.13) and (3.14), one finds that eq.(3.14) is satisfied by $\tilde R$, given by eq.(3.11),only when

$$ \tau \sqrt{\lambda} \ge \frac{97.6 \sqrt{\pi}}{M_P} = 2.45 \times 10^{-17} {\rm GeV}^{-1}. \eqno(3.15)$$
as $tanh 12.2 = 1$ and $sech 12.2 = 0.$ For $\tau \sqrt{\lambda} < \frac{97.6 \sqrt{\pi}}{M_P} = 2.45 \times 10^{-17} {\rm GeV}^{-1}$, eq.(3.15) yields $\frac{ \ddot a}{a} = 0,$ which satisfies eq.(3.13),only when $\tau = 0.$ So, to be on a safer side, one can take $\tau \sqrt{\lambda} \ge  2.45 \times 10^{-17} {\rm GeV}^{-1}.$ As a result, eq.(3.11) yields

$$\tilde R = \frac{M_P}{4 \sqrt{2{\pi}{\lambda}}}.  \eqno(3.16)$$

With $t = - i \tau, \tilde R$ is obtained from the line-element, given by eq.(3.7) as

$$\tilde R = - 6 \eta \Big[ \frac{ a^{\prime\prime}}{a} + \Big(\frac{ a^{\prime}}{a} \Big)^2 \Big], \eqno(3.17)$$
where prime denotes differentiation with respect to $\tau$. Thus, for $\tau \sqrt{\lambda} \ge  2.45 \times 10^{-17} {\rm GeV}^{-1},$ one obtains from eqs.(3.16) and (3.17)

$$\frac{ a^{\prime\prime}}{a} + \Big(\frac{ a^{\prime}}{a} \Big)^2 = -  \frac{M_P}{24 \eta \sqrt{2{\pi}{\lambda}}}.  \eqno(3.18)$$

This equation yields the solution

$$ a^2(t) = a_0^2 cos \Big[2 \Big(\frac{M_P}{24 \eta \sqrt{2{\pi}{\lambda}}} \Big)^{1/2} (\tau - \tau_0) \Big]  \eqno(3.19a)$$

with

$$ a_0^2 = \Big(\frac{24 \eta \sqrt{2{\pi}{\lambda}}}{M_P} \Big)^{1/2}   .  \eqno(3.19b)$$

Connecting eqs.(3.13) and (3.19), one finds that

$$S_E \le \frac{v \lambda^2}{4} \int_{ - \infty}^{\infty} {d \tau} \Big(\frac{24 \eta \sqrt{2{\pi}{\lambda}}}{M_P} \Big)^{3/4} \Big( \frac{M_P}{4 \sqrt{2{\pi}{\lambda}}}\Big)^4 sech^4\frac{M_P \tau}{8\sqrt{\frac{\lambda}{2 \pi}}}\Big)  $$

$$ = 0.62 \times {\pi}^{-9/8} {\lambda}^{7/8} M_P^{9/4}l^3.$$

Using $\tau = it$ in eq.(3.19a), we get
$$ a(t) = a_0 cosh^{1/2} \Big[2 \Big(\frac{M_P}{24 \eta \sqrt{2{\pi}{\lambda}}} \Big)^{1/2} (t - t_0) \Big]  \eqno(3.20)$$

It is discussed above that eq.(3.14) is valid when $t = |\tau| \ge 2.45 \times
10^{-17} {\rm GeV}^{-1}$, so we can take $t_0 = 2.45 \times 10^{-17} {\rm
  GeV}^{-1} = 1.61 \times 10^{-41} {\rm sec.}$ in eq.(3.20). Thus $t_0$ is
well before the grand unified phase transition which is expected to occur
around $10^{-35} {\rm sec.}$. This phenamenon is called primordial inflation
\cite{je, sk87}.

.
Thus, it is seen $riccion$ behaves as an instanton too and the instanton
solution, so obtained, inspires inflationary scenario which starts much
earlier to GUT phase transition.

\vspace{0.5cm}

 \centerline{\bf 4. Emergence of riccions and riccinos from higher - derivative}

\centerline{\bf higher - dimensional gravity   }

\bigskip

Here, the line-element for (4 + D)-dimensional space-time is taken as
\cite{skp97,sk92}

$$ dS^2 = g_{\mu\nu} dx^{\mu}dx^{\nu} - \rho_1^2 d{\theta}_1^2 - \rho_2^2 d{\theta}_2^2 - \cdots - - \rho_D^2 d{\theta}_D^2 , \eqno(4.1)$$
where $\mu , \nu = 0,1,2,3$ and $y_1 = \rho_1 \theta_1 ,y_2 = \rho_2 \theta_2, \cdots ,y_D = \rho_D \theta_D (0 \le \theta_1, \theta_2, \cdots, \theta_D$ $ \le 2 \pi)$ as $T^D$ is topologically equivalent to product of $D$ copies of circles with different radii $\rho_1, \rho_2, \cdots , \rho_D.$

In this case, the gravitational action for higher-dimensional higher-derivative gravity is taken as \cite{skp97}

$$ S_g = \int{d^4x}{d^Dy}{\sqrt{|g_{4+D}|}}\Big[ 
{\frac{R_{4+D}}{16{\pi}G_{4+D}}}  +
{\tilde\alpha}R^2 + {\tilde\beta}{\Box_{4+D}}R_{4+D}],  \eqno(4.2a)$$
where $G_{4+D} = (2 \pi)^D \rho_1 \rho_2 \cdots \rho_D G, g_{4+D}$ is the determinant of $g_{MN}$ $(M , N = 0,1,2, \cdots ,(D+3), {\tilde\alpha} = \alpha \times[(2 \pi)^D \rho_1 \rho_2 \cdots \rho_D]^{-1}$ and ${\tilde\beta} = \beta \times[(2 \pi)^D \rho_1 \rho_2 \cdots \rho_D]^{-1}$ $\alpha$ and $\beta$ are dimensionless coupling constants in the four-dimensional theory) and

$${\Box_{4+D}} = \frac{1}{\sqrt{| g_{4+D}|}}\frac{\partial}{\partial x^M}
\Big[
\sqrt{| g_{4+D}|} g^{MN} \frac{\partial}{\partial x^N}\Big].
  \eqno(4.2b)$$

The action(4.2) yields gravitational field equations  
$${\frac{1}{16{\pi}G_{(4+D)}}}(R_{MN} -{\frac{1}{2}}g_{MN}R_{(4+D)}) + 
{\tilde\alpha}( 2R_{;MN} - 2 g_{MN}{\Box_{(4+D)}}R_{(4+D)} $$
 $$- {\frac{1}{2}}
g_{MN}R^2_{(4+D)} + 2 R_{(4+D)} R_{MN} )     + {\tilde\beta}
( 2 {\Box_{(4+D)}}R_{MN}  $$ $$- {\frac{1}{2}} g_{MN}{\Box_{(4+D)}}R_{(4+D)})
 = 0 , \eqno(4.3)$$
From the line-element, given by eq.(4.1), $R_{(4+D)} = R_4.$ So, taking trace of eqs.(4.3), one obtains

$${\Box}R + \lambda R^2 + m^2 R = 0, \eqno(4.4a)$$
where $R_4 \equiv R, {\Box_4} \equiv {\Box}.$

$$ m^2 = \frac{(D+2)}{16 \pi G [4 (D+3) \alpha + D \beta]}  \eqno(4.4b)$$
and
$$ \lambda = \frac{D \alpha}{16 \pi G [4 (D+3) \alpha + D \beta]}  \eqno(4.4c)$$
subject to the condition $[4 (D+3) \alpha + D \beta] > 0.$ This condition is imposed to avoid the ghost problem as above. The operator ${\Box_{(4+D)}}$ effectively reduces to ${\Box_4}$ as $R$ is independent of $y$- coordinates.

Thus , from eqs.(4.4),equation for $\tilde R$ is obtained  multiplying it by $\eta$ (which is used for dimensional correction as explained in section 2, as 

$${\Box}{\tilde R} + \lambda R {\tilde R} + m^2 {\tilde R} = 0, \eqno(4.5)$$
as a result of spontaneous compactification. It is interesting to note that
$\lambda = 0$ for $D = 0,$ which corresponds to the case of 4-dimensional
space-time geometry, discussed in sections 2 and 3. So, the term $\lambda R {\tilde R}$ is a manifestation of higher-dimensional geometry in four-dimensional space-time.

In what follows, two cases are discussed (i) when $\lambda = 1/4$ (ii) when $\lambda \ne 1/4$. The first case is possible only when $\beta \ne 0$ i.e. in the presence of ${\Box_{(4+D)}}R_{(4+D)}$ in $S_g$, given by eq.(4.1). In this case, eq.(4.5) looks like

$$ \Big({\Box} + \frac{1}{4} R  + m^2 \Big) {\tilde R} = 0, \eqno(4.6)$$
with $m^2$ given in eq.(4.4a).

One can express the scalar field $\tilde R$ as ${\bar \psi} \psi$ ( where $\psi$ is a Dirac spinor, $\bar \psi = \psi^{\dagger}{\tilde \gamma}^0$ with $\psi^{\dagger}$ as hermitian conjugate and ${\tilde \gamma}^0$ as a Dirac matrix in flat space-time). Physically, it means that the riccion can be envisaged as a composite of spin-1/2 fermion and anti-fermion \cite{pdb}. For this, it is necessary for eq.(4.6) to decouple into Dirac equations for $\psi$ and $\bar \psi$, which is possible under a condition derived below.

Equation (4.6) can be written in the matrix form as
 
$$ AB = Z \eqno(4.7a)$$

with $$A =\Big({\Box} + \frac{1}{4} R  + m^2 \Big)I,\eqno(4.7b)$$  $$B = \tilde R I\eqno(4.7c)$$ and $Z$ is the zero matrix which is a $4 x 4$ matrix with all elements $0$. $I$ is the $4 x 4$ unit matrix.

The operator $\Big({\Box} + \frac{1}{4} R  \Big)I$ can be written as
$\gamma^{\mu}\nabla_{\mu}\gamma^{\nu}\nabla_{\nu}$ \cite{ilb, cdew, ndb}. So, the matrix operator $A$ looks like

$$A  = \Big(\gamma^{\mu}\nabla_{\mu}\gamma^{\nu}\nabla_{\nu}+ m^2I\Big) \eqno(4.7d)$$

Here $\gamma^{\mu}$ are Dirac matrices in curved space-time related to $\tilde \gamma^a$, Dirac matrices in flat space-time as 

$$\gamma^{\mu} = e^{\mu}_a \tilde \gamma^a .  \eqno(4.8)$$

$\gamma^{\mu}$ and $\tilde \gamma^a$ satisfy anticommutation rules

$$ \{\gamma^{\mu} , \gamma^{\nu} \} = 2 g^{\mu\nu}  \eqno(4.9a)$$
and

$$ \{\tilde \gamma^a , \tilde \gamma^b \} = 2 \eta^{ab},  \eqno(4.9b)$$
where $\eta^{ab}$ $(a,b = 0,1,2,3)$ are components of Minkowskian metric tensor. $e^{\mu}_a$, in eq.(4.8) are tetrad components defined as

$$ g^{\mu\nu} = e^{\mu}_a e^{\nu}_b \eta^{ab}.  \eqno(4.10)$$
The operator $\nabla_{\mu}$ (covariant derivative in curved space - time), given above, is defined as

$$\nabla_{\mu} = \partial_{\mu} - \Gamma_{\mu},   \eqno(4.11a)$$
where

$$\Gamma_{\mu} = - \frac{1}{4}(\partial_{\mu}e^{\rho}_a + \Gamma^{\rho}_{\sigma\mu}e^{\sigma}_a ) g_{\nu\rho}e^{\nu}_b \tilde \gamma^b \tilde \gamma^a  \eqno(4.11b)$$
with $\Gamma^{\rho}_{\sigma\mu}$ being Riemann-Christoffel's symbols.

Using $\tilde R = {\bar \psi}\psi$ in eq.(4.7a) and algebra of Dirac matrices, one obtains

$$ {\bar \psi}(\gamma^{\mu}\nabla_{\mu} - i m)(\gamma^{\nu}\nabla_{\nu} + i m)\psi = 0 \eqno(4.12a)$$
provided that

$$\nabla_{\mu}\nabla_{\nu}{\bar \psi} \gamma^{\mu}\gamma^{\nu}\psi = 0.   \eqno(4.12b)$$
This condition can also be expressed as

$$\nabla_{\mu}\nabla_{\nu}{\bar \psi}_L \gamma^{\mu}\gamma^{\nu}\psi_R = - \nabla_{\mu}\nabla_{\nu}{\bar \psi}_R \gamma^{\mu}\gamma^{\nu}\psi_L, \eqno(4.13)$$
where $\psi_{L,R} = [(1 \pm {\tilde \gamma}^5)/2]\psi$ and ${\bar \psi}_{L,R} = {\bar \psi}[(1 \mp {\tilde \gamma}^5)/2]$ with ${\tilde \gamma}^5 = i {\tilde \gamma}^0 {\tilde \gamma}^1 {\tilde \gamma}^2 {\tilde \gamma}^3.$ The equation (4.12a) implies the Dirac equation for $\psi$ as

$$(\gamma^{\mu}\nabla_{\mu} + i m)\psi = 0 \eqno(4.14a)$$
and its complex conjugate

$$ {\bar \psi}(\gamma^{\mu}\nabla_{\mu} - i m) = 0. \eqno(4.14b)$$

This shows that eq.(4.7a) implies eqs.(4.14) under the condition provided by eq.(4.13),which shows the breaking of left- right symmetry. Physically, it means that a riccion (with $\lambda = 1/4$) can decouple into a fermion and anti-fermion pair if left - right symmetry breaks \cite{skp97}. In other words, riccion can decompose as

$$ {\tilde R} \to \psi + {\tilde \psi}, \eqno(4.15a)$$
provided that parity is violated. Eqs.(4.14) indicate that spin-1/2 particles, so obtained, are electrically neutral showing that $\psi$ and ${\tilde \psi}$ will be invariant under charge conjugation ( C-symmetry). So, CPT-invariance implies that the decoupling of riccion to  pair of riccinos  and anti-riccinos (fermionic and anti-fermionic counterpart of a riccion) will not respect T-symmetry (time-reversal)\cite{lhr}. It means that riccinos  and anti-riccinos can not combine to form riccions i.e. it is not possible to have back reaction

$$  \psi + {\tilde \psi} \to  {\tilde R}, \eqno(4.15b)$$

Here CP violation is taking place , so it may explain excess of baryons over
anti-baryons if riccions decay to observed baryons.
 
If $\lambda \ne 1/4$, the operator $({\Box} + \lambda R)$ can not be expressed
as $\gamma^{\mu}\nabla_{\mu} \gamma^{\nu}\nabla_{\nu}$, so eq.(4.6) can not be
written as (4.12a). So, in this case, decoupling of a riccion to a riccino and
an anti-riccino is not possible.

According to the above theory, it is tempting to speculate that our physical
universe might have emerged from higher-dimensional geometry in the extreme
past through decay riccions to riccions and anti-riccinos, when parity is
violated.

\vspace{0.5cm}

 \centerline{\bf 5. One - loop renormalization of riccion obtained}
 
 \centerline{\bf from the higher-dimensional space-time   }

\bigskip

In what follows,lagrangian density of the riccion with a quartic self interaction potential is obtained from higher-derivative gravitational action in (4+D)-dimensional space-time.

The topology of the space-time is taken as $M^4 \bigotimes T^D$ ($M^4$ as a 4-dimensional space-time and $T^D$ as D-dimensional torus as an extra-dimensional compact manifold) with the distance function  defined as
$$ d S^2 = g_{\mu\nu} d x^{\mu} d x^{\nu} - \rho_1^2 d\theta_1^2 -
\rho_2^2 d\theta_2^2 - \cdots - - \rho_D^2 d\theta_D^2 ,  \eqno(5.1)$$
where  $g_{\mu\nu} (\mu ,\nu = 0, 1, 2, 3)$ are components of the
metric tensor, $\rho_i(i = 1,2, \cdots ,D)$ are radii of circle components
of $T^D (D \ne 2)$ and $0 \le \theta_1, \theta_2, \cdots , \theta_D$  $ \le
2\pi.$ As usual, the space-time manifold is taken to be
$C^{\infty}$-connected, Hausdorff and paracompact without boundary \cite{sk99,
  sks99,sk00, sk03}.

Here action for the higher-derivative gravity is taken as \cite{sk00}
$$ S_g^{(4+D)} = \int {d^4 x} {d^D y}  \sqrt{- g_{(4+D)}} \quad
\Big[\frac{R_{(4+D)}}{16 \pi
G_{(4+D)}} + {\alpha_{(4+D)}} R_{(4+D)}^2 + \gamma_{(4+D)} ( R_{(4+D)}^3   $$ 
$$- \frac{6(D+3)}{(D-2)}{\Box}_{(D+4)}R^2_{(D+4)}) 
\Big], \eqno(5.2a)$$ 
where  $G_{(4+D)} = G V_D,\quad \alpha_{(4+D)} = \alpha V_D^{-1},\quad
\gamma_{(4+D)} = \frac{\eta^2}{3! (D-2)} V_D^{-1},$
$$ V_D = (2\pi)^D \rho_1 \rho_2 \cdots \rho_D \quad ( D \ne 2), \quad R_D = 0 $$and $g_{(4+D)}$ is the determinant of the
metric tensor $g_{MN} (M,N = 0,1,2, \cdots,$   $ (4+D).$ Here $G_{(4+D)}$ is
the (4+D)-dim. gravitational constant and $\alpha_{(4+D)}$ as well as
$\gamma_{(4+D)}$ are coupling constants. $\alpha$ is a dimensionless
coupling constant, $R_{(4+D)}$ is the (4+D)-dim. Ricci scalar and $G$ is
the 4-dim. gravitational constant.

From the action (5.2a), gravitational field equations are obtained as
$$(16 \pi G_{(4+D)})^{-1} (R_{MN} - {1 \over 2} g_{MN} R_{(4+D)} )
+ {\alpha_{(4+D)}} H^{(1)}_{MN} + {\gamma_{(4+D)}} H^{(2)}_{MN}
\eqno(5.2b)$$ 
where
$$ H^{(1)}_{MN} = 2 R_{; MN} - 2 g_{MN} {\Box}_{(4+D)} R_{(4+D)} - {1
\over 2} g_{MN} R^2_{(4+D)} + 2 R_{(4+D)} R_{MN},
\eqno(5.2c)$$  and
$$ H^{(2)}_{MN} = 3 R^2_{; MN} - 3 g_{MN} {\Box}_{(4+D)} R^2_{(4+D)} - \frac{6(D+3)}{(D-2)}\{- { 1
\over 2} g_{MN}{\Box}_{(4+D)} R^2_{(4+D)}$$  $$+ 2{\Box}_{(4+D)}R_{(D+4)}R_{MN} + R^2_{; MN}\} - {1
\over 2} g_{MN} R^3_{(4+D)} + 3 R^2_{(4+D)} R_{MN}  \eqno(5.2d) $$
with semi-colon (;) denoting curved space covariant derivative and
$${\Box}_{(4+D)} = {1\over \sqrt{-g_{(4+D)}}}{\frac{\partial}{\partial
x^M}}\Big( \sqrt{-g_{(4+D)}}\quad g^{MN} \frac{\partial}{\partial x^N}\Big)$$

Trace of these field equations is obtained as \cite{sk00}
$$\Big[-\frac{D+2}{32 \pi G_{(4+D)}}\Big] R_{(4+D)}  -   {\alpha_{(4+D)}} [
2(D+3) {\Box}_{(4+D)} R_{(4+D)}  +  {1 \over2} D R^2_{(4+D)} ] $$
$$ - \gamma_{(4+D)} [  +  {1 \over2} (D - 2)
R^3_{(4+D)} ]  = 0 .   \eqno(5.2e)$$

Using the method of earlier sections, the equation for riccion is obtained as

$$[{\Box} + {1 \over2} \xi R + m^2 + \frac{\lambda}{3!} {\tilde R}^2]
{\tilde R} = 0  \eqno(5.3a)$$
with 
\begin{eqnarray*}
 \xi & = & \frac{D}{2(D+3)}\\ m^2  & = & \frac{(D + 2)}{64 \pi G \alpha
(D+3)} \\ \lambda  & = & \frac{1}{4(D+3)\alpha},
 \end{eqnarray*} 
 $$   \eqno(5.3b,c,d)$$
 where $\alpha > 0$ to avoid the ghost problem \cite{aas}. As done in earlier sections, the corresponding action for ${\tilde R}$ is obtained as
$$ S_{\tilde R} = \int {d^4 x}  \sqrt{- g}
\Big[{1 \over2} g^{\mu\nu} \partial_{\mu}{\tilde R}
\partial_{\nu}{\tilde R} 
-  \Big({1 \over 3! \eta} \xi {\tilde R}^3 + {1 \over 2} m^2 {\tilde R}^2 +
\frac{\lambda}{4!} {\tilde R}^4 \Big) \Big].  \eqno(5.4)$$

\bigskip

\noindent \underline{5(a).One-loop quantum correction and renormalization}

The $S_{\tilde R} $ with the lagrangian density, given by eq.(5.4), can 
be expanded around the classical minimum ${\tilde R}_0$ in powers of
quantum fluctuation 
${\tilde R}_q = {\tilde R} - {\tilde R}_0 $ as
$$ S_{\tilde R} =  S_{\tilde R}^{(0)} + S_{\tilde R}^{(1)} +  S_{\tilde
R}^{(2)} + \cdots ,$$
where
\begin{eqnarray*}
S_{\tilde R}^{(0)} & = &  \int {d^4 x}  \sqrt{- g}
\Big[{1 \over2} g^{\mu\nu} \partial_{\mu}{\tilde R}_0
\partial_{\nu}{\tilde R}_0
-  \Big({1 \over 3! \eta} \xi {\tilde R}^3_0 + {1 \over 2} m^2 {\tilde R}^2_0 +
\frac{\lambda}{4!} {\tilde R}^4_0 \Big] \\ S_{\tilde R}^{(2)} & = &  \int
{d^4 x}  \sqrt{- g} {\tilde R}_q[{\Box} +  \xi
R + m^2 + \frac{\lambda}{2!} {\tilde R}^2_0] {\tilde R}_q
\end{eqnarray*}
and
$$ S_{\tilde R}^{(1)} = 0$$ 
as usual, because this term contains the classical equation.

The effective action of the theory is expanded in powers of $\hbar$ (with
$\hbar = 1$) as 
$$ \Gamma({\tilde R}) = S_{\tilde R} + \Gamma^{(1)} + \Gamma^{\prime}$$
with one-loop correction given as 
$$ \Gamma^{(1)} = {i \over 2} ln Det (D/{\mu}^2),   \eqno(5.5a)$$
where
$$ D \equiv \frac{{\delta}^2 S_{\tilde R}}{{\delta}{\tilde R}^2}
\Big|_{{\tilde R}={\tilde R}_0} = {\Box} +  \xi
R + m^2 + \frac{\lambda}{2!} {\tilde R}^2_0   \eqno(5.5b)$$
and  $\Gamma^{\prime}$  is  a term for higher-loop quantum corrections. In
eq.(5.5), $\mu$ is a mass parameter to keep $\Gamma^{(1)}$ dimensionless.

To evaluate $\Gamma^{(1)}$,  the  operator regularization method \cite{rbm} is
used. Up to 
adiabatic order 4 (potentially divergent terms are expected  up to this
order only in 
a 4-dim. theory), one-loop correction is obtained as
\begin{eqnarray*}
\Gamma^{(1)} & = & (16 {\pi}^2 )^{-1} \frac{d}{d s} \Big[ \int {d^4 x}
\sqrt{-g (x)} 
\Big(\frac{{\tilde M}^2}{{\mu}^2}\Big)^{-s} \Big\{\frac{{\tilde M}^4}{(s -
2)(s - 1)} \\ && + \frac{{\tilde M}^2}{(s - 1)}\Big(\frac{1}{6} - \xi
\Big) R + \Big[ {1 \over 6} \Big(\frac{1}{5} - \xi \Big){\Box} R + {1
\over 180} 
R^{\mu\nu\alpha\beta} R_{\mu\nu\alpha\beta}\\ &&
-{1 \over 180} R^{\mu\nu} R_{\mu\nu} + {1 \over 2}\Big(\frac{1}{6} - \xi
\Big)^2 R^2
\Big]\Big\}\Big] \Big|_{s=0},
\end{eqnarray*}
$$   \eqno(5.6a)$$
where $$ {\tilde M}^2 = m^2  +  (\lambda/2) {\tilde R}^2_0 .  \eqno(5.6b)$$
Here it is important to note that matter as well as geometrical both
aspects of the Ricci scalar are used in eq.(5.6). The matter aspect is
manifested by ${\tilde R}$ and the geometrical aspect by $R,$ Ricci tensor
components  $R_{\mu\nu}$ and curvature tensor components
$R_{\mu\nu\alpha\beta}$ as it is mentioned above also.

After some manipulations, the lagrangian density in ${\Gamma}^{(1)}$ is
obtained as 
\begin{eqnarray*}
\Gamma^{(1)} & = & (16 {\pi}^2 )^{-1}  \Big[ (m^2  +  (\lambda/2) {\tilde
R}^2_0 )^2 
\Big\{{3 \over 4} - {1 \over 2} ln \Big(\frac{m^2  +  (\lambda/2) {\tilde
R}^2_0 
}{{\mu}^2} \Big)\Big\} \\ &&  - \Big({1 \over 6} - \xi \big)  R ( m^2  +
(\lambda/2) {\tilde 
R}^2_0 ) \Big\{1 - ln \Big(\frac{m^2  +  (\lambda/2) {\tilde
R}^2_0}{{\mu}^2} \Big) 
\Big\} \\ &&- ln \Big(\frac{m^2  +  (\lambda/2) {\tilde R}^2_0}{{\mu}^2} \Big)
\Big\{ {1 \over 6}\Big(\frac{1}{5} - \xi \Big) {\Box}R + \frac{1}{180}
R^{\mu\nu\alpha\beta} R_{\mu\nu\alpha\beta} \\&&
-{1 \over 180} R^{\mu\nu} R_{\mu\nu} + {1 \over
2}\Big(\frac{1}{6} - \xi \Big)^2 R^2
\Big\}\Big] .
\end{eqnarray*}
$$   \eqno(5.7)$$

Now the renormalized form of lagrangian density can be written as
\begin{eqnarray*}
L_{\rm ren} &=& {1 \over 2} g^{\mu\nu} {\partial}_{\mu} {\tilde R}_0
{\partial}_{\nu} {\tilde R}_0  - \frac{\xi}{3! \eta} {\tilde R}_0^3- {1
\over 2}m^2 {\tilde R}^2_0  - 
{\lambda \over 4!} {\tilde R}^4_0 + \Lambda \\ &&  + {\epsilon}_0 R + {1
\over 2} 
{\epsilon}_1 R^2 + {\epsilon}_2 R^{\mu\nu} R_{\mu\nu} + {\epsilon}_3
R^{\mu\nu\alpha\beta} R_{\mu\nu\alpha\beta} \\&& + {\epsilon}_4 {\Box} R +
\Gamma^{(1)} + 
L_{\rm ct}
\end{eqnarray*}
$$  \eqno(5.8a)$$
with bare coupling constants $\lambda_i \equiv ( m^2, \lambda, \Lambda, \xi,
\epsilon_0, \epsilon_1, \epsilon_2, \epsilon_3, \epsilon_4 ), \Gamma^{(1)} $ 
given by eq.(5.7) and $L_{\rm ct}$ given as
\begin{eqnarray*}
L_{\rm ct} &=&  - {1 \over 2} \delta\xi R {\tilde R}^2_0 - {1 \over 2}
\delta m^2 {\tilde R}^2_0  -
{\delta\lambda \over 4!} {\tilde R}^4_0 + \delta\Lambda  +
{\delta\epsilon}_0 R + {1 
\over 2}{\delta\epsilon}_1 R^2 \\ && + {\delta\epsilon}_2 R^{\mu\nu}
R_{\mu\nu} + 
{\delta\epsilon}_3
R^{\mu\nu\alpha\beta} R_{\mu\nu\alpha\beta} + {\delta\epsilon}_4 {\Box} R .
\end{eqnarray*}
$$  \eqno(5.8b)$$
In eq.(5.8b), $\delta\lambda_i \equiv (\delta m^2, \delta \lambda,
\delta\Lambda, \delta\xi,
\delta\epsilon_0, \delta\epsilon_1, \delta\epsilon_2, \delta\epsilon_3, \delta 
\epsilon_4 )$ are counter-terms, which are calculated using the following
renormalization conditions \cite{blh}
\begin{eqnarray*}
\Lambda & = &  L_{\rm ren} |_{{\tilde R}_0 = {\tilde R}_{(0)0} , R=0}\\ 
\lambda & = &  - \frac{{\partial}^4}{\partial {\tilde R}^4_0} L_{\rm ren}
\Big|_{{\tilde R}_0 = {\tilde R}_{(0)1} , R=0} \\
 m^2 & = &  - \frac{{\partial}^2}{\partial {\tilde R}^2_0} L_{\rm ren}
\Big|_{{\tilde R}_0 = 0 , R=0} \\
\xi & = &  - \eta \frac{{\partial}^3}{{\partial R}{\partial {\tilde
R}^2_0}} L_{\rm ren} 
\Big|_{{\tilde R}_0 = {\tilde R}_{(0)2} , R=0} \\
\epsilon_0 & = &  \frac{\partial}{\partial R} L_{\rm ren}
\Big|_{{\tilde R}_0 = 0 , R=0} \\
\epsilon_1 & = &  \frac{{\partial}^2}{\partial R^2} L_{\rm ren}
\Big|_{{\tilde R}_0 = 0 , R = R_5} \\
\epsilon_2 & = &   \frac{\partial}{\partial( R^{\mu\nu}R_{\mu\nu})} L_{\rm ren}
\Big|_{{\tilde R}_0 = 0 , R = R_6} \\
\epsilon_3 & = &   \frac{\partial}{\partial(R^{\mu\nu\alpha\beta
}R_{\mu\nu\alpha\beta})} L_{\rm ren}
\Big|_{{\tilde R}_0 = 0 , R = R_7} \\
\epsilon_4 & = &  \frac{\partial}{\partial ({\Box} R)} L_{\rm
ren}\Big|_{{\tilde R}_0 = 0 , R = R_8} .
\end{eqnarray*}
$$\eqno(5.9a,b,c,d,e,f,g,h,i)$$
 As ${\tilde R} = {\eta} R,$ so when $R = 0, {\tilde R}_{(0)0} =
{\tilde R}_{(0)1} = 
{\tilde R}_{(0)2} = 0 \quad {\rm  and}\quad  R_5 = R_6 = R_7 = R_8 = 0
\quad {\rm when}\quad {\tilde R}_0 = 0 .$

Eqs.(5.8) and (5.9) yield counter-terms as
\begin{eqnarray*}
16 {\pi}^2 \delta\Lambda & = & {m^4 \over 2} ln (m^2/{\mu}^2) - \frac{3}{4}m^4\\
16 {\pi}^2 \delta\lambda & = & - 3 \lambda^2 ln (m^2/{\mu}^2)\\
16 {\pi}^2 \delta m^2 & = & - \lambda m^2 ln (m^2/{\mu}^2)\\
16 {\pi}^2 \delta \xi& = & - 3 \lambda \Big( \xi - \frac{1}{6} \Big) ln
(m^2/{\mu}^2)\\ 
16 {\pi}^2 \delta \epsilon_0& = &   m^2 \Big( \xi - \frac{1}{6} \Big) ln
(m^2/{\mu}^2)\\ 
16 {\pi}^2 \delta \epsilon_1& = &  \Big( \xi - \frac{1}{6} \Big)^2 ln
(m^2/{\mu}^2)\\ 
16 {\pi}^2 \delta \epsilon_2& = &  - {1 \over 180} ln (m^2/{\mu}^2)\\
16 {\pi}^2 \delta \epsilon_3& = &  {1 \over 180} ln (m^2/{\mu}^2)\\
16 {\pi}^2 \delta \epsilon_4& = &  {1 \over 6}\Big( \frac{1}{5} - \xi
\Big) ln (m^2/{\mu}^2)\\ 
\end{eqnarray*}
$$  \eqno(5.10a,b,c,d,e,f,g,h,i)$$

\bigskip

\noindent \underline{(b) Renormalization group equations and their solutions}

\bigskip

The effective renormalized lagrangian can be improved further by solving
renormalization group equations for coupling constants
$\lambda_{i(R)}\quad({\rm suffix}\quad R\quad$ stands for renormalization,
which is dropped onwards). For
this purpose one-loop $\beta$-functions, defined by the equation \cite{ilb,
  blh,ee}
$$ \beta_{\lambda_i} = \mu \frac{d}{d\mu}(\lambda_{i} +
\delta\lambda_i)\Big|_{\lambda_i}   \eqno(5.11)$$
with counter-terms $\delta\lambda_i$ from eqs.(5.10), are obtained 
 as
\begin{eqnarray*}
\beta_{\Lambda} &=& - \frac{m^4}{16 {\pi}^2}\\
\beta_{\lambda} &=&  \frac{3 {\lambda}^2}{8 {\pi}^2}\\
\beta_{m^2} &=& \frac{{\lambda} m^2}{8 {\pi}^2}\\
\beta_{\xi} &=&  \frac{3\lambda\Big( \xi - \frac{1}{6} \Big)}{8 {\pi}^2}\\
\beta_{\epsilon_0} &=&  - \frac{m^2 \Big( \xi - \frac{1}{6} \Big)}{8
{\pi}^2}\\ 
\beta_{\epsilon_1} &=& - \frac{\Big( \xi - \frac{1}{6} \Big)^2}{8 {\pi}^2}\\
\beta_{\epsilon_2} &=&  \frac{1}{1440 {\pi}^2}\\
\beta_{\epsilon_3} &=& - \frac{1}{1440 {\pi}^2}\\
\beta_{\epsilon_4} &=& - \frac{1}{48 {\pi}^2}\Big( \frac{1}{5} - \xi \Big)
\end{eqnarray*}
$$  \eqno(5.12a,b,c,d,e,f,g,h,i)$$
using the fact that $ \mu \frac{d}{d\mu}{\lambda_i} = 0 $ for bare
coupling constants $\lambda_i.$

The renormalization group equations are given as
$$ \frac{d {\lambda}_i}{dt} = \beta_{\lambda_i},  \eqno(5.13)$$
where $ t = {1 \over 2} ln (m^2_c/{\mu}^2)$ with $\mu$ being a mass
parameter  and $m_c$ being a reference mass scale such that $\mu \ge m_c.$
Using $\beta$-functions for different coupling constants given by
eqs.(5.12), solutions of differential equations (5.13) are derived as
\begin{eqnarray*}
\Lambda & = & \Lambda_0 + \frac{m^4_0}{2\lambda_0} \Big[\Big(1 - \frac{3
\lambda_0 t}{8 \pi^2}\Big)^{1/3} - 1 \Big] \\
\lambda & = & \lambda_0 \Big[1 - \frac{3 \lambda_0 t}{8 \pi^2} \Big]^{-1}\\
 m^2 & = & m^2_0 \Big[1 - \frac{3 \lambda_0 t}{8 \pi^2} \Big]^{-1/3}\\
\xi & = &  {1 \over 6}  + \Big(\xi_0 - \frac{1}{6} \Big) \Big[1 - \frac{3
\lambda_0 t}{8 \pi^2} \Big]^{-1}\\
\epsilon_0 & = & \epsilon_{00} + \frac{m^2_0 \Big(\xi_0 - \frac{1}{6}
\Big) }{\lambda_0} 
\Big[1 - \Big(1 - \frac{3 \lambda_0 t}{8 \pi^2} \Big)^{-1/3}  \Big]\\
\epsilon_1 & = & \epsilon_{10} +   \frac{ \Big(\xi_0 - \frac{1}{6}
\Big)^2 }{\lambda_0} 
\Big[1 - \Big(1 - \frac{3 \lambda_0 t}{8 \pi^2} \Big)^{-1}  \Big]\\
\epsilon_2 & = & \epsilon_{20} + \frac{ t}{1440 \pi^2}+ \\
\epsilon_3 & = & \epsilon_{30} - \frac{ t}{1440 \pi^2} \\
\epsilon_4 & = & \epsilon_{40} - \frac{ t}{1440 \pi^2} - \frac{ \Big(\xi_0
- \frac{1}{6} \Big) }{18\lambda_0} ln \Big(1 - \frac{3 \lambda_0 t}{8 \pi^2}
\Big) , 
\end{eqnarray*}
$$  \eqno(5.14a,b,c,d,e,f,g,h,i)$$
where $\lambda_{i0} = \lambda_i (t=0)\quad {\rm and}\quad  t = 0 \quad {\rm at}\quad
 \mu
= m_c $ according to definition of $t$ given above.

These results show that as $\mu \to \infty ( t \to  - \infty )\quad
\lambda \to  0 \quad {\rm and}\quad
m^2 \to 0$ and $\xi \to \frac{1}{6}.$ Thus it follows from these
expressions that in the limit $\mu \to \infty,$ theory is asymptotically free.

Further it is assumed that $D = 0$ at energy mass scale $\mu = m_c.$ Now
recalling the definition of $\xi$ from eq.(5.3b), one obtains $\xi_0 = 0$
and 

$$ D = \frac{3 \Big[ 1 - (1 - \frac{3 \lambda_0 t}{8 \pi^2}\Big)^{-1}
\Big]}{\Big[ 2 + (1 - \frac{3 \lambda_0 t}{8 \pi^2}\Big)^{-1} \big]}.
\eqno(5.15)$$ 

It is interesting to see from this result that $D$ increases with
increasing energy mass scale $\mu$ and it is equal to 1 when $\mu =
1.8\times 10^{34} m_c$ provided that $\lambda_0 = 1.$ Moreover, $D$ is not
necessarily an integer but a real number. This result suggests that
dimension of the space-time will be equal to 4 at $\mu = m_c$ and will
increase continuously with increasing energy mass scale. Non-integer
values of the dimension indicate that the space-time above the energy mass
scale $\mu = m_c$ should be fractal \cite{ln89, ln93}, as fractal dimensions need
not be integers like topological dimensions.

Eqs.(5.3d) and (5.14b) yield
$$ \alpha = {1 \over 12} \Big[ 1 - \frac{t}{4 \pi^2} \Big]  \eqno(5.16)$$

and $\alpha_0 = \frac{1}{12}, \quad{\rm if} \lambda_0 = 1.$

Recalling the definition of $m^2$ from eq.(2.8c) and using eqs.(5.15)-(5.16)
with $\lambda_0 = 1,$ one obtains

$$ G = G_N  \frac{\Big[ 7 \Big(1 - \frac{3 t}{8 \pi^2} \Big) - 1 \Big]}{ 6
\Big[1 - \frac{t}{4 \pi^2} \Big] \Big[1 - \frac{3 t}{8 \pi^2}
\Big]^{2/3}}, \eqno(5.17)$$
where  $G_0 = G_N (G_N$ is the Newtonian gravitational constant). Now from
eq.(5.3c)

$$ m_0^2 = \frac{1}{8 \pi G_N} = \frac{M_P^2}{8 \pi}  \eqno(5.18)$$
as $D = 0$ at $t = 0 \quad{\rm or} \mu = m_c \quad {\rm and}\quad G_N =
M_P^{-2} (M_P$ is the Planck's constant). Thus eqs.(5.14a) and (5.18) imply
$$ \Lambda - \Lambda_0 = \frac{M_P^4}{2 (8 \pi)^2} \Big[ \Big(1 - \frac{3
t}{8 \pi^2} \Big)^{1/3} - 1 \Big].   \eqno(5.19)$$

The equation (5.19) implies that vacuum energy density $\Lambda$ increases with
increasing energy mass scale $\mu$ ( or decreasing $t$ ). Taking 
$\lambda_0 = 1$, it
is obtained that $(\Lambda - \Lambda_0) \to \infty 
\quad{\rm
when}\quad \mu \to \infty ( t \to - \infty ).$ In the following table  $( \Lambda -
\Lambda_0 )$ is exhibited  for different values of $\mu.$

\centerline{\bf Table no.1}

\bigskip

$(\Lambda - \Lambda_0), G/G_N$ $and$ $ dimension$ $ of$ $ the$ $ space-time$ $(4+D)$ $ are $
 $tabulated $  $ below $  $ against $ $\mu/m_c$ $ with $ $m_c = 1.76\times 10^{16}{\rm GeV}$
 $taking $ $\lambda_0 = 1.$

\bigskip

$$
\begin{array}{llll}
\hline
\mu/m_c  & (\Lambda - \Lambda_0)  &  G/G_N   &  (4+D) \\

         &   in \quad{\rm GeV}^4  & \omit & \omit    \\
\hline

1        &     0   &  1 & 4 \\

1+  10^{-6} &  9.58 \times 10^{64} & 1 & 4 + 3.6\times 10^{-8} \\

1+ 2\times 10^{-6} &  2.04 \times 10^{65} & 1 & 4 + 7.7\times 10^{-8} \\

1+ 3\times 10^{-6} &  2.99 \times 10^{65} & 1 & 4 + 1.1\times 10^{-7} \\

1+ 4\times 10^{-6} &  4.07 \times 10^{65} & 1 & 4 + 1.5\times 10^{-7} \\
 
 1+  10^{-5} &  1.01 \times 10^{66} & 0.9999999 & 4 + 3.8\times 10^{-7} \\
 
  1+  10^{-4} &  1.01 \times 10^{67} & 0.9999998 & 4 + 3.8\times 10^{-6} \\
  
   1+  10^{-3} &  1.004 \times 10^{68} & 0.9999993 & 4 + 3.8\times 10^{-5} \\
   
    1+  10^{-2} &  9.995 \times 10^{68} & 0.9999389 & 4 + 3.78\times 10^{-4} \\
    
1.1 &  9.56 \times 10^{69} & 0.9993944 & 4 + 3.6\times 10^{-3} \\

2 &  6.903 \times 10^{70} & 0.9995533 & 4 + 2.6\times 10^{-2} \\

10 &  2.23 \times 10^{71} & 0.9846903 & 4 + 8.2\times 10^{-2} \\

10^5 &  1.02 \times 10^{72} & 0.9199941 & 4.339 \\

10^{10} &  1.998 \times 10^{72} & 0.825 & 4.587 \\

10^{20} &  3.3 \times 10^{72} & 0.705 & 4.826 \\

10^{30} &  4.35 \times 10^{72} & 0.618 & 4.966 \\

\infty & \infty & 0  & 5.5 \\
\hline
\end{array}
$$

Here, it is interesting 
to note
that when $\mu$ comes down from infinity to $1.000001 m_c$ ,
$(\Lambda - \Lambda_0)$ decreases slowly. But when it comes down from 
$ 1.000001 m_c$ to $ m_c,$ there is a sudden
drop in the value of $( \Lambda - \Lambda_0 )$
from $9.58 \times 10^{64} {\rm GeV}^4$ to zero. It means that all of a sudden,
a huge amount of energy ( with density equal to $9.58 \times 10^{64} {\rm GeV}
^4$ ) is
released at $\mu = 1.000001 m_c,$ which is sufficient to heat the
universe up to a temperature $1.76\times 10^{16} {\rm GeV}.$ It corresponds to 
the
energy scale $1.76\times 10^{16} {\rm GeV}$ in natural units ( as Boltzmann
constant
$\kappa_B = 1$ in these units ). Sudden release of energy indicates a phase
transition at $1.76\times 10^{16} {\rm GeV}.$  According to the standard model
of
grand unified theories, strong and electro-weak interactions unite at $10^{15}
{\rm
GeV}.$ But gravity maintains its identity different from these interactions 
up to this
energy scale also. So unification of these interactions are expected above 
$10^{15}
{\rm GeV}$ \cite{pdb}. Thus symmetry breaking due to phase transition at $1.76\times 
10^{16}{\rm GeV}$ ( discussed above ) is expected to be the energy mass
scale where gravity decouples from strong-electroweak interactions. In
other words, these results  suggest that unification of gravity with
strong-electroweak interactions should take  
place at 
$1.76\times 10^{16} {\rm GeV}.$ It implies that
$$ m_c \simeq 1.76\times 10^{16} {\rm GeV} .  \eqno(5.20)$$

As discussed in the preceding section, another interesting result is given
by eq.(5.15) which suggests that above the energy mass scale $\mu = 1.000001
m_c \simeq 1.76\times 10^{16} {\rm GeV},$ space-time should be fractal
. Moreover, according to this result, dimension of the space-time
increases from 4 to 5.5 with the increasing mass scale.

The equation (5.17) implies that the gravitational constant $G$ decreases
with increasing $\mu$ with $G = G_N$ at $\mu = 1.76\times 10^{16} {\rm GeV}.$

Eqs.(5.14) also show that $\alpha, \epsilon_0,\epsilon_1,
\epsilon_2,\epsilon_3 $ and $\epsilon_4$ increase with increasing $\mu$
showing that higher- derivative terms grow stronger with increasing energy
mass scale.

 \vspace{0.5cm}
\centerline{\bf 6. Cosmological models inspired by dual role of the Ricci Scalar}
 \centerline{\bf and particle production}

\bigskip
 
 \centerline{\bf 6.1. Homogeneous as well as inhomogeneous} 

\centerline{\bf models of the early universe   }

\smallskip

Here, the gravitational action with thermal 
correction is taken as \cite{sk98,sks99}
$$ S_g = \int{d^4x}{\sqrt{-g}}\Big[ -
{\frac{R}{16{\pi}G}} + {\frac{1}{2}}{\lambda}T^2({\tilde\alpha} +
3{\tilde \beta})R + {\tilde\alpha} R_{mu\nu} R^{\mu\nu} +$$ $$
{\tilde\beta}R^2 - 2{\lambda}{\eta}^2({\tilde\alpha} +
3{\tilde\beta})(R^3 + 9 {\Box}R^2)\Big],  \eqno(6.1.1)$$
where   ${\tilde\alpha},{\tilde\beta}$ and ${\lambda}$ are dimensionless
coupling constants. Here $T$ is the temperature as well as 
${\eta}$ is another constant with 
length dimension used for dimensional correction \cite{skp93,skps93, skp96,
  skp97, sk98, sk99, sks99, sk00, sk03},as given in earlier sections. 

 $S_g$ ( given by eq.(6.1.1)) yields gravitational field equations  
$$\Big[{\frac{1}{2}}{\lambda}T^2({\tilde\alpha} + 3{\tilde\beta}) -
{\frac{1}{16{\pi}G}}\Big](R_{\mu\nu} -{\frac{1}{2}}g_{\mu\nu}R) + 
{\tilde\alpha}( 2R^{\rho}_{\mu;\nu\rho} -{\Box}R_{\mu\nu} $$ $$- {\frac{1}{2}}
g_{\mu\nu}{\Box}R + 2 R^{\rho}_{\mu}R_{\rho\nu} - {\frac{1}{2}}
g_{\mu\nu}R^{\rho\sigma}R_{\rho\sigma}) + {\tilde\beta}( 2 R_{;\mu\nu} -
2 g_{\mu\nu}{\Box}R  $$ $$- {\frac{1}{2}} g_{\mu\nu} R^2 + 2 R R_{\mu\nu})
- 2{\lambda}{\eta}^2({\tilde\alpha} +
3{\tilde\beta})(12 R^2_{;\mu\nu} - \frac{15}{2}g_{\mu\nu}{\Box}R^2 $$
$$ + 18{\Box}R^2  -{\frac{1}{2}}g_{\mu\nu}R^3 + 3R^2R_{\mu\nu}) = 0 , \eqno(6.1.2)$$

Trace of these field equations is obtained as
 $${\Box} R = ( m^2 -
{\frac{1}{4}}{\lambda}T^2)R^2 - {\lambda}{\eta}^2 R^3 +
 , \eqno(6.1.3)$$ where
$$m^2 = [32{\pi}({\tilde\alpha} + 3{\tilde\beta})G]^{-1}  \eqno(6.1.3b)$$

Using the method ,used in section 4, eq.(2.3) is re-written as

  $${\Box} R = ( m^2 -
{\frac{1}{4}}{\lambda}T^2) R - {\lambda}{\eta}^2 R^3  \eqno(6.1.4)$$

Like earlier sections, here also this equation is multiplied by $\eta$ and ${\eta}R$ is recognized as ${\tilde R}.$   As a result,the equation (6.1.4) looks like
 
$${\Box}{\tilde R} = ( m^2 -
{\frac{1}{4}}{\lambda}T^2) {\tilde R} - {\lambda}{\eta}^2 {\tilde R}^3 ).
\eqno(6.1.5)$$ 

The action for ${\tilde R}$, yielding the equation (6.1.5), is obtained here, following the method of earlier sections, as

$$S_{\tilde R} =  \int{d^4x}
{\sqrt{-g}}\quad\Big [{1 \over2} g^{\mu\nu}{\partial_{\mu}{\tilde
R}}{\partial_{\nu}{\tilde R}} - V^T({\tilde R})\Big], \eqno(6.1.6a)$$
where

$$V^T({\tilde R}) = - {\frac{1}{2}}m^2({\tilde R}^2 + {\frac{1}{12}}T^2) +
{\frac{1}{4}}{\lambda}{\tilde R}^4 + {\frac{1}{8}}{\lambda}T^2{\tilde R}^2 -
{\frac{{\pi}^2}{90}}T^4 . \eqno(6.1.6b)$$

Spontaneous symmetry breaking can be discussed with the Higgs - like
potential  $V^T({\tilde R})$ given by eq.(6.1.6b). Vacuum states are
obtained from \cite{skp93, sks99, pdb}

$$\frac{{\partial}V^T}{{\partial}{\tilde R}} = 0,$$
or $$-m^2{\tilde R} + {\lambda}{\tilde R}^3 +
{\frac{1}{4}}{\lambda}T^2{\tilde R} = 0 , \eqno(6.1.7)$$ 
 yielding
turning points of $V^T({\tilde R})$ as $$\tilde R = 0  \eqno(6.1.8a)$$
and
$$\tilde R = \pm {\frac{1}{2}}{\sqrt(T^2_c - T^2)},  \eqno(6.1.8b)$$ where $$
T_c = {\frac{2m}{\sqrt{\lambda}}}  \eqno(6.1.8c)$$ 

These states  can be written in terms of vacuum
expectation value of $\tilde R$ as $<{\tilde R}> = 0 $ and $<{\tilde R}> =
(1/2){\sqrt{(T^2_c - T^2)}}.$ It shows that as long as $T\ge T_c$ , the
field $\tilde R$ remains confined to the state $<{\tilde R}> = 0.$ But
when temperature goes down such that $T \ll T_c$ , $\tilde R$ tunnels through
the temperature barrier $ T = T_c$ and settles in the state $<{\tilde R}>
= (1/2)T_c$ when $T\ll T_c.$ The state $<{\tilde R}> = 0 $ is called the
false vacuum and $<{\tilde R}>
= (1/2)T_c$ corresponds to the true vacuum state . Moreover , symmetry is
intact in the state  $<{\tilde R}> = 0, $  but it is broken spontaneously at
 $<{\tilde R}>
= \pm(1/2)T_c .$ As a result, energy is released as latent heat in the form
of radiation with density given as 
$$ V^T (<{\tilde R}> = 0) - V^T (<{\tilde R}> = {1 \over 2} T_c) =
\frac{m^4}{4 \lambda}. \eqno(6.1.8d)$$

\smallskip

\noindent \underline{\bf (a) Homogeneous models of the universe}

\smallskip

In the homogeneous case, the line-element of the cosmological model of the early universe can be taken as \cite{skp93}

$$ dS^2 = dt^2 - a^2(t)[ dx^2 +  dy^2 +  dz^2 ], \eqno(6.1.9)$$
where $a(t)$ is the scale factor and $t$ is the cosmic time. The 4-components of a time-like vector $u^{\mu}$ obey the normalization condition

$$ u^{\mu}u_{\mu} = + 1 . \eqno(6.1.10)$$

For the model, given by eq.(6.1.9), $\tilde R$ is obtained as

$$\tilde R = 6 \eta \Big[\frac{\ddot a}{a} + \Big(\frac{\dot a}{a}\Big)^2 \Big], \eqno(6.1.11)$$
where dot(.) stands for derivative with respect to the cosmic time $t$. In the symmetric state $<{\tilde R}> = 0 $, when $T\ge T_c$, the equation (6.1.11) reduces to

$$\frac{a^{\prime\prime}}{a} + \Big(\frac{a^{\prime}}{a}\Big)^2 = 0, \eqno(6.1.12)$$
where prime $(\prime)$ means derivation with respect to $(t/t_P),$ with $t_P$ being Planck's time. 

Solution of the equation (6.1.12) is obtained as

$$ a^2(t) = a^2_0  + \frac{|t|}{t_P} \eqno(6.1.13)$$
with $a(t = 0) = a_0.$ It shows that the model expands as $(a^2_0  + |t|/t_P)^{1/2}$ in the symmetric state above the critical temperature $T_c.$ Now, it is important to decide whether $a_0 = 0$  at $t = 0.$ Answer to this question is obtained as follows.

Components of the energy-momentum tensor for riccion , obtained from the action $S_{\tilde R}$, given by eq.(6.1.6), are defined as

$$ T_{\mu\nu} = \frac{2}{\sqrt{- g}} \frac{\delta S_{\tilde R}}{g^{\mu\nu}}$$
yielding \cite{skp93, sks99}

\begin{eqnarray*}
T_{\mu\nu} &=&{\partial_{\mu}{\tilde
R}}{\partial_{\nu}{\tilde R}} - g_{\mu\nu} \Big [{1 \over2} g^{\alpha\beta}{\partial_{\alpha}{\tilde
R}}{\partial_{\beta}{\tilde R}} - V^T({\tilde R})\Big]  - 2 \eta R_{\mu\nu}{\Box}{\tilde R} \\&& + 2 \eta m^2 {\tilde R} R_{\mu\nu} -  2 \eta \lambda {\tilde R}^3 R_{\mu\nu} - \frac{1}{2} \eta \lambda T^2 {\tilde R} R_{\mu\nu} + 4 \eta m^2 {\tilde R}_{; \mu\nu} -  4 \eta m^2 g_{\mu\nu}{\Box}{\tilde R} \\&& - 4 \eta \lambda {\tilde R}^3_{; \mu\nu} + 4 \eta \lambda g_{\mu\nu}{\Box}{\tilde R}- \lambda T^2 {\tilde R}_{;\mu\nu} +  \lambda T^2 g_{\mu\nu}{\Box}{\tilde R}.
\end{eqnarray*}
$$  \eqno(6.1.14)$$

Using eqs.(6.1.10) and (6.1.14), one obtains that

$$T_{\mu\nu}u^{\mu}u^{\nu} < 0  \eqno(6.1.15)$$
in the state $<{\tilde R}> = 0 ,$ implying violation of the energy condition at and above the critical temperature. This result shows that the model will bounce at $t = o$ without an encounter with singularity. As a reult, it is obtained that

$$ a_0 \ne 0 .  \eqno(6.1.16)$$

Due to expansion of the model, given by eq.(6.1.13), temperature falls down. As discussed above, when it is sufficiently below the critical temperature such that $T^2 \ll T^2_c,$ the model acquires the state

$$<{\tilde R}> = (1/2)T_c .\eqno(6.1.17)$$

Connecting eqs.(6.1.11) and (6.1.17), a differential equation is obtained as

$$\frac{\ddot a}{a} + \Big(\frac{\dot a}{a}\Big)^2 = \frac{T_c}{12 \eta} \eqno(6.1.18)$$
yielding the solution

\begin{eqnarray*}
a^2(t) &=& {\sqrt{\frac{24}{{\eta}T_c}}} sinh\Big[(t -
t_c){\sqrt{\frac{T_c}{6{\eta}}}} + D\Big] \\ & =& a^2_c [sinh
D]^{-1} sinh\Big[(t - t_c){\sqrt{\frac{T_c}{6{\eta}}}} + D\Big] \\&&
\simeq a^2_c  exp\Big[(t - t_c){\sqrt{\frac{T_c}{6{\eta}}}} \Big]
\end{eqnarray*}
$$  \eqno(6.1.19)$$ 
where $a_c$ and $t_c$ are constants such that $a_c = a(t=t_c)$ with $t_c$,
being the time when this type of expansion of the model begins as well as 
$$ D  =  sinh^{-1} \Big[a^2_c {\sqrt{\frac {{\eta}T_c}{24}}} \Big] .$$

The second  case is 
 $$ \frac{\ddot a}{a} + \Big({\frac{\dot
a}{a}}\Big)^2 = - \frac{T_c}{12 \eta} \eqno(6.1.20)$$
yields the solution
$$ a^2(t) = a^2_c sin\Big[(t -
t_c){\sqrt{\frac{T_c}{6{\eta}}}} + {\frac{\pi}{2}}\Big] , \eqno(6.1.21)$$ 
where $a_c$ and $t_c$ are constants such that $a_c = a(t=t_c)$ with $t_c$, being the time when expansion of the model follows eq.(6.1.21).

But $a(t),$ given by
eqs.(6.1.19), shows that the cosmological model will expand more rapidly in
the state  $ {\tilde R} = <{\tilde R} > = (1/2) T_c$ compared to the state
$ {\tilde R} = <{\tilde R} > = 0$. The expansion , in the state   $
{\tilde R} = <{\tilde R} > = - (1/2) T_c$, is given by eq.(6.1.21)., which
shows that $ - a^2_{10} \leq a^2 \leq   a^2_{10}$ leading to imaginary
$a(t)$ also, which is unphysical. So the state $<{\tilde R} > = - (1/2)
T_c$ will not be considered any more.

\smallskip

\noindent \underline{\bf (b) Inhomogeneous models of the universe}

\smallskip
\noindent \underline{\bf First type of inhomogeneous models}

\smallskip

Though large-scale observations, in the present epoch, support homogeneous and isotropic cosmological models, still it is believed that a realistic model of the universe should not be exactly homogeneous, at least, at small scales. This idea motivates to investigate cosmological models, being spatially inhomogeneous and anisotropic at small scales but asymptotically homogeneous and isotropic.

Spatially homogeneous and anisotropic cosmological models such as Kasner
model, Lukash and Starobinsky, Bianchi-typeI and type IX models \cite{ndb}
have been proposed from time to time in the past. But more complicated
spatially inhomogeneous and anisotropic models were investigated during the
last two decades \cite{mah, mc81, mc83, jw, af, jmm}. Due to mathematical difficulty, in solving Einstein's field equations, even these models are inhomogeneous with respect to only one spatial coordinate.

Here, cosmological models, being inhomogeneous and anisotropic with respect to all three spatial co-ordinates in the (1+3)-dimensional space-time, are derived using the method employed in section 6.1(a).

The general framework of the line-element, for cosmological models of the early universe, is taken as \cite{sk98}

$$ dS^2 = dt^2 - a^2(t) f^2(x,y,z)[ dx^2 +  dy^2 +  dz^2 ], \eqno(6.1.22)$$
where $f(x,y,z)$ is the function resposible for the inhomogeneity.

Looking at the line-element (given by eq.(6.22)), the reader may think that the function $f(x,y,z)$ can be absorbed redefining coordinates $(x,y,z)$ as new coordinates $(X,Y,Z)$ such that

$$  f^2(x,y,z)[ dx^2 +  dy^2 +  dz^2 ] = dX^2 +  dY^2 +  dZ^2. \eqno(6.1.23)$$

At this stage, it is important to mention that redefinition of coordinates is not valid at points, where $f(x,y,z) = 0,$ as such points yield spatial singularity (the singularity at the $t = constant$ hypersurface). If such points exist, these can not be ignored as ignoring these will lead to the physical inconsistency. So, absorption of $f$ in coordinates $(x,y,z)$ is possible, only when (i)$f \ne 0$ or (ii) $f < \infty$ at all points of the $t = constant$ hypersurface. It means that this question can be decided after getting the explicit definition of the function $f(x,y,z)$, which is obtained in what follows.

Using the geometrical aspect of the Ricci scalar $R$, $\tilde R$ is obtained for the line-element, given by eq.(6.1.22), as
\begin{eqnarray*}
\tilde R &=& <\tilde R> - \frac{2 \eta}{a^2 f^4}\Big[\Big\{\frac{\partial^2
 f^2}{\partial x^2} +   \frac{\partial^2 f^2}{\partial y^2} + \frac{\partial^2
 f^2}{\partial z^2} \Big\} \\&&+  \frac{3}{4 f^2} \Big\{ \Big(\frac{\partial
 f^2}{\partial x}\Big)^2 +  \Big(\frac{\partial f^2}{\partial y}\Big)^2  +
 \Big(\frac{\partial f^2}{\partial z}\Big)^2 \Big\}\Big]
\end{eqnarray*}
 $$\eqno(6.1.24a)$$
with
$$ <\tilde R> = 6 \eta  \Big[\frac{\ddot a}{a} + \Big({\frac{\dot
a}{a}}\Big)^2 \Big], \eqno(6.1.24b)$$
where $ <\tilde R>$ is the homogeneous part of $\tilde R$, being its vacuum expectation value.

When $\tilde R$ acquires states $ <\tilde R> = 0$, in the symmetric state at temperature $T$ greater than or around $T_c$ as well as $  <{\tilde R} > =  (1/2) T_c$ at $T \ll T_c$ (as mentioned above ), the equation (6.1.24) implies that

$$\Big\{\frac{\partial^2 f^2}{\partial x^2} +   \frac{\partial^2 f^2}{\partial y^2} + \frac{\partial^2 f^2}{\partial z^2} \Big\} +  \frac{3}{4 f^2} \Big\{ \Big(\frac{\partial f^2}{\partial x}\Big)^2 +  \Big(\frac{\partial f^2}{\partial y}\Big)^2  + \Big(\frac{\partial f^2}{\partial z}\Big)^2 \Big\} = 0. \eqno(6.1.25)$$

The exact solution, of the partial differential equation (6.1.25), is obtained as

$$ f^2 = [(x - x_0)(y - y_0)(z - z_0)]^{4/7}. \eqno(6.1.26)$$

Thus, in the state $\tilde R = <\tilde R> = 0,$ the line-element (6.1.22) is obtained with $a^2(t)$ and $ f^2(x,y,z)$, given by eqs.(6.1.13) and (6.1.26) respectively. In the state $  <{\tilde R} > =  (1/2) T_c$, it is obtained with $a^2(t)$, given by the equation (6.1.19) and $ f^2(x,y,z)$ from eq.(6.1.26).

At this stage, it is important to test for the anisotropy and inhomogeneity of the models.

A model is isotropic, if Killing vectors exist at every point, stisfying
equations \cite{swh73}

$$ \xi_{\mu ; \nu} + \xi_{\nu ; \mu} = 0, \eqno(6.1.27)$$
where $\xi_{\mu}$ are components of the Killing vector and  semi-colon (;) stands for covariant derivatives. 

With  $ f^2(x,y,z)$, given by eq. (6.1.26), it is found that at the point $
(x_0,y_0,z_0)$ as well as in the planes $ x = x_0, y = y_0, z = z_0$, $
(\xi_{\mu ; \nu} + \xi_{\nu ; \mu})$ are divergent showing singularity of
these models at these points. It means that at these points isotropy of the
model is broken leading to inhomogeneity also.

Further, it is also important to see whether these singular points of the model are real physical singularities. This aspect can be tested by finding the scalar polynomials of the curvature tensor \cite{swh73}. For the models, obtained above, it is found that 
\begin{eqnarray*}
R^{\mu\nu}R_{\mu\nu} &=& \frac{3 {\ddot a}}{a} - \frac{8}{49 a^2}[(x - x_0)(y - y_0)(z - z_0)]^{-4/7}\\ && \times \Big[\frac{3}{(x - x_0)^2} + \frac{3}{(y - y_0)^2} + \frac{3}{(z - z_0)^2}\\ &&  + \frac{1}{(x - x_0)(y - y_0)} + \frac{1}{(y - y_0)(z - z_0)} + \frac{1}{(z - z_0)(x - x_0)}\Big]\\ && - a^{-2}\Big[3 (a {\ddot a} + {\dot a}^2 ) + 2 a{\dot a} \Big\{\frac{1}{(x - x_0)} + \frac{1}{(y - y_0)} +  \frac{1}{(z - z_0)} \Big\} \Big]
\end{eqnarray*}

$$\eqno(6.1.28a)$$
and

$$R^{\mu\nu\rho\sigma}R_{\mu\nu\rho\sigma} = a^{-4} [\{ (x - x_0)(y - y_0)\}^{-8/7}(z - z_0)^{-36/7} + \cdots ]. \eqno(6.1.28b)$$

Divergence of $R^{\mu\nu}R_{\mu\nu}$ and $R^{\mu\nu\rho\sigma}R_{\mu\nu\rho\sigma}$ at above mentioned points shows that these points are physical singularities.

It is interesting to see that models become homogeneous at large scales, when $x \gg x_0 , y \gg y_0 $ and $z \gg z_0.$

\smallskip
\noindent \underline{\bf Second type of in
homogeneous models}

\smallskip

Normally, two types of models are used for manifestation of gravitational interactions in the Nature (i) static space-times representing geometry around compact objects e.g. Swarzschild space-time, Riessner - Nordstr$\ddot {\rm o}$m space -time etc. and (ii) cosmological models exhibiting expansion of the universe. Former type models ignore expansion of the universe, compact objects exist and latter type of models do not account for the gravitational effect of these compact object treating these as extremely tiny particles. So, it is very natural to think of cosmological models of the universe exhibiting local gravitational effect of compact objects as well as expansion of the universe simultaneously. Motivated by this idea, the second type of models are derived here \cite{sks99} using the method demonstrated above in this section.

Here, the line-element for cosmological models of the early universe is taken
as \cite{sks99} 

$$ dS^2 = dt^2 - a^2(t) {\bf f}^2(r)[ dr^2 + r^2 ( d \theta^2 + \theta^2 d \phi^2 )], \eqno(6.1.29)$$
where $0 \le r \le \infty, 0 \le \theta \le \pi$ and $0 \le \phi \le 2 \pi.$

In this case also, ${\bf f}^2(r)$ can not be absorbed in $[ dr^2 + r^2 ( d \theta^2 + \theta^2 d \phi^2 )]$ at points, where ${\bf f}^2(r)$ vanishes or it is divergent. So, like the case of first type of models, this question is resolved after getting explicit form of ${\bf f}^2(r)$, which is obtained as follows.

Here,  $\tilde R$ is obtained for the line-element, given by eq.(6.1.29), as

$$\tilde R = <\tilde R> - \frac{2 \eta}{a^2 r^2 {\bf f}^4}\Big[\Big(\frac{\partial^2 {\bf f}^2}{\partial r^2} +  \frac{2}{r} \frac{\partial {\bf f}^2}{\partial r} \Big) +  \frac{3}{4 {\bf f}^2}  \Big(\frac{\partial {\bf f}^2}{\partial r}\Big)^2            \Big] \eqno(6.1.30)$$
with $<\tilde R>$, the homogeneous part of $\tilde R$, given by eq.(6.1.24b).

So, in the vacuum states $\tilde R =  <\tilde R> = 0$,  at temperature $T$ greater than or around $T_c$ (in the symmetric state) as well as $ \tilde R = <{\tilde R} > =  (1/2) T_c$ at $T \ll T_c$ (when symmetry is broken ), it is obtained from the equation (6.1.30)  that

$$\Big(\frac{\partial^2 {\bf f}^2}{\partial r^2} +  \frac{2}{r} \frac{\partial {\bf f}^2}{\partial r} \Big) +  \frac{3}{4 {\bf f}^2}  \Big(\frac{\partial {\bf f}^2}{\partial r}\Big)^2   =  0,     
\eqno(6.1.31)$$
which is integrated to
$$ {\bf f}^2(r) = \Big[1 - \frac{r_0}{r} \Big]^{4/7}. \eqno(6.1.32)$$

Thus, in the state $\tilde R = <\tilde R> = 0,$ the line-element (6.1.29) is obtained as

$$ dS^2 = dt^2 - \Big[a^2_0  + \frac{|t|}{t_P} \Big]\Big[1 - \frac{r_0}{r} \Big]^{4/7}[ dr^2 + r^2 ( d \theta^2 + \theta^2 d \phi^2 )], \eqno(6.1.33)$$
when the temperature $T \sim T_c$.
 
In the state $\tilde R = <\tilde R> = T_c/2,$ the line-element (6.1.29) is obtained as

$$ dS^2 = dt^2 - a^2_c  exp\Big[(t - t_c){\sqrt{\frac{T_c}{6{\eta}}}}\Big] \Big[1 - \frac{r_0}{r} \Big]^{4/7}[ dr^2 + r^2 ( d \theta^2 + \theta^2 d \phi^2 )], \eqno(6.1.34)$$
when the temperature $T \ll T_c$.
 
In these models , it is obtained that
\begin{eqnarray*}
 R^2 &=& \frac{144 r^2_0}{49 r^6}\Big[1 - \frac{r_0}{r} \Big]^{- 36/7} + \cdots  \\ R^{\mu\nu}R_{\mu\nu}   &=& \frac{144 r^2_0}{49 r^6}\Big[1 - \frac{r_0}{r} \Big]^{- 36/7} + \cdots.
\end{eqnarray*}
$$\eqno(6.1.35 a,b)$$
These equations exhibit divergence of scalar curvature polynomial of curvature tensor (which will contain $ R^2$ and $R^{\mu\nu}R_{\mu\nu}$) at  $r=0$ and $r=r_0.$ It means that models, given by eqs.(6.1.33) and (6.1.34), possess real physical singularities at the point $r=0$ and the surface $r=r_0$ \cite{swh73}.

\bigskip
 
 \centerline{\bf 6.2. Creation of spinless particles in homogeneous models}

\smallskip

Manifestation of the Ricci scalar as a physical field also, motivates to study its interaction with another scalar field $\phi$. The action for such a theory is

$$S =  \int{d^4x}
{\sqrt{-g}}\quad\Big [{1 \over2} g^{\mu\nu}{\partial_{\mu}{\tilde
R}}{\partial_{\nu}{\tilde R}} - V^T({\tilde R}) + {1 \over2} g^{\mu\nu}{\partial_{\mu}{\phi}}{\partial_{\nu}{\phi}} - {1 \over2}(\xi\eta^{-1}{\tilde R} + \lambda {\tilde R}^2 )\phi^2  \Big], \eqno(6.2.1)$$
where $\xi\eta^{-1}{\tilde R} = \xi R $ with  $\xi$ being the dimensionless non-minimal coupling constant, $V^T({\tilde R})$ is given by eq.(6(i).6b) and ${1 \over2}\lambda {\tilde R}^2\phi^2$ is the ${\tilde R}-\phi$ interaction term with the dimensionless coupling constant $\lambda$. As the model is homogeneous, ${\tilde R} = <{\tilde R}>$, which is the vacuum expectation value.

The invariance of $S$, under transformation $\phi \to \phi + \delta\phi$, yields the Klein-Gordon equation for $\phi$ as
$$({\Box} + \xi\eta^{-1}{\tilde R} + \lambda {\tilde R}^2 )\phi = 0. \eqno(6.2.2)$$

For the purpose of the second quantization, the general solution of the equation(6.2.2) can be written as

$$\phi = \sum_k [a_k f_k(t) exp( ik_{\hat a} x^{\hat a}) + a_k^{\dagger} f^{*}_k(t) exp(- ik_{\hat a} x^{\hat a}) ]. \eqno(6.2.3)$$

Particles are produced due to conformal symmetry breaking through the mass term . In eq.(6.2.2), the mass term $(\xi\eta^{-1}{\tilde R} + \lambda {\tilde R}^2 )$ vanishes for the state $<{\tilde R}> = 0.$ So, no production of particles are expected in this state.

In the true vacuum state $<{\tilde R}> = (1/2)T_c$, the geometry of the model is given by the line-element(6.1.19). Connecting eqs.(6.1.19), (6.2.2) and (6.2.3), the differential equation for the $k$th mode is obtained as

$${\ddot f_k} + {\sqrt{\frac{27 T_c}{2{\eta}}}}{\dot f_k} + \Big[\frac{k^2}{a_c^2}exp\Big\{- (t - t_c){\sqrt{\frac{T_c}{6{\eta}}}}\Big\} + \frac{1}{2}\xi\eta^{-1}T_c + \frac{1}{4}\lambda T_c^2 \Big]f_k = 0, \eqno(6.2.4)$$
which can be re-written as

$${\ddot v_k} + \Big[\frac{k^2}{a_c^2}exp\Big\{- (t - t_c){\sqrt{\frac{T_c}{6{\eta}}}}\Big\} + \frac{1}{2}\xi\eta^{-1}T_c + \frac{1}{4}\lambda T_c^2 - \frac{27 T_c}{2{\eta}} \Big]v_k = 0 \eqno(6.2.5)$$ 
with $$f_k = v_k exp(- {\sqrt{\frac{27 T_c}{2{\eta}}}}(t - t_c)).\eqno(6.2.6)$$

The $in-state$ is obtained in the extreme past i.e., when $t \to - \infty$ and the $out-state$ lies in the extreme future i.e., when $t \to + \infty$. So, $\phi$ can be written in terms of $in-$ and $out-states$ as 
\begin{eqnarray*}
\phi &=& \sum_k [a^{in}_k f^{in}_k(t) exp( ik_{\hat a} x^{\hat a}) +
a^{\dagger in}_k f^{*in}_k(t) exp(- ik_{\hat a} x^{\hat a}) ]\\ &=& \sum_k
[a^{out}_k f^{out}_k(t) exp( ik_{\hat a} x^{\hat a}) + a^{\dagger out}_k f^{*out}_k(t) exp(- ik_{\hat a} x^{\hat a}) ].
\end{eqnarray*}
$$ \eqno(6.2.7a,b)$$

The normalized solution of eq. (6.2.5), in the $in-state$, is obtained as
$$ v_k^{in} = exp(+ i\nu_k(t - t_c))/{\sqrt{2 a^3_c \nu_k}} \eqno(6.2.8a)$$ 
 with
$$nu^2_k =  \frac{1}{2}\xi\eta^{-1}T_c + \frac{1}{4}\lambda T_c^2 - \frac{27 T_c}{2{\eta}} \eqno(6.2.8b)$$ 
It yields 
$$\phi_k^{in} = exp(+ i[\nu_k - i{\sqrt{\frac{27 T_c}{2{\eta}}}}](t - t_c) + ik_{\hat a} x^{\hat a}))/{\sqrt{16 \pi^3 a^3_c \nu_k}}\eqno(6.2.8c)$$ 
connecting eqs.(6.2.6), (6.2.7) and (6.2.8a).

Similarly, in the $out-state$, the normalized solution of eq. (6.2.5) is obtained as
$$ v_k^{out} = exp(- i\nu_k(t - t_c))/{\sqrt{2 a^3_c \nu_k}} \eqno(6.2.9a)$$ 
 with $\nu_k$ given by eq.(6.2.8b), which yields
 
$$\phi_k^{out} = exp(- i[\nu_k - i{\sqrt{\frac{27 T_c}{2{\eta}}}}](t - t_c) + ik_{\hat a} x^{\hat a}))/{\sqrt{16 \pi^3 a^3_c \nu_k}}  \eqno(6.2.9b)$$ 
on connecting eqs.(6.2.6), (6.2.7) and (6.2.8a).  

The $in-$ and $out-states$ belong to the same Fock - space, $\phi_k^{out}$ can be represented as linear combination of $\phi_k^{in}$ and $\phi_k^{*in}$ as well as $\phi_k^{in}$ can be written     as linear combination of $\phi_k^{out}$ and $\phi_k^{*out}$. Hence \cite{ndb},

$$\phi_k^{out} = \alpha_k \phi_k^{in} + \beta_k \phi_k^{*in}, \eqno(6.2.10a)$$
where Bogolubov coefficients $\alpha_k$ and $\beta_k$ are given as
$$ \alpha_k = (\phi_k^{out} , \phi_k^{in}) = - i \int_{t = t_c} {d^3 x} a^3(t) [\phi_k^{out}(\partial_t\phi_k^{*in}) - (\partial_t\phi_k^{out})\phi_k^{*in}]  \eqno(6.2.10b)$$

and

$$ \beta_k = (\phi_k^{out} , \phi_k^{*in}) =  i \int_{t = t_c} {d^3 x} a^3(t) [\phi_k^{out}(\partial_t\phi_k^{in}) - (\partial_t\phi_k^{out})\phi_k^{in}]  \eqno(6.2).10c)$$

employing the scalar product $(\phi_1 , \phi_2)$ defined as
$$(\phi_1 , \phi_2) = - i \int_{t = constant}{d^3 x} \sqrt{-g_{t = constant}} [\phi_1(\partial_t\phi_2^{*}) - (\partial_t\phi_1)\phi_2^{*}]$$ 
with
$$ (\phi_k^{in} , \phi_k^{in}) = 1 , (\phi_k^{*in} , \phi_k^{*in}) = - 1 .$$

Moreover, Bogolubov coefficients satisfy the relationship

$$ |\alpha_k|^2 - |\beta_k|^2 = 1 . \eqno(6.2.11)$$

Using eqs.(6.2.8c) and (6.2.9b) in (6.2.10b) and (6.2.10c), it is obtained that

$$\alpha_k = \frac{i[\nu_k + i \sqrt{\frac{27 T_c}{2{\eta}}}]}{\nu_k} \quad {\rm and} \quad  \beta_k = 1 . \eqno(6.2.12a,b)$$

Eqs.(6.2.11) and (6.2.12) imply

$$ \nu_k^2 = \frac{27 T_c}{2{\eta}} \eqno(6.2.13a)$$
on using eqs.(6.2.12) in eq.(6.2.11). Further, connecting eqs.(6.2.8b) and (6.2.13a), the result is obtained as

$$ T_c = \frac{(108 - 2 \xi)}{ \eta \lambda}. \eqno(6.2.13b)$$

The relative probability of production of a particle - antiparticle pair, in mode $k$ is given by

$$\omega_k = |\beta_k/\alpha_k|^2 = \frac{\nu_k^2}{\nu_k^2 + ({27 T_c}{2{\eta}})} = \frac{1}{2} \eqno(6.2.14)$$
employing eq.(6.2.13b).

This result is true for all modes, being independent of $k$. So, the absolute
probability of creation $1,2,3, \cdots $ particle - antiparticle pairs is
given by \cite{lp, em}
\begin{eqnarray*}
\sum^{\infty}_{n = 0} |_{out}<0|n|0>_{in}|^2 &=& \sum^{\infty}_{n = 1}
\prod^{\infty}_{k = - \infty} \omega_k^n |\alpha_k|^{-2}\\ &=&
\prod^{\infty}_{k = - \infty}\sum^{\infty}_{n = 0}\frac{1}{2^{(n+1)}} \\ &=& 1
\end{eqnarray*} 
$$ \eqno(6.2.15)$$
on using eqs.(6.2.12a),(6.2.13a) and (6.2.14).

It means that production of infinitely many particles, in all modes, is
certain. The energy density of these particles can be caculated as
$${\bar T}^0_0 = _{in}<0|T^0_0|>_{in} -  _{out}<0|T^0_0|>_{out}, \eqno(6.2.16a)$$
where 
$$ T^{\nu}_{\nu} = \partial^{\nu}\phi^{*} \partial_{\nu}\phi - \frac{1}{2}[{\partial^{\mu}{\phi^{*}}}{\partial_{\mu}{\phi}} - (\xi\eta^{-1}{\tilde R} + \lambda {\tilde R}^2 )\phi^{*}\phi. \eqno(6.2.16b)$$
Anamolous terms get cancelled in eq.(6.2.16a), as ${\bar T}^0_0$ is the difference of $T^0_0$ in $in-$ and $out-$ states .

\bigskip
 
 \centerline{\bf 6.3. Creation of spin-1/2 particles in homogeneous models}

\smallskip

Due to the physical nature of the Ricci scalar, its interaction with spin-1/2 particles is also important.

$$ S =  \int{d^4x}
{\sqrt{-g}}\quad\Big [{1 \over2} g^{\mu\nu}{\partial_{\mu}{\tilde
R}}{\partial_{\nu}{\tilde R}} - V^T({\tilde R}) + {1 \over2}({\bar \psi}i\gamma^{\mu}\bigtriangledown_{\mu}\psi - g{\tilde R}{\bar \psi}\psi + c.c.) \Big], \eqno(6.3.1)$$
where $g{\tilde R}{\bar \psi}\psi$ is the ${\tilde R}-\psi$ interaction term
with the dimensionless coupling constant $g$ and $V^T({\tilde R})$ is given by
eq.(6.1.6b). As the model is homogeneous, ${\tilde R} = <{\tilde R}>$, which
is the vacuum expectation value. Here $\bigtriangledown_{\mu} = \partial_{\mu} -
\Gamma_{\mu}.$ Dirac matrices $\gamma^{\mu}$ as well as $\Gamma_{\mu}$ are
defind in eqs.(4.8), (4.9a,b), (4.10) and (4.10a,b).

Invariance of $S$ (given by eq.(6.3.1)) under transformation ${\bar \psi} \to {\bar \psi}
 + \delta {\bar \psi}$ yields the Dirac equation
$$ \Big( i \gamma^{\mu}D_{\mu}  -  M_f  \Big) \psi = 0,
\eqno(6.3.2a)$$
where
 
$$ M_f = g {\tilde R}.\eqno(6.3.2b)$$

\smallskip
 
 \noindent{\bf (a) Solution of the Dirac equation}

\smallskip

In general, solution of the Dirac equation $\psi$ can be written as
$$ \psi = {\sum_{s=\pm1}}{\sum_k} \Big( b_{(k,s)} \psi_{(I k,s)} +
d^{\dagger}_{(-k,-s)} \psi_{(II k,s)} \Big)   , \eqno(6.3.3a)$$
$$ \psi^{\dagger} = {\sum_{s=\pm1}}{\sum_k} \Big(  {\bar \psi}_{(Ik,s)} \gamma^0 b_{(k,s)}^{\dagger}+
 {\bar \psi}_{(II k,s)} \gamma^0 d_{(k,s)} \Big)   , \eqno(6.3.3b)$$
where
$$\psi_{(I k,s)} =    f_{(k  ,s)} (t) e^{- i k_a x^a} u_s   \eqno(6.3.3c)$$
and
$$\psi_{(II k,s)} =    g_{(k ,s)} (t) e^{ ik_a x^a} {\hat u}_s .  \eqno(6.3.3d)$$
with $g_{(k ,s)} (t) = f_{(- k  ,s)} (t)$   

In eqs.(6(iii).8c,d)
$$ u^{T}_1 = (1 0 0 0), \quad u^{T}_{-1} = (0 1 0 0)       $$
$$ {\hat u}^{T}_1 = (0 0 1 0), \quad {\hat u}^{T}_{-1} = (0 0 0 1) ,
\eqno(6.3.4a,b,c,d)      $$ 
where the index $(T)$ stands for transpose of the column matrices $u_{\pm
s}$ and ${\hat u}_{\pm s}$. In eqs.(6.3.3a,b), $b_{k,s} (b^{\dagger}_{k,s})$ are creation (annihilation)
operators for the positive - energy particles and
$d_{-k,s}(d^{\dagger}_{-k,s}) $  are creation (annihilation)
operators for the negative - energy particles (anti-particles).
 
The geometry of the model, under consideration is given by the distance function, given by eq.(6.1.9) with $a^2(t)$ defined by the equation (6.1.19). The equation (6.1.9) is re-written in terms of conformal time as
$$ dS^2 = A^2({\bar \tau}) [d{\bar \tau}^2 -  dx^2 -  dy^2 -  dz^2 ], \eqno(6.3.5a)$$
where
$$ {\bar \tau} = \int^t \frac{dt^{\prime}}{a(t^{\prime})} \quad {\rm and}\quad A({\bar \tau}) = a(t) . \eqno(6.3.5b,c)$$

Connecting eqs.(6.3.4) and (6.3.5), tetrad components are obtained as
$$ e^0_0 = e^1_1 = e^2_2 = e^3_3 =  \frac{1}{A({\bar \tau})}. \eqno(6.3.6)$$
As a result, Dirac matrices $\gamma^{\mu}$ are written as
$$\gamma^{\mu} = \frac{1}{A({\bar \tau})}{\tilde \gamma}^a \delta^{\mu}_a \eqno(6.3.7a)$$

and the Dirac equation(6.3.2a) reduces to
$$[i({\tilde \gamma}^0\partial_0 + {\tilde \gamma}^1\partial_1 + {\tilde \gamma}^2 \partial_2 + {\tilde \gamma}^3\partial_3 - g A({\bar \tau}){\tilde R}]\psi = 0.  \eqno(6.3.7b)$$ 

Employing eqs.(6.3.3) to the equation(6.3.7b), following equations are obtained

$$[i({\tilde \gamma}^0\partial_0 + {\tilde \gamma}^a\partial_a) - \frac{1}{2} g A({\bar \tau})T_c ]\psi_{(I k,s)} = 0.  \eqno(6.3.8a)$$ 

$$[i({\tilde \gamma}^0\partial_0 + {\tilde \gamma}^a\partial_a) - \frac{1}{2} g A({\bar \tau})T_c ]\psi_{(II k,s)} = 0.  \eqno(6.3.8b)$$ 
for mode $k$ and spin $s$ in the state $<{\tilde R}> = (1/2)T_c$, when $T \ll T_c.$

Operation of $[- i({\tilde \gamma}^0\partial_0 + {\tilde \gamma}^a\partial_a) - \frac{1}{2} g A({\bar \tau})T_c ]$ on equations(6.3.8) yields ordinary differential equations

$$ \frac{d^2 f_{(k,s)}}{d\eta^2} + \Big( k^2 + \frac{i \epsilon g}{2} T_c \partial_0 A({\bar \tau}) + \frac{g^2 T^2_c}{4} A^2({\bar \tau}) \Big)f_{(k,s)} = 0  \eqno(6.3).9a)$$ 

$$ \frac{d^2 g_{(k,s)}}{d\eta^2} + \Big( k^2 + \frac{i \epsilon g}{2} T_c \partial_0 A({\bar \tau}) + \frac{g^2 T^2_c}{4} A^2({\bar \tau}) \Big)g_{(k,s)} = 0  \eqno(6.3.9b)$$ 
using 
$${\tilde \gamma}^0 u_s = \epsilon u_s , {\tilde \gamma}^0 {\hat u}_s = \epsilon {\hat u}_s \quad {\rm and} \quad  \epsilon = \pm 1. \eqno(6.3.9c,d,e)$$ 

As $g_{(k,s)} = f_{(-k,s)}$, it is better to concentrate on one of the two
equations(6.3.9a,b). So eq.(6.3.9a) is chosen for further analysis. Connecting
eqs.(6.1.19) and (6.3.7)

$$ {\bar \tau} = \int^t \frac{dt^{\prime}}{a_c}exp\Big[- (t - t_c){\sqrt{\frac{T_c}{24{\eta}}}}\Big]  = - \frac{e^{[- H(t - t_c)]}}{a_c H}\eqno(6.3.10a)$$
with $H^2 = {T_c}/{24{\eta}}$ and 
 $$ A({\bar \tau}) = a(t) = - \frac{1}{H \tau}  . \eqno(6.3.10b)$$

Using $ A({\bar \tau})$, from eq.(6.3.10b), eq.(6.3.9a) is re-written as

$$ \frac{d^2 f_{(k,s)}}{d{\bar \tau}^2} + \Big( k^2 + \frac{i \epsilon gT_c}{2 H{\bar \tau}^2}   + \frac{g^2 T^2_c}{4H^2{\bar \tau}^2 } \Big)f_{(k,s)} = 0  \eqno(6.3.11)$$ 

 Eq.(6.3.10a)implies that the open interval $- \infty < t < \infty$ corresponds to the open interval $- \infty < \tau < 0.$ The equation(6.3.16) yields  solutions

$$f_{(k, +1)} = {\bar \tau}^{\nu} e^{\pm ik {\bar \tau}} _1F_1(\nu; 2\nu; \mp 2ik{\bar \tau}) \eqno(6.3.12a)$$ 

$$f_{(k, -1)} = {\bar \tau}^{\nu} e^{\pm ik{\bar \tau}}(\mp 2ik{\bar \tau}^{(1 - 2\nu)} _1F_1(1 - \nu; 1 - 2\nu; \mp 2ik{\bar \tau}) \eqno(6.3.12b)$$ 

where
\begin{eqnarray*}
\nu &=& \frac{1}{2} \Big[1 \pm \Big\{1 - 4\Big( \frac{i \epsilon gT_c}{2 H\tau^2} + \frac{g^2 T^2_c}{4H^2\tau^2 } \Big)\Big\}^{1/2}\Big] \\ & \simeq & \frac{1}{2}\Big[1 \pm \Big\{ \frac{ gT_c}{ H} + i \Big\} \Big] 
\end{eqnarray*}
$$\eqno(6.3.12c)$$ 
as $gT_c >> H.$ Here $_1F_1(a; b; x)$ is the confluent hypergeometric
function, which is connected to the Bessel's function through an identity \cite{etw}

$$J_l(z) = \frac{(z/2)^l}{(1 + l)^{1/2}}_1F_1(1/2 + l; 1 + 2l; \mp 2iz). \eqno(6.3.13)$$ 

Using the identity (6.3.13) in eqs.(6.3.12), the solutions are obtained as

$$f_{(k, +1)} = (\pi/4)^{1/2} e^{(\pm1 - i)(\pi/4) \mp i\pi g T_c/(4H)} {\bar
  \tau}^{1/2} J_{\pm (g T_c/(2H + i/2)}(k {\bar \tau})  \eqno(6.3.14a)$$

$$f_{(k, -1)} = (\pi/4)^{1/2} e^{(\mp1 - i)(\pi/4) \pm i\pi g T_c/(4H)}{\bar \tau}^{1/2} J_{\mp (g T_c/(2H + i/2)}(k {\bar \tau})  \eqno(6.3).14b)$$

Using the asymptotic form of the Bessel's functions \cite{lap}

\[J_l(z) @>>{z \rightarrow \infty}> \Big(\frac{\pi z}{2} \Big)^{-1/2} cos
\Big(z - \frac{\pi l}{2} - \frac{\pi }{4}\Big) \],
$$\eqno(6.3.15a)$$
it is obtained that
\[ 
\lim_{\tau \to -\infty}f_{(k, +1)} = 
\begin{cases}
(2k)^{-1/2} e^{- ik {\bar \tau}} &\text{for  $k > 0$;} \\
 (2k)^{-1/2} e^{+ ik {\bar \tau}} &\text{for $k < 0$,}
\end{cases}
\]
which are normalized. It shows that the solution(6(iii).19a) is normalized for ${\bar \tau} \to -\infty$.  The wronskian of this solution(6.3.14a) is constant. It implies that the solution(6.3.14a)is not only normalized for ${\bar \tau} \to -\infty$, but for all ${\bar \tau}$ in the interval $(-\infty, 0]$. The same result is obtained for$f_{(k, -1)}$.

 For small value of $z$,

$$J_l(z) \simeq \frac{(z/2)^l}{(1 + l)^{1/2}}. \eqno(6(iii).20b)$$

For analyzing asymptotics of solutions for $ t \to \pm \infty$, eqs.(6.3.14) are written as
\[ 
f_{(k, +1)} = 
\begin{cases}
 A_1 {\bar \tau}^{1/2}J_{(\frac{gT_c}{H} + i)/2}(k {\bar \tau})  &\text{as  $t \to \infty$}
\\       A_1 {\bar \tau}^{1/2}J_{- (\frac{gT_c}{H} + i)/2}(k {\bar \tau}) &\text{as  $t \to -\infty,$}
\end{cases}
\]
$$ \eqno(6.3.16a)$$

are written as
\[ 
f_{(k, -1)} = 
\begin{cases}
 A_2 {\bar \tau}^{1/2}J_{(\frac{gT_c}{H} + i)/2}(k {\bar \tau})  &\text{as  $t \to \infty$}
\\       A_2 {\bar \tau}^{1/2}J_{- (\frac{gT_c}{H} + i)/2}(k {\bar \tau}) &\text{as  $t \to -\infty,$}
\end{cases}
\]
$$ \eqno(6.3.16b)$$

Also using $g_{(k, s)} = f_{(- k, s)}$,
\[ 
g_{(k, +1)} = 
\begin{cases}
 A_1 {\bar \tau}^{1/2}J_{(\frac{gT_c}{H} + i)/2}(- k {\bar \tau})  &\text{as  $t \to
 \infty$} \\       A_1{\bar \tau}^{1/2}J_{- (\frac{gT_c}{H} + i)/2}(- k {\bar \tau}) &\text{as  $t \to -\infty,$}
\end{cases}
\]
$$\eqno(6.3.16c)$$
\[ 
g_{(k, -1)} = 
\begin{cases}
 A_2 {\bar \tau}^{1/2}J_{(\frac{gT_c}{H} + i)/2}(- k {\bar \tau})  &\text{as  $t \to \infty$}
\\       A_2 {\bar \tau}^{1/2}J_{- (\frac{gT_c}{H} + i)/2}(- k {\bar \tau}) &\text{as  $t \to -\infty,$}
\end{cases}
\]
$$\eqno(6.3.16d)$$
In eqs.(6.3.16a,b,c,d),
$$A_1 = \Big(\frac{\pi}{4}\Big)^{1/2} e^{(\pm 1 - i[\pi \pm \frac{gT_c}{H}]/4)} \eqno(6.3.16e)$$
and 
$$A_2 = \Big(\frac{\pi}{4}\Big)^{1/2} e^{(\mp 1 + i[\pi \pm \frac{gT_c}{H}]/4)}. \eqno(6.3.16f)$$

Thus the normalized solutions $\psi_{(I k,s)}$ and $\psi_{(II k,s)}$ are obtained as

$$ \psi_{(I k,s)} = \frac{i}{V^{3/2}} \Big(\frac{T_c}{24 \eta^3} \Big)^{1/8}f_{(k, s)}e^{(-i k_a x^a)} u_s \eqno(6.3.17a)$$
and

$$ \psi_{(II k,s)} = \frac{i}{V^{3/2}} \Big(\frac{T_c}{24 \eta^3} \Big)^{1/8}g_{(k, s)}e^{(i k_a x^a)} {\hat u}_s \eqno(6.3.17b)$$
in the three-space volume $V$, using the normalizing prescription \cite{ndb}

$$({\bar \psi}_{(k,s)}, \psi_{(k^{\prime},s^{\prime})}) = \int_{t = constant}
{d^3 x} \sqrt{-g_{t = constant}}{\bar \psi}_{(k,s)}{\tilde \gamma}^0
\psi_{(k^{\prime},s^{\prime})} = \delta_{ss^{\prime}} \delta_{kk^{\prime}}. \eqno(6.3.17c)$$

\smallskip
 
 \noindent{\bf (b) Production of particles}

\smallskip

For the study of creation of spin-1/2 particles, it is better to go to Fock-space formulation, where the $in$-state ($t \to - \infty)$ as well as $out$-state ($t \to - \infty)$ is defined as

$$b^{in}_{(k,s)}|0>_{in} =  b^{in}_{(-k,s)}|0>_{in} = 0 \quad {\rm and}\quad _{in}<0|0>_{in} = 1 . \eqno(6.3.18a,b,c)$$
and

$$b^{out}_{(k,s)}|0>_{out} =  b^{out}_{(-k,s)}|0>_{out} = 0 \quad {\rm and}\quad _{out}<0|0>_{out} = 1 . \eqno(6.3.19a,b,c)$$

The decomposed form of $\psi$ is written in the $in$-region as well as $out$-region as
\begin{eqnarray*}
\psi & = &  {\sum_{s=\pm1}}{\sum_k} \Big( b^{in}_{(k,s)} \psi^{in}_{(I k,s)} +
(d^{in}_{(-k,-s)})^{\dagger} \psi^{in}_{II(- k,-s)} \Big)\\ & = &  {\sum_{s=\pm1}}{\sum_k} \Big( b^{out}_{(k,s)} \psi^{out}_{(I k,s)} +
(d^{out}_{(-k,-s)})^{\dagger} \psi^{out}_{II (- k,-s)} \Big)
\end{eqnarray*}
$$\eqno(6.3.20a,b)$$
and

\begin{eqnarray*}
\psi^{\dagger} & = &  {\sum_{s=\pm1}}{\sum_k} \Big[(\psi^{in}_{(I k,s)})^{\dagger}( b^{in}_{(k,s)})^{\dagger}  + (\psi^{in}_{II( k,s)})^{\dagger}
d^{in}_{(-k,-s)}  \Big]\\ & = &   {\sum_{s=\pm1}}{\sum_k} \Big[(\psi^{out}_{(I k,s)})^{\dagger}( b^{out}_{(k,s)})^{\dagger}  + (\psi^{out}_{II( k,s)})^{\dagger}
d^{out}_{(-k,-s)}  \Big]
\end{eqnarray*}
$$\eqno(6.3.20c,d)$$

Bogolubov transformations for Fermi-fields are written as \cite{bsd}
\begin{eqnarray*}
b^{out}_{(k,s)}& = & b^{in}_{(k,s)} \alpha_{(k,s)} + d^{in}_{(-k,-s)}\beta_{(k,s)}      \\ ( b^{out}_{(k,s)})^{\dagger} & = & \alpha^{*}_{(k,s)}( b^{in}_{(k,s)})^{\dagger} + \beta^{*}_{(k,s)}( d^{in}_{(k,s)})^{\dagger} \\ (d^{out}_{(-k,-s)})^{\dagger}& = & b^{in}_{(k,s)} \alpha_{(k,s)} + d^{in}_{(-k,-s)}\beta_{(k,s)}      \\  d^{out}_{(-k,-s)} & = & \alpha^{*}_{(k,s)}( b^{in}_{(k,s)})^{\dagger} + \beta^{*}_{(k,s)}( d^{in}_{(k,s)})^{\dagger}
\end{eqnarray*}
$$\eqno(6.3.21a,b,c,d)$$
Bogolubov coefficients $\alpha_{(k,s)}$ and $\beta_{(k,s)}$ satisfy the condition [48]

$$ |\alpha_{(k,s)}|^2  +  |\beta_{(k,s)}|^2 = 1 , \eqno(6.3.22a)$$
which are given as

$$ \alpha_{(k,s)} = \int_{t = constant} {d^3 x}\sqrt{-g_{t = constant}}\psi^{in}_{(I k,s)} ( \psi^{out}_{(I k,s)})^{\dagger}\eqno(6.3.22b)$$

and
$$ \beta_{(k,s)} = \int_{t = constant} {d^3 x}\sqrt{-g_{t = constant}}\psi^{in}_{II(- k, -s)} ( \psi^{out}_{II(- k, -s)})^{\dagger}. \eqno(6.3.22c)$$

Connecting eqs.(6.3.21), (6.3.22) and (6.3.27),
$$ \alpha_{(k,s)} =  \frac{(\pi/4) e^{s \pi/2}[-\{(24\eta)T_c\}^{1/4}
  e^{-H(t_1-t_c)}]^{1-i}}{[1 - \frac{1}{2}(i + gT_c/H)]^{1/2}[1 +
  \frac{1}{2}(i + gT_c/H)]^{1/2}} \Big(\frac{k}{2} \Big)^{-i}\eqno(6.3.23a)$$

$$ \beta_{(k,s)} =  \frac{(\pi/4) e^{- s \pi/2}[-\{(24\eta)T_c\}^{1/4}
  e^{-H(t_1-t_c)}]^{1-i}}{[1 - \frac{1}{2}(i + gT_c/H)]^{1/2}[1 +
  \frac{1}{2}(i + gT_c/H)]^{1/2}} \Big(\frac{- k}{2} \Big)^{-i}\eqno(6.3.23b)$$

Eqs.(6.3.23) yield

$$|\alpha_{(k,s)}|^2 = \frac{(\pi/4)^2 e^{ s \pi}[\{(24\eta)T_c\}^{1/4}
  e^{-H(t_1-t_c)}]^2}{([(1 - g\sqrt{6\eta T_c})^2 + 1/4][(1 + g\sqrt{6\eta T_c})^2 + 1/4]^{1/4}}\eqno(6.3.24a)$$

$$|\beta_{(k,s)}|^2 = \frac{(\pi/4)^2 e^{- s \pi}[\{(24\eta)T_c\}^{1/4}
  e^{-H(t_1-t_c)}]^2}{([(1 - g\sqrt{6\eta T_c})^2 + 1/4][(1 + g\sqrt{6\eta T_c})^2 + 1/4]^{1/4}}\eqno(6.3.24b)$$

The absolute probability of no creation of particles are given as \cite{lp, em}
\begin{eqnarray*}
 | _{out}<0|0>_{in} |^2 &=& \prod_{k,s}|\alpha_{(k,s)}|^{-2} \\ &=& \prod_k
   |\alpha_{(k,+1) }|^{-2} |\alpha_{(k,-1) }|^{-2} \\ &=& exp\Big[\sum_k
   ln(|\alpha_{(k,+1) }|^{-2} |\alpha_{(k,-1) }|^{-2}) \Big] \\ &=&
   \frac{(\pi/4)^4[\{(24\eta)T_c\}^{1/4} e^{-H(t_1-t_c)}]^4}{([(1 -
   g\sqrt{6\eta T_c})^2 + 1/4][(1 + g\sqrt{6\eta T_c})^2 + 1/4]^{1/2}} \\  &=& \frac{(\pi/4)^4[(24\eta)T_c e^{-4 H(t_1-t_c)}]}{[\frac{25}{16} + 36g^4\eta^2 T_c^2 + 51g^2\eta T_c]^{1/2}}\\  & \simeq & (\pi/4)^4(4/g^2)e^{-4 H(t_1-t_c)}.
\end{eqnarray*}
$$ \eqno(6.3.25a)$$

The decay of the $in$-state per unit time per unit volume is given by
\begin{eqnarray*}
{\tilde \Gamma}&=& - V_4^{-1} ln| _{out}<0|0>_{in} |^2 \\ & \simeq & a_c^{-3}e^{[-3 H(t-t_c)]}[4 H(t-t_c) - ln(\pi/4)^4(4/g^2)].
\end{eqnarray*}
$$ \eqno(6.3.25b)$$
It shows that when $t > t_c$, the decay of the $in$-state is very fast. As a result, particle-antiparticle pairs will get created.

\bigskip
 
 \centerline{\bf 6.4. Creation of spinless particles in inhomogeneous models}

\smallskip

Here, inhomogeneous models, given by distance functions (6..33) and
(6.1.34), are considered for states $<{\tilde R}> = 0$ and $<{\tilde R}> =  T_c/2$ respectively.

Action for the scalar field is taken as \cite{sk05} 

$$ S_{\phi} =  \int {d^4 x} \Big\{\sqrt{- g} [{1 \over2}
g^{\mu\nu}{\partial_{\mu}\Phi}{\partial_{\nu}\Phi^*}  -
\frac{1}{2}\Lambda {\tilde R}^2 \Phi\Phi^* \Big] \eqno( 6.4.1a)$$

The Klein - Gordon equation, obtained from this action is

$$[{\Box} + M^2_b ] \Phi = 0 ,   \eqno(6.4.1b)$$ 
where 
$$ M^2_b = \Lambda <{\tilde R}>^2 .
\eqno(6.4.2)$$ 

In the background geometry of the line - element (6.1.33), the
equation (6.4.1b) is re-written as
$$\frac{\partial^2 \Phi}{\partial t^2}  + \frac{3{\dot a}}{a}
\frac{\partial \Phi}{\partial t} - {1 \over a^2}\Big[ 1 - \frac{r_0}{r}
\Big]^{-4/7} \frac{\partial^2 \Phi}{\partial r^2}   - {1 \over r^2 a^2}\Big[ 1
- \frac{r_0}{r} \Big]^{-6/7} \frac{\partial}{\partial r}\Big[ r^2 \Big( 1
- \frac{r_0}{r}\Big)^{2/7}  \Big] \frac{\partial \Phi}{\partial r}$$
$$- {1 \over a^2 r^2}\Big[ 1 - \frac{r_0}{r}
\Big]^{-4/7} \Big[\frac{1}{sin \theta} \frac{\partial }{\partial \theta}
\Big({sin \theta} \frac{\partial }{\partial \theta} \Big) + \frac{1}{sin^2
\theta} \frac{\partial^2 }{\partial \psi^2} \Big] \Phi + M^2_b \Phi = 0,
\eqno(6.4.3)$$ 
where dot (.) over the variable denotes the derivative with respect to time.

The solution for this equation is taken as
$$\Phi = \Big[ (2 \pi )^{1/2} r \Big( 1 - \frac{r_0}{r} \Big)^{3/7} \Big]^{-1}
{\sum_{k,l,m}}  \Big[ A_{k l m} \Psi_{k l m} (t) Y_{l m}(
\theta, \psi) e^{ i k.r} + c.c. \Big],  \eqno(6.4.4)$$
where $r> r_0, m = - l \cdots +l, l = 1,2,3, \cdots and -\infty < k < \infty.$
$Y_{l m}(\theta, \psi)$ satisfies the equation
$$\Big[\frac{1}{sin \theta} \frac{\partial }{\partial \theta}
\Big({sin \theta} \frac{\partial }{\partial \theta} \Big) + \frac{1}{sin^2
\theta} \frac{\partial^2 }{\partial \psi^2} \Big] Y_{l m} = - l (l + 1)
Y_{l m} \eqno(6.4.5a)$$
with normalization as
$$ \int sin \theta { d \theta}{d \psi} Y_{l m} Y_{l^{\prime} m^{\prime}} =
\delta_{l l^{\prime}} \delta_{m m^{\prime}}. \eqno(6.4.5b)$$

Connecting eq.(6.4.3) with eq.(6.4.4) and using eq.(6.4.5), it is obtained that
$$ e^{i k.r} \Big[{\ddot \Psi_{k l m}} + \frac{3 {\dot a}}{a} {\dot
\Psi_{k l m}} + \Big(\Big\{\frac{k^2}{a^2} + \frac{ l (l + 1)}{a^2 r^2} +
 \frac{ 6 r_0 ( 2 r - r_0 )}{7 a^2 r^2 ( r - r_0 )^2}\Big\} \Big[ 1
- \frac{r_0}{r} \Big]^{-4/7}$$ $$ + M^2_b \Big) \Psi_{k l m} \Big]  = 0
.  \eqno(6.4.6)$$

Using the convolution theorem \cite{jm} and integrating over $r$, eq.(6.4.6) looks like
$$ {\int^{\infty}_{r_0 + \epsilon}}e^{i k.r} {dr} \Big[{\ddot \Psi_{k l m}} +
\frac{3 {\dot a}}{a} {\dot 
\Psi_{k l m}} + \Big\{\frac{k^2}{a^2 \eta } {\int^{\infty}_{r_0 + \epsilon}}
e^{ - i k.y} \Big[ 1 - \frac{r_0}{y}
\Big]^{- 4/7}{dy} $$ $$+ \frac{ l (l + 1)}{a^2 \eta} {\int^{\infty}_{r_0 +
\epsilon}} \frac{e^{ - i k.y}}{y^2} \Big[ 1 - \frac{r_0}{y}
\Big]^{-4/7} {dy} + \frac{6r_0}{7 a^2 \eta}{\int^{\infty}_{r_0 +
\epsilon}} \frac{( 2 y - r_0 )}{ y^2 ( y - r_0 )^2} \Big[ 1 
- \frac{r_0}{y} \Big]^{-4/7} {d y}$$ $$+ M^2_b \Big\} \Psi_{k l m} \Big]  = 0
.  \eqno(6.4.7)$$ 
Here $\epsilon$ is an extremely small positive real number.

Details of the evaluation of integrals with respect to $y$ are given in
the Appendix B. Using these results, in eq.(6.4.7), one obtains
$${\ddot \Psi_{k l m}} + \frac{3 {\dot a}}{a} {\dot \Psi_{k l m}} + \Big[
\frac{X_{k l m}}{a^2} + M^2_b \Big] \Psi_{k l m} = 0 , \eqno(6.4.8a)$$
where
$$X_{k l m} = ( 1/a^2 \eta ) \epsilon^{3/7} (r_0 + \epsilon)^{4/7}
cos \{ k (r_0 + \epsilon)\} 
\Big[  k^2 + \frac{ l ( l + 1)}{ ( r_0 + \epsilon )^2}$$  $$ -
\frac{12}{7} r_0 \epsilon^{-2} (r_0 + \epsilon)^{-1} + \frac{6}{ 7} r_0^2
\epsilon^{-2} ( r_0 + \epsilon )^{-2} \Big] . \eqno(6.4.8b)$$

\bigskip

\noindent \underline{Case 1: The case of state $ < {\tilde R}> = 0$}

 \bigskip
 
Using $a(t)$ from the line-element (6.1.33) for this state, the equation (6.4.8a) is
written as

$${\ddot \Psi_{k l m}} + \frac{3 }{ 2 ( t + a_{10}^2 t_P)} {\dot \Psi_{k l
m}} + \frac{X_{k l m} t_P}{ (t + t_P a_{10}^2)}  \Psi_{k l m} = 0 ,
\eqno(6.4.9)$$ 
which yields the solution for $t>0$ as
$$ \Psi_{k l m} = \tau^{ - 1/2} \Big[ A J_{-1/2} (\tau) + B Y_{-1/2}
(\tau) \Big] , \eqno(6.4.10a)$$
where $A$ and $B$ are integration constants. Here $\tau$ is defined as
$$ \tau =  \sqrt{t_P X_{k l m}  ( t + t_P a_{10}^2 )}  \eqno(6.4.10b)$$
showing that $\tau \to \infty$ when $t \to \infty.$

For large $\tau$,
$$J_{-1/2} (\tau) \simeq \frac{ cos (\tau)}{\sqrt{\pi\tau/2}} \quad{\rm
and} \quad Y_{-1/2} (\tau) \simeq \frac{ sin (\tau)}{\sqrt{\pi\tau/2}}.$$
So, when $\tau$ is large,
\begin{eqnarray*}
\Psi_{k l m} &=&  [ \pi \tau^2/2]^{-1/2} [ A cos \tau  +  B sin \tau] \\ &
= & \Big[ \frac{\tau_1^3}{2 t_P X_{k l m}}\Big]^{1/2} \tau^{-1} \Big[(1 + i)
e^{-i \tau} + (1 - i) e^{i \tau} \Big]
\end{eqnarray*}
$$ \eqno(6.4.11)$$   
using the normalization condition
$$ (\Phi_{k l m} , \Phi_{k l m}) = 1  = - (\Phi^*_{k 
l m} , \Phi^*_{k l m}), \eqno(6.4.12a)$$ 
where 
$$\Phi_{k l m} = \Big[( 2 \pi \eta^{-1} W)^{1/2} r \Big( 1  - \frac{r_0}{r} \Big)^{1/7}
\Big]^{-1} \Psi_{k l m} (t) e^{ i k.r} Y_{lm} \eqno(6.4.12b)$$  
and the scalar product is defined as
$$ (\Phi_{k l m} , \Phi_{k l m}) = -
i {\int_{r_0 + \epsilon}^{\infty}}{\int_0^{\pi}}
{\int_0^{2\pi}}\sqrt{ - g_{\Sigma}}  {d{\Sigma}} [ 
\Phi_{k l m} (\partial_t \Phi^*_{k^{\prime}l^{\prime} m^{\prime}})$$  $$ -
(\partial_t \Phi_{k l m}) \Phi^*_{k^{\prime}l^{\prime} m^{\prime}}] |Y_{l m}(
\theta, \psi)|^2 , 
\eqno(6.4.12c)$$ 
where $\Sigma$ is the $t = t_1$ hypersurface, $\sqrt{ - g_{\Sigma}} = r^2
\Big[1 - \frac{r_0}{r} \Big]^{6/7}$  and $ d\Sigma = sin \theta dr
d\theta d\psi.$ 

Thus
$$\Psi^{\rm out}_{k l m} = \Big[ \frac{\tau_1^3}{2 t_P X_{k l
m}}\Big]^{1/2} \tau^{-1} (1 + i) e^{-i \tau}.  \eqno(6.4.13)$$  

For $t<0,$ eq.(6.4.9) yields the solution
$$ \Psi_{k l m} = {\tilde \tau}^{ - 1/2} \Big[ A^{\prime} J_{-1/2}
({\tilde \tau}) 
+ B^{\prime} Y_{-1/2} ({\tilde \tau}) \Big] , \eqno(6.4.14a)$$
where $A^{\prime}$ and $B^{\prime}$ are integration constants. Here $\tilde \tau$
is defined as $$ {\tilde \tau} = -  \sqrt{t_P X_{k l m} (- t + t_P a_0^2
)}  \eqno(6.4.14b)$$ 
showing that $\tilde \tau \to - \infty$ when $t \to - \infty.$

Using above approximations for Bessel's functions and normalization
prescription, it is obtained that for 
$t \to - \infty$
$$\Psi^{\rm in}_{k l m} =  \Big[ \frac{{\tilde \tau}_1^3}{2 t_P X_{k l
m}}\Big]^{1/2} {\tilde \tau}^{-1} (1 + i) e^{+i {\tilde \tau}}.
\eqno(6.4.15)$$   

Since $\Phi^{\rm out}_{k l m}$ and $\Phi^{\rm in}_{k l m}$ both belong to
the same Hilbert space, so one write
\begin{eqnarray*}
\Phi &=& {\sum_{k,l,m}}  \Big[ A^{\rm in}_{k l m} \Phi^{\rm in}_{k
l m}  Y_{l m}( \theta, \psi)  +   A^{{\rm in}\dagger}_{k l m} \Phi^{*{\rm
in}}_{k l m}  Y_{l m}( \theta, \psi)\Big] \\ &=&  
{\sum_{k,l,m}}  \Big[ A^{\rm out}_{k l m} \Phi^{\rm out}_{k l m}
 Y_{l m}( \theta, \psi)  +   A^{{\rm out}\dagger}_{k l m} \Phi^{*{\rm
out}}_{k l m}  Y_{l m}( \theta, \psi)\Big] .
\end{eqnarray*}

As a result, one obtains
$$\Phi^{\rm out}_{k l m} = \alpha_{k l m} \Phi^{\rm in}_{k l m}  +
\beta_{k l m} \Phi^{*{\rm in}}_{k l m} , \eqno(6.4.16a)$$  
where $\alpha_{k l m}$ and $\beta_{k l m}$ are Bogoliubov coefficients
satisfying the condition \cite{pdb, ndb, bsd}
$$|\alpha_{k l m}|^2  - |\beta_{k l m}|^2  = 1. \eqno(6.4.16b)$$

The in- and out- vacuum states are defined as
$$A^{\rm in}_{k l m}|{\rm in}> = 0 = A^{\rm out}_{k l m}|{\rm out}>
.\eqno(6.4.17a,b)$$ 
Moreover,
$$A^{\rm out}_{k l m} = \alpha_{k l m} A^{\rm in}_{k l m}  +
\beta_{k l m} A^{*{\rm in}}_{k l m} , \eqno(6.4.17c)$$

Connecting eqs.(6.4.11) - (6.4.17c), it is obtained that
\begin{eqnarray*}
\alpha_{k l m} & =& {1 \over 2} e^{ - i ( \tau + {\tilde \tau})}\Big[
\Big( \frac{ t_1 + a_{10}^2 t_P}{ - t_1 + a_{10}^2 t_P}\Big)^{1/4} + \Big(
\frac{ - t_1 + a_{10}^2 t_P}{  t_1 + a_{10}^2 t_P}\Big)^{1/4}\Big] \\
\beta_{k l m} & =& {1 \over 2} e^{ - i ( \tau + {\tilde \tau})}\Big[ 
\Big( \frac{ t_1 + a_{10}^2 t_P}{ - t_1 + a_{10}^2 t_P}\Big)^{1/4} - \Big(
\frac{ - t_1 + a_{10}^2 t_P}{  t_1 + a_{10}^2 t_P}\Big)^{1/4}\Big]  .
\end{eqnarray*}
$$\eqno(6.4.18a,b)$$ 
These results yield
$$ \beta_{k l m} \simeq i\frac{a_{10}^2 t_P}{ t_1} e^{-i(\tau_1 + {\tilde
\tau}_1)}   \eqno(6.4.19a)$$ 
implying
$$ |\beta_{k l m}|^2 \simeq \Big[\frac{a_{10}^2 t_P}{ t_1}\Big]^2 =
\Big[\frac{5 t_P}{ t_1}\Big]^2.  
\eqno(6.4.19b)$$

 Eq.(6.4.19b) shows that when $t_1$ is sufficiently larger than $5 t_P$, $
 |\beta_{k l m}|^2 = 0.$ It means that spinless particles will  be created in the state
$<{\tilde R}> = 0$ only when $t_1 \le 5 t_P$. For $t_1$ sufficiently larger
 than $5 t_P$, there will be no production of scalar particles, in this state.

Anamolous terms emerge when vacuum expectation value of the
energy-momentum tensor components $T_{\mu\nu}$ is evaluated for $\Phi$, if
dimension of the space-time is an even integer. These terms get cancelled
for the difference of vacuum 
expectation values $T_{\mu\nu}$ in  in- and out-states. Thus $T_{\mu\nu}$
for created particles is defined as \cite{em, sk05}
$$<T_{\mu\nu}> =   <{\rm out}|
T_{\mu\nu}|{\rm out}> - <{\rm in}| T_{\mu\nu}|{\rm in}> .  \eqno(6.4.20a)$$

 $T_{\mu\nu}$ for $\phi$ are obtained from the  action ( 6.4.1a) as
$$T_{\mu\nu} = {\partial_{\mu}\Phi^*}{\partial_{\nu}\Phi}  - 2 \eta
{\Lambda}[R_{\mu\nu} + \{ ({\tilde R}\Phi^* \Phi)_{; \mu\nu} - g_{\mu\nu}
{\Box}({\tilde R}\Phi^* \Phi) \} $$
$$  - {1 \over2}g_{\mu\nu}[{\partial^{\rho}\Phi^*}{\partial_{\rho}\Phi} - {\Lambda}{\tilde R}^2 \Phi^* \Phi ]  \eqno( 6.4.20b)$$
with its trace
$$ T = - ( 1 -  12 \eta {\Lambda}{\tilde R}
){\partial^{\mu}\Phi^*}{\partial_{\mu}\Phi}  - 12 \eta \Lambda^2 <{\tilde R} >^3 \Phi^* \Phi  . \eqno( 6.4.20c)$$

Now trace of $T_{\mu\nu}$ for created partcle-antiparticle pairs is obtained
as

$$<T> =<{\rm out}| T |{\rm out} > - <{\rm in}| T |{\rm in} > \eqno( 6.4.20d)$$

Eqs.( 6.4.12b), ( 6.4.13), ( 6.4.15) and ( 6.4.20 d) lead to trace of created particles in the state $<{\tilde R}> = 0$ as

$$ <T_1(b)> = 0 . \eqno( 6.4.21)$$
 
\bigskip

\noindent \underline{Case 2: The case of state $ < {\tilde R}> =  {1 \over
2} T_c$}

 \bigskip

Using $a(t)$ from the line-element ( 6.1.34) for this state, the equation ( 6.4.8) is
written as

$${\ddot \Psi_{k l m}} + \frac{3 }{ 2} \sqrt{\frac{T_c}{6 \eta}} {\dot
\Psi_{k l m}} + \Big[ \frac{X_{k l m}}{a_0^2} e^{ - (t -
t_0)  \sqrt{T_c/6 \eta}}  + M^2_b \Big]  \Psi_{k l m} = 0 ,
\eqno( 6.4.22)$$ 
where $M^2_b = {1 \over 4}\Lambda T_c^2$ using the definition of $M_b$ from eq.( 6(iv).2). The equation ( 6.4.22) yields the
solution for $t>0$ as
$$ \Psi_{k l m} = \Big[-\frac{\pi}{sin h 2 \pi \alpha}\Big]^{1/2}\Big[
\frac{6 \eta }{ T_c}\Big]^{1/4}
e^{ - \frac{3}{4} (t - t_{10}) \sqrt{ T_c/6 \eta}}\times$$ $$ J_{\pm 2i \alpha} \Big(
\gamma_{k l m} e^{ - {1 \over 2} (t - t_{10}) \sqrt{T_c/6 \eta}} \Big),
\eqno( 6.4.23a)$$ 
where 
$$\alpha^2 = \frac{6 \eta}{T_c} \Big[{1 \over 4}\Lambda T_c^2 - \frac{3
T_c}{32 \eta} \Big] \eqno( 6.4.23b)$$
and
$$\gamma^2_{k l m} = \frac{24 \eta X_{k l m}}{ a_0^2 T_c} .\eqno( 6.4.23c)$$ 

For $t<0,$ the equation ( 6.4.22) yields the
solution  as
$$ \Psi_{k l m} =  \Big[-\frac{\pi \alpha}{\eta sin h (2 \pi \alpha)}\Big]^{1/2}\Big[
\frac{6 \eta }{ T_c}\Big]^{1/4} e^{\frac{3}{4} (t - t_{10}) \sqrt{ T_c/6
\eta}}\times $$ 
$$J_{\mp 2i \alpha} \Big( \gamma_{k l m} e^{  {1 \over 2}(t - t_{10}) \sqrt{T_c/6
\eta}} \Big). \eqno( 6.4.24)$$

Using the approximation of
 $$J_n (x) \simeq  \frac{x^n}{ 2^n \Gamma (1 + n)}$$
for small $x$ in eqns.( 6.4.23) and ( 6.4.24) ,
for $t \to \infty$
$$ \Psi^{\rm out}_{k l m} =  \Big[- \frac{\pi \alpha}{\eta sin h (2 \pi
\alpha})\Big]^{1/2}\Big[ \frac{6 \eta }{ T_c}\Big]^{1/4}
\Big(\frac{1}{\Gamma(1\pm 2 i \alpha)}\Big) 
\Big(\frac{\gamma_{k l m}}{2}\Big)^{\pm 2 i \alpha}\times$$ $$
e^{- (t-t_{10}) (\frac{3}{4} \mp  i \alpha )\sqrt{ T_c/6 \eta}} , 
\eqno(6.4.25a)$$ 
and for $t \to -\infty$
$$ \Psi^{\rm in}_{k l m} =   \Big[-\frac{\pi \alpha}{\eta sinh 2 \pi
\alpha}\Big]^{1/2}\Big[  \frac{6 \eta }{ T_c}\Big]^{1/4}
 \Big(\frac{1}{\Gamma(1\mp 2 i \alpha)}\Big)
\Big(\frac{\gamma_{k l m}}{2}\Big)^{\mp 2 i \alpha} \times$$ $$ e^{(t -
t_{10})(\frac{3}{4} \mp  i \alpha )\sqrt{ T_c/6 \eta} }, 
\eqno(6.4.25b)$$ 

Using $ \Psi^{\rm out}_{k l m} $ and $ \Psi^{\rm in}_{k l m}$, one obtains
$$\alpha_{k l m} = - i \Big(\frac{3}{2} \mp 2 i
\alpha  \Big)   \eqno(6.4.26a)$$  
and
$$\beta_{k l m} = - \frac{3i}{2}  \eqno(6.4.26b)$$ 
implying that $$|\beta_{k l m}|^2  = \frac{9}{4} , \eqno(6.4.26c)$$ 
which ensures creation of spinless particles.
Connecting eqs.(6.4.16c) and (6.4.26b)

$$|\alpha_{k l m}|^2 = \frac{13}{4} \eqno(6.4.26d)$$ 
Moreover, the absolute probability for no particle creation is given as
$$|<{\rm out}|{\rm in}>|^2 = {\prod_{k l m}} |\alpha_{k l m}|^{-2} =
\frac{4 }{9 + 16 \alpha^2}.  \eqno(6.4.27)$$ 

Eqs.(6.4.26 d) and (6.4.27) yield

$$ \alpha^2 = 1/4 .   \eqno(6.4.28)$$ 

Connecting eqs.(6.4.2), (6.4.23b) and (6.4.28), it is obtained that

$$ M_b^2 = \frac{13}{16} \Big( \frac{T_c}{6 \eta} \Big). \eqno(6.4.29)$$ 

Using the convolution theorem, eq.(6.4.12b) can be written as

\begin{eqnarray*}
\Phi_{klm}(t, r, \theta, \phi) &=& \Big[ (2 pi)^{-1/2} e^{- ikr} \int_{r_0 +
  \epsilon}^{\infty} \Big\{ y \Big(1 - \frac{r_0}{y} \Big) \Big\}^{-1} e^{iky}
  {dy} \Big] \Psi_{klm} (t) Y_{lm} (\theta, \phi) \\ &=& (2 pi)^{-1/2} e^{-
  ikr}\frac{\epsilon^{4/7}}{(r_0 +  \epsilon)^{11/7}} cos[k(r_0 +  \epsilon)] \Psi_{klm} (t) Y_{lm} (\theta, \phi).
\end{eqnarray*}
$$ \eqno(6.4.30)$$

Connecting eqs. (6.4.20 d), (6.4.25) and (6.4.30) as well as taking average
over $\theta$ and $\phi$, trace of the energy-momentum tensor for created
particles, in the  state $ {\tilde R} = (1/2) T_c,$ is obtained at $t =
t_{2e}$ as

\begin{eqnarray*}
T_{2(b)} &=& \sum_{klm}(1/2 \pi) \Big(\frac{6 \eta}{T_c} \Big)^{1/2} sinh
[(3/4)\sqrt{\frac{T_c}{6 \eta}} (t_{2e} - t_{1e}) ] \\&& \times\frac{\epsilon^{8/7}}{( r_0 + \epsilon )^{22/7}} cos^2 k(r_0 + \epsilon ) \Big[(13/16) \frac{T_c}{6  \eta} ( 6\eta \Lambda T_c - 1 ) - (3/2) \eta \Lambda^2 T_c^3 \Big].
\end{eqnarray*}
$$ \eqno(6.4.31)$$

\bigskip
 
 \centerline{\bf 6.5. Creation of spin-1/2 particles in inhomogeneous models}

\smallskip

In the homogeneous model (6.1.33) and (6.1.34), the general
solution of the Dirac equation (6.3.7a) is taken as $\psi$ and 
 $ \psi^{\dag}$, given by eqs.(6.3.8a) and (6.3.8b) with

$$\psi_{I k,s} =  \Big[ (2 \pi \eta)^{1/2} r \Big( 1 - \frac{r_0}{r}
\Big)^{1/7} \Big]^{-1} {\sum_{l,m}}  f_{k l m ,s} (t) Y_{l m}(
\theta, \phi) e^{ i k.r} u_s   \eqno(6.5.1a)$$
and
$$\psi_{II k,s} =  \Big[ (2 \pi  \eta)^{1/2} r \Big( 1 - \frac{r_0}{r}
\Big)^{1/7} \Big]^{-1} {\sum_{l,m}}  g_{k l m ,s} (t) Y_{l m}(
\theta, \phi) e^{-i k.r} {\hat u}_s .  \eqno(6.5.1b)$$
with $g_{k l m ,s} (t) = f_{- k l m ,s} (t)$.

In eqs.(6.5.1a, b), $u_s$ and ${\hat u}_s$ are given by eqs.(6.3.9a,
b,c,d).Moreover, $ m = -l, \cdots,+l ; l = 1,2,3, \cdots$  and  $ - \infty <
k < \infty.$

Connecting eqs.(6.3.7a) and (6.5. 1a,b), it is obtained that
$$ \Big( i \gamma^{\mu}D_{\mu}  -  M_f  \Big) \psi_{I k,s} = 0,
 \eqno(6(v).2a)$$  
$$ \Big( i \gamma^{\mu}D_{\mu}  -  M_f  \Big) \psi_{II k,s} = 0, 
\eqno(6(v).2b)$$  

Now using the operator $ \Big(- i \gamma^{\mu}D_{\mu}  -  M_f  \Big)$
 from the
left in eq.(6.5.2a) as
$$\Big(- i \gamma^{\mu}D_{\mu}  -  M_f  \Big) \Big( i \gamma^{\mu}D_{\mu}  -
M_f  \Big) \psi_{I k,s} = 0, $$   
it is obtained that \cite{sks99}
$$ \Big({\Box} + (4 \eta)^{-1} <{\tilde R}> + g^2 <{\tilde R}>^2
\Big)\psi_{I k,s} = 0, \eqno(6.5.3a)$$

Similarly,  from eq.(6.5.2b), it is obtained that
$$ \Big({\Box} + (4 \eta)^{-1} <{\tilde R}> + g^2 <{\tilde R}>^2
\Big)\psi_{II k,s} = 0. \eqno(6.5.3b)$$

When temperature $T$ is not very much below $T_c $
one obtains the vacuum state $<{\tilde R}> = 0,$ where eqs.(6.5.9a,b)
reduce to
$$ {\Box}\psi_{I k,s} = 0 = {\Box}\psi_{II k,s} \eqno(6.5.4a,b) $$

Particles are created due to conformal symmetry breaking, which is caused by the mass term. In the state  $<{\tilde R}> = 0,$ there is no term to break this symmetry. So, production of spin-1/2 particles are not possible in this state.

As given in  section 6.1, one obtains the state
 $<{\tilde R}> =
{1 \over 2} T_c$ when the temperature falls sufficiently below
 $T_c.$ In this
state, the universe obeys the geometry given by the line-element (6.1.34) .

In what follows, creation of spin-1/2 particles is investigated 
in the
model given by this line-element. In the background geometry of this model,
 for
every $k, l, m $ and $s,$ eq.(6.5.3a,b) look like

$$ e^{i k.r} \Big[{\ddot f_{k l m,s}} + \frac{3 {\dot a}}{a} {\dot
f_{k l m,s}} + \Big(\Big\{\frac{k^2}{a^2} + \frac{ l (l + 1)}{a^2 r^2} +
 \frac{ 6 r_0 ( 2 r - r_0 )}{7 a^2 r^2 ( r - r_0 )^2}\Big\} \Big[ 1
- \frac{r_0}{r} \Big]^{-4/7}$$ $$ + {\tilde M}^2_f \Big) f_{k l m,s} \Big]  = 0
  \eqno(6.5.5a)$$ 
connecting eqs.(6.5.1a) and (6.5.3a).. Here
$${\tilde M}^2_f = (4\eta)^{-1} <{\tilde R}> + g <{\tilde R}>^2$$
$$ = \frac{T_c}{8\eta} + \frac{g^2 T_c^2}{4}.   \eqno(6.5.5b)$$ 

Using the convolution theorem \cite{jm} as above and integrating over $r$,
 one obtains that
$$ {\int^{\infty}_{r_0 + \epsilon}}e^{i k.r} {dr}
 \Big[{\ddot f_{k l m,s}} +
\frac{3 {\dot a}}{a} {\dot 
f_{k l m,s}} + \Big\{\frac{k^2}{a^2 \eta } {\int^{\infty}_{r_0
 + \epsilon}}
e^{ - i k.y} \Big[ 1 - \frac{r_0}{y}
\Big]^{- 4/7}{dy} $$ $$+ \frac{ l (l + 1)}{a^2 \eta} 
{\int^{\infty}_{r_0 +
\epsilon}} y^{-2} \Big[ 1 - \frac{r_0}{y}
\Big]^{-4/7}e^{ - i k.y} {dy} + \frac{6r_0}{7 a^2 \eta} \times $$
 $${\int^{\infty}_{r_0 +
\epsilon}} \frac{( 2 y - r_0 )}{ y^2 ( y - r_0 )^2} \Big[ 1 
- \frac{r_0}{y} \Big]^{-4/7} e^{ - i k.y} {d y}+ {\tilde M}^2_f \Big\} f_{k
l m,s} \Big]  = 0 .  \eqno(6.5.6)$$ 

Details of the evaluation of integrals with respect to $y$ are
 given in
the Appendix B. Using these results, in eq.(6.5.7), one obtains
$${\ddot f_{k l m,s}} + \frac{3 }{2} \sqrt{\frac{T_c}{6\eta}}
 {\dot f_{k l
m,s}} + \Big[ \frac{X_{k l m}}{a^2_{10}} e^{-(t - t_{10})
\sqrt{T_c/6\eta}} +
{\tilde M}^2_f \Big] f_{k l m,s} = 0 , \eqno(6.5.7a)$$
where
$$X_{k l m,s} = ( 1/a^2 \eta ) \epsilon^{3/7}
 (r_0 + \epsilon)^{4/7}
Cos \{ k (r_0 + \epsilon)\} 
\Big[  k^2 + \frac{ l ( l + 1)}{ ( r_0 + \epsilon )^2}$$  $$ -
\frac{12}{7} r_0 \epsilon^{-2} (r_0 + \epsilon)^{-1} + \frac{6}{ 7} r_0^2
\epsilon^{-2} ( r_0 + \epsilon )^{-2} \Big] . \eqno(6.5.7b)$$ 

Eqs.(6.5.8a,b) yield the solution for $t>0$,
$$f_{k l m,s} = C_1 e^{\pm \frac{3}{4} (t - t_{10})
 \sqrt{T_c/6\eta}}
J_{\pm 2 i \alpha}\Big(\gamma_{k l m, s} 
 e^{ - \frac{1}{2} (t - t_{10})
\sqrt{T_c/6\eta}} \Big),  \eqno(6.5.8a)$$ 
where $C_1$ is a normalization constant,
$$\gamma^2_{k l m, s} = \frac{ 24 \eta X_{k l m, s}}{a^2_{10} T_c}
\eqno(6.5.8b)$$  
and
$$ \alpha^2 = \frac{6 \eta}{T_c} \Big[{\tilde M}^2_f
 - \frac{3 T_c}{32 \eta}
\Big] = \frac{6 \eta}{T_c} \Big[\frac{1}{4} \sigma^2 T_c^2 +
\frac{5T_c}{32 \eta} \Big]  \eqno(6(v).8c)$$  
using the definition of ${\tilde M}^2_f$ from eq.(6.5.5b).

Connecting eqs.(6.3.9a,b,c,d) and (6.5.8a,b)

$$\psi_{I k,s} =  \Big[ (2 \pi \eta)^{1/2} r \Big( 1 - \frac{r_0}{r}
\Big)^{1/7} \Big]^{-1} {\sum_{l,m}} C_1
 e^{- \frac{3}{4} (t - t_{10})
\sqrt{T_c/6\eta}} \times$$
$$J_{\pm 2 i \alpha}\Big(\gamma_{k l m, s} 
 e^{ - \frac{1}{2} (t - t_{10})
\sqrt{T_c/6\eta}} \Big)  Y_{l m}(\theta, \phi)
 e^{ i k.r} u_s . \eqno(6.5.9)$$

Normalization of $\psi_{I k,s}$ is done using the prescription (6.3.22c)
 at the $t =
t_c$ hypersurface denoted as $\Sigma$ onwards. As a result,
 $C_1$ is
obtained as 
$$ C_1 = \Big| J_{\pm 2 i \alpha}(\gamma_{k l m, s} ) \Big|^{-1}.
\eqno(6.5.10)$$ 
Thus
$$\psi_{I k,s} =  \Big[ (2 \pi \eta)^{1/2} r \Big( 1 - \frac{r_0}{r}
\Big)^{1/7} \Big]^{-1} {\sum_{l,m}} \Big| J_{\pm 2 i \alpha}
(\gamma_{k l
m, s} ) \Big|^{-1}  e^{- \frac{3}{4} (t - t_c)
\sqrt{T_c/6\eta}} \times$$
$$J_{\pm 2 i \alpha}\Big(\gamma_{k l m, s} 
 e^{ - \frac{1}{2} (t - t_c)
\sqrt{T_c/6\eta}} \Big)  Y_{l m}(\theta, \phi) e^{ i k.r} u_s .
 \eqno(6.5.11)$$

Similarly  
$$\psi_{II k,s} =  \Big[ (2 \pi \eta)^{1/2} r \Big( 1 - \frac{r_0}{r}
\Big)^{1/7} \Big]^{-1} {\sum_{l,m}}
 \Big| J_{\pm 2 i \alpha}(\gamma_{-k l
m, s}) \Big|^{-1}  e^{- \frac{3}{4} (t - t_c)
\sqrt{T_c/6\eta}} \times$$
$$J_{\pm 2 i \alpha}\Big(\gamma_{-k l m, s} 
 e^{ - \frac{1}{2} (t - t_c)
\sqrt{T_c/6\eta}} \Big)  Y_{l m}(\theta, \phi)
 e^{- i k.r} {\hat u}_s .
\eqno(6.5.12)$$ 

Asymptotics of these solutions, when $t\to \infty$, yield
$$\psi^{\rm out}_{I k,s} = 
 \Big[ (2 \pi \eta)^{1/2} r \Big( 1 - \frac{r_0}{r}
\Big)^{1/7} \Big]^{-1} {\sum_{l,m}}
 \Big| J_{\pm 2 i \alpha}(\gamma_{k l
m, s} ) \Big|^{-1}  e^{- \frac{3}{4} (t - t_c)
\sqrt{T_c/6\eta}} \times$$
$$ \Big[\Gamma{(1 - 2 i \alpha)}\Big]^{-1}
 \Big(\frac{\gamma_{k l m,
s}}{2}\Big)^{-2i\alpha}  e^{ + i \alpha (t - t_c) 
\sqrt{T_c/6\eta}}   Y_{l m}(\theta, \phi) e^{ i k.r} u_s 
. \eqno(6.5.13a)$$
 and
 $$\psi^{\rm out}_{II k,s} =  \Big[ (2 \pi \eta)^{1/2} r
 \Big( 1 - \frac{r_0}{r}
\Big)^{1/7} \Big]^{-1} {\sum_{l,m}} \Big| J_{\pm 2 i \alpha}
(\gamma_{-k l
m, s} ) \Big|^{-1}  e^{- \frac{3}{4} (t - t_c)
\sqrt{T_c/6\eta}} \times$$
$$ \Big[\Gamma{(1 - 2 i \alpha)}\Big]^{-1}
 \Big(\frac{\gamma_{-k l m,
s}}{2}\Big)^{-2i\alpha}  e^{ + i \alpha (t - t_c) 
\sqrt{T_c/6\eta}}   Y_{l m}(\theta, \phi)
 e^{- i k.r} {\hat u}_s .
\eqno(6.5.13b)$$ 
For $t\to - \infty$, one obtains
$$\psi^{\rm in}_{I k,s} = 
 \Big[ (2 \pi \eta)^{1/2} r \Big( 1 - \frac{r_0}{r}
\Big)^{1/7} \Big]^{-1} {\sum_{l,m}}
 \Big| J_{\pm 2 i \alpha}(\gamma_{k l
m, s}) \Big|^{-1}  e^{ \frac{3}{4} (t - t_c)
\sqrt{T_c/6\eta}} \times$$
$$ \Big[\Gamma{(1 + 2 i \alpha)}\Big]^{-1} 
\Big(\frac{\gamma_{k l m,
s}}{2}\Big)^{-2i\alpha}  e^{ - i \alpha (t - t_c) 
\sqrt{T_c/6\eta}}   Y_{l m}(\theta, \phi) e^{ i k.r} u_s .
 \eqno(6.5.14a)$$
 and
 $$\psi^{\rm in}_{II k,s} = 
 \Big[ (2 \pi \eta)^{1/2} r \Big( 1 - \frac{r_0}{r}
\Big)^{1/7} \Big]^{-1} {\sum_{l,m}}
 \Big| J_{\pm 2 i \alpha}(\gamma_{-k l
m, s}) \Big|^{-1}  e^{ \frac{3}{4} (t - t_c)
\sqrt{T_c/6\eta}} \times$$
$$ \Big[\Gamma{(1 + 2 i \alpha)}\Big]^{-1}
\Big({\frac{\gamma_{-klm,s}}{2}}\Big)^{-2i\alpha} 
 e^{ - i \alpha (t -
t_c)\sqrt{T_c/6\eta}}  Y_{l m}(\theta, \phi) e^{- i k.r} {\hat
u}_s . \eqno(6.5.14b)$$  
In eqs.(6.5.13a,b) - (6.5.14a,b), $\Gamma{(x)}$ stands for the gamma function
 of $x.$

The decomposed form of $\psi$ and $\psi^{\dag}$ are given by
eqs.(6.3.25a,b) and eqs.(6.3.25c,d) respectively.

Connecting eqs.(6.5.13a), (6.5.14a) and (6.3.27b)
\begin{eqnarray*}
\alpha_{k.s} &=& (1/2)\Big[\Gamma{(1 - 2 i \alpha)}\Big]^{-2}
\Big(\frac{\gamma_{k l m, s}}{2}\Big)^{-4i\alpha} \Big| J_{\pm 2 i
\alpha}(\gamma_{k l m, s} ) \Big|^{-2} \\ & =& \frac{sinh (2 \pi \alpha)} {(4
\pi \alpha)}\Big[\Gamma{(1 - 2 i \alpha)}\Big]^{-2}.
\end{eqnarray*}
 $$ \eqno(6.5.15a)$$ 
Similarly eqs.(6.5.13b), (6.5.14b) and (6.3.27c)
\begin{eqnarray*}
\beta_{k.s} &=& (1/2)\Big[\Gamma{(1 + 2 i \alpha)}\Big]^{-2}
\Big(\frac{\gamma_{k l m, s}}{2}\Big)^{-4i\alpha} \Big| J_{\pm 2 i
\alpha}(\gamma_{k l m, s} ) \Big|^{-2} \\ & =& \frac{sinh (2 \pi \alpha)} {(4
\pi \alpha)}\Big[\Gamma{(1 + 2 i \alpha)}\Big]^{-2}.
\end{eqnarray*}
 $$ \eqno(6.5.15b)$$

Conditions (6.3.27a) and eqs.(6.5.15a,b) yield an useful result
 $$ |\alpha_{k.s}|^2 = |\beta_{k.s}|^2 = \frac{1}{4} \eqno(6.5.16)$$
suggesting that $\alpha$ should be very small. 
using the symmetry $k \to -k$ caused by symmetry of the model under
 $r \to -r.$

The relative probability of creation of a particle - antiparticle
 pair is given as 
$$\omega_{k l m,s} = \Big| \beta_{k.s}/\alpha_{k.s} \Big|^2 = 1.
\eqno(6.5.17)$$  

Absolute probability of the creation of particle - antiparticle
 pairs
requires the total probability of creating $0,1,2, \cdots$ pairs
 to be
unity \cite{yab71}, which means that
$$N_{k l m,s}(1 + \omega_{k l m,s} + \omega_{k l m,s}^2 + \cdots)
 = 1, $$
where $N_{k l m,s}$ is the probability of creation of no pair of
particle-antiparticle. Using eq.(6(v).17), it is obtained that
$$N_{k l m,s} = \frac{1}{1 + 1 + 1 + \cdots} = 0. $$
It shows that probability of vacuum to remain vacuum is
$$|<{\rm out}| {\rm in}>|^2 = {\prod_{k l m,s}}N_{k l m,s} = 0.
\eqno(6.5.18)$$     
implying certainity of creation of particle-antiparticle pairs.

The action $S$, given by eq.(6.3.1), yields components of energy - momentum
tensor for $\psi$ field  as \cite{bsd, skb}
\begin{eqnarray*}
T_{\mu\nu} &=& \frac{1}{\sqrt{-g}}
 \frac{\delta S}{\delta g^{\mu\nu}} \\ &=& i{\bar \psi}  \gamma_{\mu} D_{\nu}
 \psi - \sigma \eta R_{\mu\nu}{\bar \psi}\psi - {1 \over 2}
g_{\mu\nu} {\bar \psi} \Big( i  \gamma^{\rho} D_{\rho}  - \sigma {\tilde R}  \Big)
 \psi \\&& - \sigma \eta ( D_{\mu} D_{\nu} - g_{\mu\nu} {\Box} ){\bar \psi}\psi + c.c. 
\end{eqnarray*}
$$ \eqno(6.5.19)$$

Connecting eqs.(6.5.20a) and (6.5.19), energy density of created particles in the state ${\tilde R} = 1/2 T_c$ is obtained as
\begin{eqnarray*}
\rho = < T^0_0 > & = & < {\rm in}|- i {\bar \psi}\gamma^0 \partial_0   -
\sigma \eta R^0_0 {\bar \psi}\psi + 2 \sigma {\tilde R}{\bar \psi}\psi +
c.c.|{\rm in}>\\&& - < {\rm out}|- i {\bar \psi}\gamma^0 \partial_0   - \sigma
\eta R^0_0 {\bar \psi}\psi + 2 \sigma {\tilde R}{\bar \psi}\psi + c.c.|{\rm
  out}> \\ &=& < {\rm in}|- i {\bar \psi}\gamma^0 \partial_0   + 1/2 \sigma
T_c{\bar \psi}\psi + c.c.|{\rm in}> \\&& - < {\rm out}|- i {\bar \psi}\gamma^0 \partial_0    + 1/2 \sigma T_c{\bar \psi}\psi + c.c.|{\rm out}>
\end{eqnarray*}
$$ \eqno(6.5.20)$$ 
using $R^0_0 = 3 {\ddot a}/a = T_c / {2 \eta}.$

In- and out- $\psi$, given by eqs.(6.5.13a,b) and (6.5.141,b), recast eq.(6.5.20) as

$$ \rho = \sum_{klm}\frac{8 \alpha}{\eta} \frac{\epsilon^{8/7}}{(r_0 + \epsilon)^{22/7}}
cos^2 [k(r_0 + \epsilon)] sinh[(3/2)(t - t_c) \sqrt{T_c/6\eta}] \eqno(6.5.21)$$
using convolution theorem for $\psi$ as given in the preceding section.

Similarly trace of stress - tensor for created spin-1/2 particles is given as

$$ < T > = <{\rm in}| T_{\mu\nu} |{\rm in} > - <{\rm out}| T_{\mu\nu} |{\rm
  out} > = 0  , \eqno(6.5.22)$$
using In- and out- solutions for $\psi$, given by eqs.(6(v).13a,b) and (6.5.14a,b).

Eq.(6(v).16) shows number density  of created spin-1/2 particles as

$$|\beta_k|^2 = \frac{1}{2} \eqno(6.5.23)$$
yielding energy density of created particles also as

$$ \rho = \frac{1}{2} M_f . \eqno(6.5.24)$$

 Here summation is done with the help of the Riemann
zeta function defined as $\zeta(s) = {\sum_{n=1}^{\infty}} n^{-s}.$  Moreover,
\begin{eqnarray*}
{\sum_{k l m}}cos^2 [k(r_0 + \epsilon)] &=& {\sum_{k l m}}[1 - k^2 (r_o + \epsilon)^2 +
\cdots]\\ & = & {\sum_{k l m}} 1 = {\sum_{k l }}(2 l + 1)\\ &=&{\sum_{k  }}[2 \zeta(-1) + \zeta(0)] = - (4/3)  \zeta(0) = 2/3
\end{eqnarray*}
as $\zeta (-2m) = 0$. Also $\zeta (0) = - {1 \over 2}$ and $\zeta (-1 ) = - {1 \over 12}$ obtained through the analytic continuation.

\vspace{0.5cm}
\centerline{\bf 7. Contribution of riccion to dark energy and dark matter}
\centerline{\bf of the universe}

\bigskip
Experimental probes, like luminosity measurements of high redshift supernova
\cite{ar}, anisotropies in cosmic microwave background \cite{dn} and observational
gravitational clustering \cite{ja}, strongly indicate late time accelerated
expansion of the universe. Theoretically, accelerated expansion of the
universe can be obtained either by modifying left hand side of Einstein's
equations, like introduction of C-field in $steady-$ $state$ $theory$ adhered
to the Perfect Cosmological Principle \cite{fh} or by using energy momentum tensor,having dominance of $exotic$ matter, with negative pressure violating the
$strong$ $energy$ $condition$. This kind of matter is known as $dark $
$energy$ (vacuum energy), which has drawn much attention of cosmologists
today. Past few years have witnessed concerted efforts to propose different
$dark $ $energy$ models. Important efforts, in this direction, are scalar
field models like (i) quintessence \cite{br88}, (ii) k-essence \cite{ca},
(iii) tachyon scalar fields \cite{as, mr, ea, gb, ps, az} and models based on
quantum particle production, Chaplygin gas \cite{vs00, vs02}. In these models,
a scalar field acts as a source of are $dark $ $energy$ and plays crucial role
in the dynamical universe. But these models have $no$ answer to  the question
``Where are these scalars coming from?'' It is like Higg's fields in GUTs as
well as inflaton in inflationary models of the early universe.  In the
cosmology probed here, no scalar field is required to be incorporated, from
outside the theory, to discuss cosmic $dark $ $energy.$ Following arguments in
\cite{vs02, brp},here also, vacuum energy is recognized as DE.

All these dark energy models (mentioned above) try to explain rolling down of
DE density from a very high value in the early universe to an extremely small
value in the current universe as it is suggested by astronomical
observations. But its initial value, in the early universe, is different in
different contexts. For example, it is  $\sim
10^{76} {\rm GeV}^4$ at Planck scale, $\sim 10^{60} {\rm GeV}^4$ at GUT phase
transition and quantum chromodynamics yields its value $\sim 10^{-3} {\rm
  GeV}^4$. In what follows, one-loop renormalization of riccion \cite{sk02, sk04} contributes
initial value equal to  $ 10^{6} {\rm
  GeV}^4$ for slowly varying time-dependent DE density to the dynamical
universe beginning with a phase transition at the electro-weak scale $M_{\rm
  ew} = 100 {\rm GeV}$.

 Here, there is no role of any other
scalar field in the cosmological dynamics except riccion which emerges from
higher-dimensional high-derivative gravity. Thus, unlike other models, dark
energy is contributed by the gravitational sector. Geometry of the space-time
is considered to be $(4 + D)$-dimensional with topology $M^4 \otimes S^D$,
where $S^D$ is a $D$-dimensional sphere being the $hidden$ extra-dimensional
compact space. Further, it is obtained that $ D \to 6$ which makes effective
dimension of the space-time equal to $10$. The observable universe is a
4-dimensional hypersurface of the higher-dimensional world. So far sharpest
experiments could probe gravity upto 0.1 mm.and, in the matter sector, probe
could be possible  upto electroweak scale $M_{\rm ew} = 100 {\rm GeV}$, having
wavelength as short as $\sim 1.97 \times 10^{-16}$cm. It means that it is
difficult to realize 4-dimensional gravity for length scales less than 0.1
mm. So, it is reasonable to think higher-dimensional gravity for length scales
smaller than this scale. This idea is parallel to the brane-world gravity,
where it is assumed that gravity is stronger in higher-dimensional space-time,
called $bulk$ and only a small part of it is realized in the observable universe \cite{rm}.

Thus, in the proposed cosmological scenario, our dynamical universe begins
with the initial $dark$ $energy$ density (vacuum energy
density) $\rho_{\Lambda_{\rm ew}} = 10^{6} {\rm GeV}^4$ and background
radiation with temperature $T_{\rm ew} = 78.5 {\rm GeV} = 9.1 \times
10^{14}K$, contributed by $riccion$ at $M_{\rm ew}$. The background radiation
is caused by a phase transition at $M_{\rm ew}$. This event is recognized as
$big-bang$. As usual, dynamical universe grows in the $post$ $big-bang$
era. In the present theory, $riccion$ has effective role in the $pre-$
$big-bang$ era only and it remains $passive$ in the $post$ $big-bang$ times.

\bigskip

\centerline{\bf 7.1.Riccion from $M^4 \otimes S^D$ geometry of the space-time}

\smallskip

Here ,geometry of the space-time is given by The distance function 

$$ d S^2 = g_{\mu\nu} d x^{\mu} d x^{\nu} - l^2 d\Omega^2  \eqno(7.1.1a)$$
with
$$ d\Omega^2 = d\theta_1^2 + sin^2\theta_1 d\theta^2_2 + \cdots +
sin^2\theta_1 \cdots sin^2\theta_{(D-1)} d\theta^2_D, \eqno(7.1.1b)$$
where  $g_{\mu\nu} (\mu ,\nu = 0, 1, 2, 3)$ are components of the
metric tensor in $M^4$, $l$ is radius of the sphere $S^D$ which is independent of coordinates $x^{\mu}$ and $0 \le \theta_1,\theta_2, \cdots , \theta_{(D-1)} \le \pi$  and $0 \le \theta_D \le 2\pi.$

Theory begins with the gravitational action 

$$ S = \int {d^4 x} {d^D y}  \sqrt{- g_{(4+D)}} \quad
\Big[ \frac{M^{(2+D)}R_{(4+D)}}{16 \pi
} + {\alpha_{(4+D)}} R_{(4+D)}^2 + \gamma_{(4+D)} ( R_{(4+D)}^3 $$
$$  - \frac{6(D+3)}{(D-2)}{\Box}_{(D+4)}R^2_{(D+4)})\Big], \eqno(7.1.2a)$$ 
where  $G_{(4+D)} = M^{- (2+D)}$ ($M$ being the mass scale ),$ \alpha_{(4+D)}
= \alpha V_D^{-1},$ $\gamma_{(4+D)} = \frac{\eta^2}{3! (D-2)} V_D^{-1}, $ and
$ R_D = \frac{D(D-1)}{l^2}.$ Here $V_D$, being the volume of $S^D$ ,is given as
$$ V_D = \frac{2 \pi^{(D+1)/2}}{\Gamma(D+1)/2}l^D. \eqno(7.1.2b)$$ 
 $g_{(4+D)}$ is  determinant of the metric tensor $g_{MN} (M,N = 0,1,2, \cdots, (3+D)$ and $R_{(4+D)} = R + R_D.$  $\alpha$ is a dimensionless coupling constant, $R$ is the  Ricci scalar in $M^4$ and $G_N = G_{(4+D)}/V_D.$ 

Invariance of $S$ under transformations $g_{MN} \to  g_{MN}
+ {\delta} g_{MN}$ yields \cite{sk98,sk99}

$$ \frac{M^{(2+D)}}{16 \pi} (R_{MN} - {1 \over 2} g_{MN} R_{(4+D)} ) + {\alpha_{(4+D)}} H^{(1)}_{MN} + {\gamma_{(4+D)}} H^{(2)}_{MN} = 0,\eqno(7.1.3a)$$ 
where

$$ H^{(1)}_{MN} = 2 R_{; MN} - 2 g_{MN} {\Box}_{(4+D)} R_{(4+D)} - {1\over 2} g_{MN} R^2_{(4+D)} + 2 R_{(4+D)} R_{MN}, \eqno(7.1.3b)$$  
and
$$ H^{(2)}_{MN} = 3 R^2_{; MN} - 3 g_{MN} {\Box}_{(4+D)} R^2_{(4+D)} - \frac{6(D+3)}{(D-2)}\{- { 1 
\over 2} g_{MN}{\Box}_{(4+D)} R^2_{(4+D)}$$  $$+ 2{\Box}_{(4+D)}R_{(D+4)}R_{MN} + R^2_{; MN}\} - { 1 \over 2} g_{MN} R^3_{(4+D)} + 3 R^2_{(4+D)} R_{MN}  \eqno(7.1.3c) $$

Trace of these field equations is obtained as

$$- \Big[\frac{(D+2)M^{(2+D)}}{32 \pi }\Big] R_{(4+D)}  +   {\alpha_{(4+D)}} [2(D+3) {\Box}_{(4+D)} R_{(4+D)}  +  {1 \over2} D R^2_{(4+D)} ] $$
$$ + {1 \over2}\gamma_{(4+D)} (D - 2)R^3_{(4+D)}  = 0 .   \eqno(7.1.4)$$ 

In the space-time described by the distance function defined in eq.(7.1.1a,b),

$$ {\Box}_{(4+D)} R_{(4+D)} = {\Box} R = {1\over \sqrt{-g}}{\frac{\partial}{\partial x^{\mu}}}\Big( \sqrt{-g}\quad g^{\mu\nu} \frac{\partial}{\partial x^{\nu}}\Big) R ,   \eqno(7.1.5)$$
using the definition of $R_{(4+D)}$ given in eq.(7.1.2a).

Connecting eqs.(7.1.4)- (7.1.5) as well as using $R_{(4+D)}, \alpha_{(4+D)}$  and $\gamma_{(4+D)}$ from eq.(7.1.2a), one obtains in $M^4$
$$- \Big[\frac{(D+2)M^{(2+D)}V_D}{32 \pi }\Big](R + R_D)  +   \alpha [2(D+3)
{\Box} R $$
$$ +  {1 \over2} D (R + R_D)^2 ]  + \frac{\eta^2}{12}(R + R_D)^3  = 0 ,
\eqno(7.1.6)$$ 
 which is re-written as

 $$[{\Box} + {1 \over2} \xi R + m^2 + \frac{\lambda}{3!} \eta^2 R^2] R  + \eta^{-1} \vartheta = 0 ,  \eqno(7.1.7)$$
 where 

 \begin{eqnarray*}
 \xi & = & \frac{D}{2(D+3)} +  \eta^2 \lambda R_D\\ m^2  & = & - \frac{(D + 2) \lambda M^{(2+D)}V_D}{16 \pi} + \frac{D R_D}{2(D+3)} + \frac{1}{2} \eta^2 \lambda R_D^2 \\ \lambda  & = & \frac{1}{4(D+3)\alpha}, \\ \vartheta & = & - \eta \Big[- \frac{(D + 
2) \lambda M^{(2+D)}V_D}{16 \pi} + \frac{D  R_D^2}{4 (D+3)} + \frac{1}{6} \eta^2 \lambda R_D^3 \Big],
 \end{eqnarray*} 

 $$   \eqno(7.1.8 a,b,c,d)$$
 where $\alpha > 0$ to avoid the ghost problem.

A scalar field, representing a spinless particle, has unit mass dimension in
existing theories. $R$, being combination of second order derivative as well
as squares of first order derivative of metric tensor components with respect
to space-time coordinates, has mass dimension 2. So, to have  mass dimension
like other scalar fields, eq.(7.1.7) is multiplied by $\eta$ and $\eta R$ is
recognized as $\tilde R$.
 As a result, this equation looks like

 $$[{\Box} + {1 \over2} \xi R + m^2 + \frac{\lambda}{3!} {\tilde R}^2]{\tilde
 R} + \vartheta  = 0.  \eqno(7.1.9)$$

Using the method given above, the corresponding riccion action is obtained as

 $$ S_{\tilde R} =\int {d^4 x}  \Big\{\sqrt{- g}
\Big[{1 \over2}  \partial^{\mu}{\tilde R}
\partial_{\mu}{\tilde R} 
-  \Big({1 \over 3! } \xi R {\tilde R}^2 + {1 \over 2} m^2 {\tilde R}^2 + 
\frac{\lambda}{4!} {\tilde R}^4 + \vartheta {\tilde R}\Big) \Big]\Big\}. \eqno(7.1.10)$$

\bigskip

\centerline{\bf 7.2.One-loop quantum correction , renormalization  of riccion}

 \centerline {\bf and solution of renormalization group equations}

\smallskip

Here also, one-loop quantum correction and renormlization of riccion is done
using the operator regularization method \cite{rbm} with the action
(7.1.10). In contrast to riccion action (5.4), the action (7.1.10) contains
another term $\vartheta {\tilde R}$. Due to this term, in this case, aditional
term and equation appear in renormalized lagrangian (5.8a, b) and
renormalization conditions(5.9). Solution of renormalization group equations
for relevant coupling constants for further use in this section are given
\cite{sk04} as

\begin{eqnarray*}
\tilde\Lambda & = & \tilde\Lambda_{\rm ew} + \frac{m^4_{\rm ew}}{2\lambda_{\rm ew}} \Big[\Big(1 - \frac{3
\lambda_{\rm ew}\tau}{8 \pi^2}\Big)^{1/3} - 1 \Big] \\
\lambda & = & \lambda_{\rm ew} \Big[1 - \frac{3 \lambda_{\rm ew} \tau}{8 \pi^2} \Big]^{-1}\\ \vartheta & = & \vartheta_{\rm ew} (constant) \\
 m^2 & = & m^2_{\rm ew} \Big[1 - \frac{3 \lambda_{\rm ew}\tau}{8 \pi^2} \Big]^{-1/3}\\
\frac{1}{2} \xi & = &  {1 \over 6}  + \Big(\frac{1}{2} \xi_{\rm ew} - \frac{1}{6} \Big) \Big[1 - \frac{3
\lambda_{\rm ew} \tau}{8 \pi^2} \Big]^{-1}
\end{eqnarray*}
$$  \eqno(7.2.1a,b,c,d,e)$$
where $\tau = \frac{1}{2} ln (M^2_{\rm ew}/\mu^2)$ with cut-off mass scale
$M_{\rm ew}$. Here, $\lambda_{i_{\rm ew}} = \lambda_i (\tau=0)\quad {\rm and}\quad  \tau = 0 \quad {\rm at}\quad  \mu = M_{\rm ew} $ according to defnition of $\tau$ given above.

These results show that as $\mu \to \infty ( \tau \to  - \infty ),\quad \lambda \to  0 \quad {\rm and}\quad m^2 \to 0$ and $\frac{1}{2} \xi \to \frac{1}{6}.$ 

Using these limits in eq.(7.1.8a), it is obtained that

 $$ D =  6.  \eqno(7.2.2 )$$

Also,eqs.(7.1.2a) and (7.2.2) imply

 $$ R_6 = \frac{30}{l^2} .   \eqno(7.2.3)$$

So, from eq.(7.2.1a)
$$\tilde\Lambda =  \tilde\Lambda_{\rm ew}   \eqno(7.2.4)$$
at $\mu = M_{\rm ew} $. The equation (7.2.1c) shows that $\vartheta$ is
independent of mass scale $\mu$. So, $\eta^{-1} \vartheta$, being true for all
$\mu$, is obtained from eq.(7.1.8d) as
$$\eta^{-1}\vartheta  = \frac{30}{l^2_{\rm ew}}\Big[m^2_{\rm ew} - \frac{5 ( 1 + 60 \lambda_{\rm ew})}{l^2_{\rm ew}} \Big], \eqno(7.2.4)$$
at $\mu = M_{\rm ew}$. Here
$$  m^2_{\rm ew} = - \frac{\lambda_{\rm ew} M^8_{\rm ew} V_{6{\rm ew}}}{2 \pi} + \frac{10 ( 1 + 450 \lambda_{\rm ew})}{l^2_{\rm ew}}  \eqno(7.2.5)$$
with 
$$ V_{6{\rm ew}} = \frac{16 \pi^3 l^6_{\rm ew}}{15} \eqno(7.2.6)$$
which is derived connecting eqs.(7.1.8b,d), taking the arbitrary parameter $\eta
= l_{\rm ew}$ and putting $D = 6.$  

Thus $\eta^{-1}\vartheta$ , given by eq.(7.2.4),is an imprint of extra
six-dimensional compact manifold $S^6$ in the 4-dimensional universe. This
term has the dimension of energy density. Moreover, it is not generated
through matter, but the geometry of extra-dimensional space. So, it is
recognized as $dark$ $energy$ density $\rho_{\Lambda_{\rm ew}}$ at $\mu =
M_{\rm  ew}$, given as
$$\rho_{\Lambda_{\rm ew}} = \eta^{-1}\vartheta  = \frac{30}{l^2_{\rm ew}}\Big[m^2_{\rm ew} - \frac{5 ( 1 + 60 \lambda_{\rm ew})}{l^2_{\rm ew}}
\Big], \eqno(7.2.7)$$

 \bigskip

\centerline {\bf 7.3.  Equation of state for dark energy and Phase transition  }

\centerline{\bf   at the eletroweak scale }

$Dark$ $energy$ density $\rho_{\Lambda_{\rm ew}}$,
obtained through renormalization of riccion at energy mass scale $\mu = M_{\rm
  ew}$ , is given by eq.(7.2.7). As riccion is emerging from geometry of the
space-time without matter,  $\rho_{\Lambda_{\rm ew}}$ can also be obtained as zero-point energy of riccion, given as

$$ \rho_{\Lambda} (t=0) = \rho_{\Lambda_{\rm ew}} = (2 \pi)^{-3} \int_0^k \sqrt{k^2 + m^2_{\rm ew}}4 \pi k^2 {dk},  \eqno(7.3.1)$$
where $m_{\rm ew}$ is the mass of riccion at the electroweak scale $M_{\rm   ew}$. Eq.(7.3.1) shows that $\rho_{\Lambda_{\rm ew}}$ diverges  as $k \to \infty.$ But $\rho_{\Lambda_{\rm ew}} $ (as given by eq.(7.2.7)), being
finite, implies that the integral in eq.(7.3.1) should be regularized upto  a
certain cut-off mode $k = k_c$. As so far experiments could be performed upto
$M_{\rm ew}$ only, so cut-off scale is taken as $k_c = M_{\rm ew}$ as
above. Moreover, at this scale riccions heavier than $M_{\rm ew}$ can not
survive, so $m_{\rm ew} \le M_{\rm ew}$. Now eq.(7.3.1) yields

$$\rho_{\Lambda_{\rm ew}} = \frac{\pi}{4(2 \pi)^3}M^4_{\rm ew}[3 \sqrt{2} -
ln(1 + \sqrt{2})]   \eqno(7.3.2)$$
taking $m_{\rm ew} = M_{\rm ew}$.
The second quantization and uncertainty relation imply that vacuum has energy
density as well as pressure \cite{vs00}. Experimental probes, like Ia supernova and 
WMAP \cite{ar, dn, ja, abl} suggest accelerated expansion of the universe, which requires negative pressure for DE. So, to have consistency with recenprobes, the isotropic vacuum pressure is calculated as 
$$p_{\Lambda_{\rm ew}} = - \frac{(2 \pi)^{-3}}{3} \int_0^{m_{\rm ew}}4 \pi k^3 {dk} = - \frac{ \pi}{3(2 \pi)^3} M^4_{\rm ew} . \eqno(7.3.3)$$
 It yields 
$$p_{\Lambda_{\rm ew}}/\rho_{\Lambda_{\rm ew}} = \omega_{\Lambda_{\rm ew}} =
- \frac{4}{3}[3 \sqrt{2} -
ln(1 + \sqrt{2})]^{-1} = - 0.397 \simeq - 0.4.   \eqno(7.3.4a)$$

Though in certain models, time-dependence of $\omega = p/\rho$ is also proposed , but normally it is taken as a constant. So, here also, this ratio is considered independent of time. As a result

$$\omega_{\Lambda} = p_{\Lambda}/\rho_{\Lambda} = p_{\Lambda_{\rm ew}}/\rho_{\Lambda_{\rm ew}} = - 0.4 .  \eqno(7.3.4b)$$

Using $l^{-1}_{\rm ew} = M_{\rm ew} = m{\rm ew} $ and connecting
eqs.(7.2.5),(7.2.6),(7.2.7) and (7.3.2), a quardatic equation for $\lambda{\rm
  ew}$ is obtained as 

$$\Big(450 - \frac{8 \pi^2}{15} \Big)\lambda_{\rm ew}^2 + \Big[ 20\Big(450 -
\frac{8 \pi^2}{15} \Big) + 30\times 2818.8 \Big]\lambda_{\rm ew} + 11375.3 = 0, \eqno(7.3.5)$$   
 which is solved to
$$\lambda_{\rm ew} = -0.013352958 .   \eqno(7.3.6)$$

Also using $M_{\rm ew} = m{\rm ew} $ in eq.(7.3.2), it is obtained that
$$\rho_{\Lambda_{\rm ew}} \simeq 10^{6} {\rm GeV}^4.  \eqno(7.3.7)$$

In field theories, Planck scale is supposed to be a fundamental scale. So, it
is proposed that energy mass scale $\mu$ falls from the Planck mass $M_P = 10^{19}{\rm GeV} $ to the cut-off scale $M_{\rm ew} = 100  {\rm GeV} $. When it
  happens so, phase transition takes place at $\mu = M_{\rm ew}$ and energy
  with density
$$ \rho_{{\rm ew}(r)} =  \tilde \Lambda - \tilde \Lambda_{\rm ew} = 2.5 \times 10^7 {\rm GeV}^4   \eqno(7.3.8a)$$ 
is released, which is obtained connecting eqs.(7.2.1a) and (7.3.6).

 In the proposed model (PM), this event is recognized as $big-bang$, being the beginning of universe like standard model(SMU) with the release of
$background$ $radiation$ having energy density $\rho_{{\rm ew}(r)}$, given by
eq.(7.3.8a). In contrast to SMU, here, released energy density at the epoch of
big-bang is finite (it is infinite in SMU).  Temperature of photons $T_{\rm ew}$ with energy density $\rho_{\rm ew}$, is obtained from

$$ \rho_{{\rm ew}(r)} =  \frac{  \pi^2}{15} T^4_{\rm ew} = 2.5 \times 10^7 {\rm GeV}^4   \eqno(7.3.8b)$$
as \cite{sk04}

$$ T_{\rm ew} = 78.5{\rm GeV} = 9.1 \times 10^{14}K .  \eqno(7.3.9)$$

\bigskip

\centerline {\bf 7.4.Proposed cosmological scenario, Dark energy and Dark matter }

\smallskip

In the standard model of the $big-bang$ theory, it is supposed that, around
$13.7 {\rm Gyrs}$ ago, there used to be a $fireball$ ( an extremely hot
object), which was termed as $primeval$ $atom$ by Lema$\hat i$tre. Our
universe came into existence, when this $primeval$ $atom$ burst out. This event is called $big-bang$.

According to the proposed cosmological picture,  the $primeval$ $atom$
contained riccions,being
contribution of 10-dimensional higher-derivative gravity to the 4-dimensional
world. One-loop renormalization of $riccions$ and solutions of
resulting group equations yield that when energy mass scale comes down to
electroweak scale $M_{\rm ew}= 100 {\rm GeV}$, $riccion$ contributes $dark$
$energy$  density $\rho_{\Lambda_{\rm ew}} = 10^{6} {\rm GeV}^4$. Moreover,  phase transition takes place at
$M_{\rm ew}$, releasing the background radiation. This radiation thermalizes
the universe upto the temperature $T_{\rm ew} = 9.1 \times 10^{14}K$. Here,
the event of phase transition is recognized as $big-bang$, which heralds our
dynamical universe having the initial temperature $T_{\rm ew} = 9.1 \times
10^{14}K$ and initial value of dark energy density $\rho_{\Lambda_{\rm ew}} =
10^{6} {\rm GeV}^4$. 

Moreover, it is important to mention that existence of $riccions$ are possible
at energy scales where higher-derivative terms in the gravitational action
(7.1.1a) has significant role compared to Einstein-Hilbert term. At energy
scales below $M_{\rm ew}$, Einstein-Hilbert term dominates higher-derivative
terms in the action (7.1.1a). So, $riccions$ has no direct role in the evolutionof proposed model of the universe, but it has two very important contributions
to the observable universe as  value of dark energy density
$\rho_{\Lambda_{\rm ew}} = 10^{6} {\rm GeV}^4$ and cosmic background radiation
temperature $T_{\rm ew} = 9.1 \times 10^{14}K$ at cosmic time $t=0.$

Astronomical observations have compelling evidences that the current universe
is dominated by dark energy(DE). The present cosmic dark energy density is
very low, but it used to be very high in the early universe. The fall of dark
energy density from very high to extremely low value can be explained if it is
time-dependent. So, like other other cosmic dark energy models, here also,
dark energy density $\rho_{\Lambda}$ is slowly varying function of
time.

In 1933, Bronstein proposed that, in the
expanding universe, $dark$ $ energy$ $ density$ decreases due its decay as a
result of emission of dark matter or radiation \cite{brp, bro}. The radiation,
so emitted, could disturb spectrum of 3K-microwave background radiation. So,
Bronstein's original idea was modified and it was introduced that dark energy
(DE) could decay to hot or cold dark matter without any harm to spectrum of
3K-microwave background radiation \cite{brp}. Following this idea,here it is demonstrated that $\rho_{\Lambda}(t)$ decays to
hot dark matter(HDM) till temperature is high and cold dark matter after
decoupling of matter from radiation. As a result, $\rho_{\Lambda}(t)$ falls
from high value $10^{6} {\rm GeV}^4$ to currently low value $0.73 \rho_{{\rm
    cr},0}$ (where $\rho_{{\rm cr},0}$ is the critical density). This approach
also provides  a solution to $cosmic$ $ coincidence$ $ problem$. 

Here, it is proposed that our 4-dimensional dynamical universe grows below  $M_{\rm  ew}$ and begins with topology having the distance function 

$$ dS^2 = dt^2 - a^2(t) [ dx^2 + dy^2 + dz^2 ] \eqno(7.4.1)$$ 
for the spatially flat model of the universe supported by recent
experiments \cite{ar,dn,ja,abl}. This space-time is a special case of hypersurface $M^4$ of the line element (7.1.1a). Here $a(t)$ is the scale factor.

It is shown above that , in the beginning, the universe was very hot due to the background radiation (released during phase transition). Radiation energy density falls as

$$\rho_r = \frac{\rho_{{\rm ew}(r)} a^4_{\rm ew}}{a^4(t)}. \eqno(7.4.2)$$
with growing scale factor.

Matter remains in thermal equillibrium with the background radiation for
sufficiently long time. According to WMAP \cite{abl}, decoupling of matter
from background radiation takes place at $t_d \simeq 386 kyr = 1.85 \times
10^{37}{\rm GeV}^{-1}$. Equation of state for radiation is $\omega = 1/3$. So,
here, it is proposed that $dark$ $ energy$ decays to hot dark matter (HDM) till decoupling time $t_d$ obeying $\omega_m = 1/3$.

When $ t > t_d$, it decays more to cold dark matter (CDM) ( which is non-baryonic and pressureless) with $\omega_m = 0$ as well as HDM. So, ratio of densities of HDM and CDM is extremely small below $t_d$.

In what follows, development of the universe is probed using these ideas.

\bigskip

\noindent {\bf (a) \underline{Decay of $dark$ $ energy$ to $dark$ $ matter$}}

\smallskip

The conservation equations

$$ T^i_{(\Lambda) j;i} + T^i_{({\rm dm}) j;i} = 0    \eqno(7.4.3)$$
yield coupled equations

$${\dot \rho}_{\Lambda} + 3 H ( 1 + {\omega}_{\Lambda}) \rho_{\Lambda} = - Q(t)   \eqno(7.4.4a)$$ 
and

$${\dot \rho}_{\rm dm} + 3 H ( 1 + {\omega}_{\rm dm}) \rho_{\rm dm} =
 Q(t)   \eqno(7.4.4b)$$ 
with $ \rho_{\Lambda}(t)$ and $\rho_{\rm dm}(t)$, being $dark$ $ energy$ density  and $dark$ $matter$ density respectively at cosmic time $t$. Here $H = {\dot a}/a,
{\omega}_{\Lambda} = p_{\Lambda}/\rho_{\Lambda} = - 0.4 $ and ${\omega}_{\rm
  dm} = p_{\rm   dm}/\rho_{\rm dm}$ . $Q(t)$ is the loss (gain) term for
DE(DM) respectively. 

Batchelor \cite{kb67} has pointed out that dissipation is natural for all material fluids,
barring superfluids . So, presence of a dissipative term for dark
matter DM is reasonable. In eq.(7.4.4b), this term is obtained by taking

$$ Q(t) = 3 n H \rho_{\rm dm} \eqno(7.4.5a)$$
(with $n$ being the real number) as in \cite{wz} without any harm to
physics. With this setting for time-dependent arbitrary function $Q(t)$, eq.(7.4.4b)is obtained as

$${\dot \rho}_{\rm dm} + 3 H (1 - n + {\rm w} ) \rho_{\rm dm} = 0 .  \eqno(7.4.5b)$$
Here $- n \rho_{\rm dm}$ given by $Q(t)$ acts as dissipative pressure
for DM. As $ Q(t)$ is proportional to $\rho_{\rm dm}$, this setting does not
disturb the perfect fluid structure as required by the `cosmological
principle'. Thus eq.(7.4.5b) yields the effective pressure for DM as

$$ p_{\rm (dm, eff)} = (- n + {\rm w}) \rho_{\rm dm}  .  \eqno(7.4.5c)$$

Now connecting eqs.(7.4.4b) and (7.4.5) and integrating, it is obtained that
$$\rho_{\rm dm} = A a^{3 (n  - {\omega}_{\rm dm} -1)}  \eqno(7.4.6a)$$
with
$$ A = 0.23 \rho_{{\rm cr},0} a_0^{- 3 (n -1)}      \eqno(7.4.6b)$$ 
using current value of $dark$ $matter$ density $\rho_{\rm dm,0} = 0.23 \rho_{{\rm cr},0}$ and $a_0 = a (t_0).$

Connecting eqs.(7.4.4a), (7.4.5) and (7.4.6a) and integrating, it is obtained that

$$\rho_{\Lambda} = \frac{D}{a^{3( 1 + {\omega}_{\Lambda})}} - \frac{n}{n +
  {\omega}_{\Lambda} - {\omega}_{\rm dm}} \rho_{\rm dm} \eqno(7.4.7a)$$ 
where
$$D = 10^{6} a_{\rm ew}^{3( 1 + {\omega}_{\Lambda})} {\rm GeV}^4  \eqno(7.4.7b)$$ 
using the initial condition $\rho_{\rm dm} = 0$ at $t = 0$.

\smallskip

\noindent {\bf (b) \underline{Expansion of the universe,  when $\rho_{\Lambda} > \rho_{\rm dm}$}}

\smallskip

Friedmann equation is given as

$$\Big(\frac{\dot a}{a} \Big)^2 = \frac{8 \pi G_N}{3} (\rho_{\Lambda} + \rho_{\rm dm} +
\rho_r) \simeq \frac{8 \pi G_N}{3} (\rho_{\Lambda} + \rho_{\rm dm} ) \eqno(7.4.8)$$
as $(\rho_{\Lambda} + \rho_{\rm dm} )$ dominates over $\rho_r$, which is clear from
eqs.(7.4.2),(7.4.6) and (7.4.7).

In this case, the equation (7.4.8) reduces to

$$H^2 = \Big(\frac{\dot a}{a} \Big)^2 \simeq \frac{8 \pi G_N D}{3a^{3( 1 +
    {\omega}_{\Lambda})}}  \eqno(7.4.9)$$
yielding the solution

$$ a(t) = a_{\rm ew}\Big[ 1 + \sqrt{\frac{8 \pi G_N D}{3}} \frac{t}{a_{\rm ew}^{9/10}}\Big]^{10/9} $$

$$= a_{\rm ew}\Big[ 1 + 2.89\times 10^{-16} t \Big]^{10/9}  \eqno(7.4.10a)$$
with ${\omega}_{\Lambda} = - 0.4$, $a_{\rm ew} = a(t=0)$ and 
$$ \sqrt{\frac{8 \pi G_N D}{3}} = 2.89 \times 10^{-16} a_{\rm ew}^{9/10} {\rm GeV}\eqno(7.4.10b)$$
( $G_N = M_P^{-2}, M_P = 10^{19} {\rm GeV}$).

 It shows an accelerated expansion of the universe from the beginning itself as ${\ddot a}/a > 0$. 

Connecting eqs.(7.4.7) and (7.4.10)

$$\rho_{\Lambda} = \frac{ 10^{6}}{{\Big[ 1 + \sqrt{\frac{8 \pi G_N D}{3}} \Big(\frac{t}{a_{\rm ew}^{0.9}}\Big) \Big]}^2} - \frac{n}{n +   {\omega}_{\Lambda} - {\omega}_m} \rho_{\rm dm}. \eqno(7.4.11)$$
But, the present universe obeys the condition

$$\Omega_{\Lambda,0} + \Omega_{{\rm m},0} = 1, \eqno(7.4.12 a)$$
where $\Omega_{{\rm m},0} = \Omega_{{\rm dm},0} + \Omega_{{\rm r},0}.$ Here $\Omega_0 = \rho_0/\rho_{{\rm cr},0}$ and $ \rho_{{\rm cr},0} = 3H_0^2/{8 \pi G_N}$ ($ H_0$  being Hubble's constant) for the present universe. With these values the present $critical$ $density$ is calculated as

$$ \rho_{{\rm cr},0} = 3H_0^2/{8 \pi G_N} \simeq 1.2 \times 10^{-47} {\rm GeV}^4 \eqno(7.4.12 b)$$
using $H_0 = h_0/t_0$ with $h_0 = 0.68$  and the present age of the universe $t_0 = 13.7 Gyr = 6.6 \times 10^{41} {\rm GeV}^{-1}$ \cite{abl}.

As, in the present universe, $\rho_m$ is dominated by $cold$ $dark$ $matter$ density, $\omega_m = 0$ is taken in eq.(6.11). Now, conecting eqs.(7.4.11) and (7.4.12), it is obtained that

$$ n = 0.47  \eqno(7.4.13)$$

Using eq.(7.4.10), the scale factor $a_d$ at the decoupling time $t_d = 1.85
\times 10^{37} {\rm GeV}^{-1}$ and the present scale factor $a_0$ are obtained
as 

$$ a_d \simeq 1.34 \times 10^{24} a_{\rm ew} \eqno(7.4.14)$$
and

 $$ a_0 \simeq 1.58 \times 10^{29} a_{\rm ew} \eqno(7.4.15)$$
using the present age of the universe given above.

The  equation(7.4.10) shows an accelerated growth of the scale factor when $\rho_{\Lambda} > \rho_{\rm dm}$ exhibiting non-adiabatic expansion of the universe. It implies non-conservation of  entropy of the universe, which is in contrast to the decelerated adiabatic expansion of SMU,  when cosmological model is radiation-dominated or matter-dominated.

In what follows, scale factor dependence of temperature and entropy is obtained in an emperical manner, based on current temperature of the microwave background radiation $T_0 = 2.73 K$,  the initial temperature $ T_{\rm ew} = 78.5{\rm GeV} = 9.1 \times 10^{14}K$ given by eq.(5.9) and $ a_0 \simeq 1.58 \times 10^{29} a_{\rm ew}$ given by eq.(7.4.15). Thus ,  the required emperical relation is obtained as

$$ \Big[ {T}/{T_{\rm ew}} \Big]^{103/50} = \frac{a_{\rm ew}}{a} \eqno(7.4.16)$$
which yields the decoupling temperature as 

$$ T_d = T_{\rm ew} \Big[\frac{a_{\rm ew}}{a_d} \Big]^{50/103} = 1740.42 K, \eqno(7.4.17)$$
 using eqs.(7.3.7) and (7.4.14). This value is much lower than $T_d =3000 K$, obtained in the standard big - bang cosmology. These drastic changes are due to dominance of the $dark$ $energy$ in the proposed model.

Using eq.(7.3.6b), entropy of the universe is calculated as

$$ S = \frac{7 \pi^2}{90} a^3 T^3 .  \eqno(7.4.18)$$

Current value of entropy $S_0$ is supposed to be $10^{87}$. ``How could so
high entropy of the universe be generated  ?'' is an old question. A solution
to this problem was suggested in a seminal paper on inflationary model of the
early universe by Guth \cite{ahg} and  modified version of the same by Linde \cite{adl} and Albrecht and Steinhardt \cite{apj}. Here, a different answer to this question is provided on the basis of results obtained above.

Connecting eq.(6.15) and eq.(6.18) and current vlues of entropy as well as temperature $T_0 = 2.73 K$ of the universe, it is obtained that 

$$ a_{\rm ew} = 0.25  \eqno(7.4.19)$$
as

$$ S_0 = 10^{87} = \frac{7 \pi^2}{90} (1.65 \times 2.73 \times 10^{29}a_{\rm ew})^3$$

Connecting eqs.(6.16) and (7.4.18) it is obtained that entropy grows with the scale factor $a(t)$ as 

$$ S = \frac{7 \pi^2}{90} T_{\rm ew}^3 a_{\rm ew}^{150/103} a^{159/103} \eqno(7.4.20 a)$$
with the initial value

$$S_{\rm ew} = 9 \times 10^{42} \eqno(7.4.20 b)$$

The equation (7.4.6) yields the rate of production of HDM

$${\dot \rho_{\rm hdm}} = - 5.9 \times 10^{-64} \frac{a_0^{1.59}}{a^{3.48}} \eqno(7.4.21)$$
using $\omega_{\rm dm} = 1/3$ for $hot$ $dark$ $matter$. This equation shows
that rate of production of HDM increases with growing scale factor. So,
production of HDM is responsible for increasing entropy in the universe. In
SMU, number of photons decide entropy of the universe. But, in PM, photons and
HDM both are responsible for entropy. As energy of HDM increases due to its
production, owing to decay of DE, entropy is generated in this model. It is
unlike SMU, where entropy remains conserved.

Connecting eqs.(7.4.6) and (7.4.16), temperature dependence of $\rho_{\rm dm}$ is obtained as

$$ \rho_{\rm dm} = 0.23 \rho_{{\rm cr},0} \frac{a_{\rm ew}^{3 (n  - {\omega}_{\rm dm} -1)}}{a_0^{ 3 (n -1)}} \Big[ {T_{\rm ew}}/{T} \Big]^{6.18(n  - {\omega}_{\rm dm} -1)}. \eqno(7.4.22)$$

Putting $\omega_{\rm dm} = 1/3 (0)$ for HDM (CDM), in eq.(7.4.22), ratio of CDM and HDM densities,$\rho_{\rm cdm}$ and $\rho_{\rm hdm}$, is obtained as

 $${\rho_{\rm cdm}}/{\rho_{\rm hdm}} = a_{\rm ew} \Big[{T_{\rm ew}}/{T} \Big]^{2.06} \eqno(7.4.23)$$
using $n = 0.42$ from eq.(7.4.13). Eq.(7.4.23) shows that creation of HDM is higher at high temperature. But as temperature falls down, creation of CDM supersedes the production of HDM ( production of $dark$ $matter$ owing to decay of $dark$ $energy$). For the current universe, the ratio of eq.(7.4.23) is obtained as

$${\rho_{\rm cdm ,0}}/{\rho_{\rm hdm, 0}} = 2.1 \times 10^{29} \eqno(7.4.24)$$
using numerical values of $a_{\rm ew}, T_{\rm ew}$ and $T_0$ (given above) in eq.(7.4.23). It shows that currently, $\rho_{\rm hdm}$ is almost negligible compared to $\rho_{\rm cdm}$.

The scale factor $a_{\rm sd}$ , upto which  $dark$ $energy$ vanishes due to
decay, are obtained as

$$ a_{\rm sd} = 6.78 a_0  \eqno(7.4.25)$$
 connecting eqs.(7.4.6),(7.4.7) and (7.4.10) as well as using ${ \rho_{\Lambda}} = 0$. Here $\omega_{\rm dm} = 0$ is taken as for $t > t_0$, $dark$ $matter$ content is expected to be dominated by CDM. 

The ratio of densities of $dark$ $energy$, $\rho_{\Lambda}$, and $dark$ $matter$, $\rho{\rm dm}$, is obtained as

$$ {\rho_{\Lambda}}/{\rho_{\rm dm}} = 10 (a_0/a )^{0.21} - \frac{47}{7}. \eqno(7.4.26)$$

It is interesting to note, from eq.(7.4.26), that the gap between  $dark$
$energy$ density and  $dark$ $matter$ density decreases with the growing
scale factor. Connecting eqs.(7.4.6) and (7.4.7) and using $ n $ from
eq.(7.4.13). This equation shows that  $\rho_{\rm dm}/\rho_{\Lambda} < 1$ for

$$ a_{\rm ew}< a < 3.44 a_0 \quad {\rm and} \quad 0 < t < 3.03 t_0. \eqno(7.4.27 a,b)$$
The result, given by  eq.(7.4.27b), solves the $cosmic$ $coincidence$ $problem$ [24]. This approach, for solution of $coincidence$ $problem$  has an advantange that no scalar field is required to represent the $dark$ $energy$, rather it is the contribution of higher-dimensional higher-derivative gravity to the observable universe. It is unlike the case of work in \cite{la,wz}, where assumed $quintessence$ scalars are used.

\smallskip

\noindent {\bf (c) \underline{The universe,in case $\rho_{\Lambda} < \rho_{\rm dm}$}}

\smallskip

Like eq.(7.4.27), employing the same procedure, it is also possible
to find that ${\rho_{\rm dm}}/{\rho_{\Lambda}} \ge 1$ for

$$  a \ge 3.44 a_0 \quad {\rm and} \quad  t \ge 3.03 t_0. \eqno(7.4.28 a,b)$$

So, for $ t \ge 3.03t_0$, the  the Friedmann equation is written as

$$\Big(\frac{\dot a}{a} \Big)^2 = \frac{8 \pi G_N}{3} (\rho_{\Lambda} + \rho_{\rm dm} + \rho_r) \simeq \frac{8 \pi G_N}{3} {\rho_{\rm dm}} \simeq 0.27 H_0^2 (a_0/a )^{1.74} \eqno(7.4.29)$$
using eqs.(7.4.6),(7.4.8), (7.4.13) and definition of $\rho_{\rm cr,0}$ (given above). Eq.(6.29) yields the solution 

$$ a(t) = \Big[ a_{\Lambda < {\rm dm}}^{87/100} + 0.43\times 10^{-43} (t - t_{\Lambda < {\rm dm}}) \Big]^{100/87}, \eqno(7.4.30)$$
where $a_{\Lambda < {\rm dm}} = 3.44 a_0$ and $t_{\Lambda < {\rm dm}} = 3.03 t_0$. Eq.(7.4.30) also shows  an accelerated growth of the scale factor, even though $\rho_{\Lambda} < \rho_{\rm dm}$. It is interesting to see that this expansion is faster than the expansion in the interval $0 < t < 3.03 t_0$.

From eq.(7.4.25), the scale factor upto which $dark$ $energy$ vanishes is $a_{\rm sd} = 6.78 a_0$. So, eq.(7.4.30) yields the corresponding time  as

$$ t_{\rm sd} = 3.7 t_0 = 50.69 {\rm Gyr}. \eqno(7.4.31)$$

It shows that decay of $dark$ $energy$ will continue in the interval $3.03 t_0 < t < 3.7 t_0$ also. So, during this time interval,  the $dark$ $matter$ will follow the rule, given by eq.(7.4.6).

Using eqs.(7.4.16),(7.4.18), (7.4.25)and (7.4.31), $T_{\rm sd}$ and $S_{\rm sd}$ are calculated as

$$ T_{\rm sd} = T_0 \Big( \frac{a_0}{a_{\rm sd}} \Big)^{50/103} = 0.395 T_0 = 1.08 K . \eqno(7.4.32 a)$$

$Dark$ $matter$ density at $t = t_{\rm sd}$ is obtained as  

$$\rho_{\rm dm(sd)} = 1.32 \times 10^{- 49} {\rm GeV}^4 \eqno(7.4.33)$$
using $a_0$ and $a_{\rm sd}$ in eq.(7.4.6).

As $dark$ $energy$ vanishes when $t \ge t_{\rm sd}$, content of the universe will be dominated by CDM ,which is   pressureless non-baryonic matter obeying the conservation equation

$$ {\dot \rho_{\rm dm}} + 3 H \rho_{\rm dm} = 0.$$
This equation yields scale factor dependence of $\rho_{\rm dm}$ as

$$\rho_{\rm dm} = \rho_{\rm dm(sd)} \Big( a_{\rm sd}/ a(t) \Big)^3 = \frac{2.54 \times 10^{39}}{a^3(t)} \eqno(7.4.34)$$
for $ t > 3.7 t_0.$ 

Beyond the age of the universe $3.7 t_0$, the Friedmann equation looks like

$$\Big(\frac{\dot a}{a} \Big)^2 = \frac{8 \pi G_N}{3} \rho_{\rm dm} =  \frac{212.8}{a^3(t)} \eqno(7.4.35)$$
using eq.(7.4.34).

Eq.(7.4.35) yields the solution

$$ a(t) = \Big[ a_{ {\rm sd}}^{3/2} \pm 14.6 (t - t_{ {\rm sd}}) \Big]^{2/3}, \eqno(7.4.36)$$
with $a_{\rm sd}$ and $t_{\rm sd}$ given by eqs.(6.25) and (7.4.31) respectively.

On taking (+) sign in eq.(7.4.36), decelerated expansion of the universe is
obtained beyond $t > 3.7 t_0$ continuing for ever. But the (-) sign, in
eq.(6.36), exhibits a contracting universe beyond $t > 3.7 t_0$. So,
ultimately the contracting universe is expected to collapse to a very small
size with scale factor, possibly equal to $a_{\rm ew} = 0.25$. From eq.(7.4.36),with (-) sign, the collapse time is calculated as

$$ t_{\rm col} = t_{\rm sd} + \frac{a^{3/2}}{14.6} = 18.08 t_0 = 247.73 {\rm
  Gyr} \eqno(7.4.37)$$
using $a_{\rm sd} = 6.78 a_0$ and $t_{\rm sd} = 3.7 t_0$.

\bigskip

\centerline{\bf 7.5. Elementary particles, Primordial nucleosynthesis and}

 \centerline{\bf Structure formation  }

\smallskip

\noindent \underline{(i) Creation of particles}

\smallskip

In SMU, it is assumed that elementary particles such as leptons, mesons,
nucleons and their anti-particles were produced at the epoch of
big-bang. Here, production of scalar and spin-1/2 particles is proposed in the very early
universe due to topological changes caused by expansion of the universe.

\smallskip

\noindent \underline{(a) Creation of spinless particles}

\smallskip

The scalar field $\phi$ obeys the Klein-Gordon equation

$$ ({\Box} + m_{\phi}^2 ) \phi = 0,    \eqno(7.5.1)$$
where $m_{\phi}$ is mass of $\phi$. Expanding $\phi$ in terms of mode $k$,
eq.(7.5.1) is written as

$$( 1 + 2.89 \times 10^{-16} t ) {\ddot \phi_k} + \frac{ 28.9}{3} \times 10^{-16}{\dot \phi} + \Big[m_{\phi_k}^2 (1 +  2.89 \times 10^{-16} t ) $$
$$- \frac{k^2}{a^2_{\rm    ew}} ( 1 +  2.89 \times 10^{-16} t )^{-11/9} \Big]
\phi_k = 0  \eqno(7.5.2)$$
using the scale factor $a(t)$, given by eq.(6.10a).

For small $t$, eq.(7.5.2) is obtained as

$$ \tau \frac{d^2 \phi^{\rm in}_{ k}}{d \tau^2} + \frac{10}{3} \frac{d
  \phi^{\rm in}_{ k}}{d \tau} +  10^{30} \Big[- \frac{20
  k^2}{9 a^2_{\rm  ew}} + \Big( m^2_{\phi} + \frac{11 k^2}{9 a^2_{\rm  ew}}
  \Big) \tau \Big] \phi^{\rm in}_{ k} = 0 , \eqno(7.5.3)$$
where
$$ \tau =  1 +  2.89 \times 10^{-16} t . \eqno(7.5.4)$$

Eq.(7.5.3) yields the normalized solution

$$\phi^{\rm in}_{ k} = [2 (2 \pi)^3 \sqrt{b_2} ]^{-1/2} e^{- i \tau
  \sqrt{b_2}} _1F_1 \Big(\frac{5}{3} + i \frac{b_1}{2 \sqrt{b_2}} , \frac{10}{3},  2 i \tau \sqrt{b_2} \Big) $$

$$ \approx [2 (2 \pi)^3 \sqrt{b_2} ]^{-1/2} e^{- i \tau
  \sqrt{b_2}} \Big[ 1 + \frac{( - 3 b_1 + 10 i \sqrt{b_2}) \tau}{10} \Big],
 \eqno(7.5.5a)$$
where
$$ b_1 = - \frac{20}{9} \frac{( 10^{15} k)^2}{ a^2_{\rm  ew}} \eqno(7.5.5b)$$
and

$$ b_2 = 10^{31}  \Big( m^2_{\phi} + \frac{11 k^2}{9 a^2_{\rm
    ew}}  \Big) . \eqno(7.5.5c)$$
Here $_1F_1 (a,b,c)$ is the confluent hypergeometric function.

For large $t$, eq.(7.5.2) reduces to

$$ \tau \frac{d^2 \phi^{\rm out}_{ k}}{d \tau^2} + \frac{10}{3} \frac{d
  \phi^{\rm out}_{ k}}{d \tau} + ( 10^{15.5} m_{\phi})^2 \tau
  \phi^{\rm out}_{ k}  = 0 , \eqno(7.5.6)$$
which integrates to

$$ \phi^{\rm out}_{ k} = \tau^{-7/6} \Big[ c_1 J_{-7/6} (A \tau) + c_2
Y_{-7/6} (A \tau) \Big] , \eqno(7.5.7a)$$

where 

$$ A = 10^{15.5} m_{\phi} . \eqno(7.5.7b)$$
and $J_p(x)$ and $Y_p(x)$ are Bessel's function of first and second kind
respectively. For large $x$, Bessel's functions can be approximated as 
$$ J_p(x) \simeq \frac{cos (x - \pi/4 - p \pi/2)}{\sqrt{\pi x/2}} \eqno(7.5.7c)$$
and
$$ Y_p(x) \simeq \frac{sin (x - \pi/4 - p \pi/2)}{\sqrt{\pi x/2}}. \eqno(7.5.7d)$$

Using these approximations, when $\tau$ is large, eq.(7.5.7a) looks like

$$ \phi^{\rm out}_{ k} \simeq \frac{\tau^{-5/3}}{\sqrt{\pi A/2}} [c_1 cos (A
\tau + \frac{\pi}{3}) + c_2 sin (A \tau + \frac{\pi}{3}) ] \eqno(7.5.8)$$

Solutions (7.5.5a) and (7.5.8) yield number of produced spinless particles (for
mode $k$) per unit volume as

$$ |\beta_k|^2 = \frac{\tau^{-10/3}}{\pi^2 A^2 b_2} |X|^2 \ne 0 , \eqno(7.5.9a)$$
where
\begin{eqnarray*}
 X &=& \Big[ \Big(- \frac{3 b_1 \sqrt{b_2} \tau}{10} - \frac{2}{3} \sqrt{b_2}
\Big) cos(A \tau) - A\sqrt{b_2} \tau sin (A \tau) \Big]  + i \Big[
\Big\{\sqrt{b_2} \tau \\&& - \frac{3 b_1}{10} + \frac{5}{3} \Big(1 -
\frac{3 b_1}{10} \Big) \tau^{-1} \Big\} cos(A \tau) + \Big(1 -
\frac{3 b_1}{10} \Big) A sin(A \tau) \Big].
\end{eqnarray*}
$$ \eqno(7.5.9b)$$
$\beta_k$ is defined in Appendix B. Non-zero $ |\beta_k|^2$ shows creation of
spinless particles.

\smallskip

\noindent \underline{(b) Creation of spin-1/2 particles}

\smallskip

The spin-1/2 field $\psi$ satisfies the Dirac equation

$$ (i \gamma^{\mu} D_{\mu} - m_f ) \psi = 0 , \eqno(7.5.10)$$ 
where $m_f$ is mass of $\psi$, $\gamma^{\mu}$ are Dirac matrices in curved space-time and $ D_{\mu}$
are covariant derivatives defind above, $\psi$ can be written as

$$ \psi = {\sum_{s = \pm 1}} {\sum_{k}} (b_{k,s} \psi_{I k,s} + d^{\dagger}_{k,s}
\psi_{II k,s})  \eqno(7.5.11)$$
with $b_{k,s}$ and $d_{k,s}$, given in Appendix B. Eqs.(7.5.10) and (7.5.11) yield

$$(i \gamma^{\mu} D_{\mu} - m_f )\psi_{I k,s}  = 0 , \eqno(7.5.12a)$$ 

$$(i \gamma^{\mu} D_{\mu} - m_f )\psi_{II k,s}  = 0 . \eqno(7.5.12b)$$ 

Now using the operator $(- i \gamma^{\mu} D_{\mu} - m_f )$ from left of
eqs.(7.5.12a) and (7.5.12b), it is obtained that

$$ ({\Box} + \frac{1}{4} R + m_f^2 ) {\tilde \psi} = 0 , \eqno(7.5.13)$$
where ${\tilde \psi} = \psi_{I k,s} (\psi_{II k,s}).$

Writing
$$ \psi_{I k,s} = f_{I k,s}(t) e^{i{\vec k}.{\vec x}}u_s$$
and
$$ \psi_{II k,s} = f_{II k,s}(t) e^{i{\vec k}.{\vec x}}{\hat u_s},$$
eq.(7.5.13) is obtained as

$$ \tau \frac{d^2 {\tilde f}}{d \tau^2} + \frac{10}{3} \frac{d {\tilde f}}{d
  \tau} +  10^{31}  \Big[m_f^2\tau -
  \frac{k^2}{\tau^{11/9}} - \frac{4.93 \times 10^{-15}}{3 \tau} \Big]{\tilde f} 
    = 0 , \eqno(7.5.14)$$
where ${\tilde f} = f_{I k,s} (f_{II k,s}).$

For small t , eq.(7.5.14) looks like

$$ \tau \frac{d^2 {\tilde f}}{d \tau^2} + \frac{10}{3} \frac{d {\tilde f}}{d
  \tau} +  10^{31}  \Big[(m_f^2 + \frac{11 k^2}{9 a_{\rm ew}^2})
  \tau - \frac{20 k^2}{9 a_{\rm ew}^2}\Big] {\tilde f} = 0 . \eqno(7.5.15)$$

This equation is like the equation (7.5.4a). So, its solution has the same
form. Using this solution in $\psi_{I k,s}$ defined above , it is obtained that
$$ \psi^{\rm in}_{I k,s} = \frac{e^{-i \tau \sqrt{b_2}}}{\sqrt{2 (2 \pi)^3
      \sqrt{b_2}}} \Big[ 1 + \frac{( - 3 b_1 + 10 i \sqrt{b_2})}{10} \tau
      \Big] e^{i {\vec k}.{\vec x}}u_s . \eqno(7.5.16)$$

For large t , eq.(7.5.14) reduces to

$$ \tau \frac{d^2 {\tilde f}}{d \tau^2} + \frac{10}{3} \frac{d {\tilde f}}{d
  \tau} +  10^{31} m_f^2 \tau {\tilde f} = 0  \eqno(7.5.17)$$
yielding

$$ \psi^{\rm out}_{II k,s} = \frac{ \tau^{-5/3}}{\pi^2 A} cos (A \tau)
e^{i{\vec k}.{\vec x}}{\hat u_s}. \eqno(7.5.18)$$

Using $\psi^{\rm in}_{I k,s}$ and $ \psi^{\rm out}_{II k,s}$ in the definition
of $\beta_{k,s}$, given in the Appendix B,

$$\beta_{k,s} = (2 \pi)^3 \tau^{-5/3} \frac{cos (A \tau)}{\pi^2 A}\frac{e^{-i \tau \sqrt{b_2}}}{\sqrt{2 (2 \pi)^3
      \sqrt{b_2}}} \Big[ 1 + \frac{( - 3 b_1 + 10 i \sqrt{b_2})}{10} \tau
      \Big] . \eqno(7.5.19)$$

So, the number of created spin-1/2 particles per unit volume is obtained from
eq.(7.5.19)as

$$ |\beta_{k,s}|^2 = \frac{(2 \pi)^2}{2 \pi^2 A \sqrt{b_2}} \tau^{-10/3} \Big[
\Big( 1 - \frac{3 b_1}{10} \tau \Big)^2 + b_2 \tau^2 \Big]. \eqno(7.5.20)$$

Eqs.(7.5.9a,b) and (7.5.20) show creation of spinless and spin-1/2 particles due tochanging gravitational field in the proposed speeded-up model. The scale
factor, given by eq.(7.4.10a) show that the significant change in the
gravitational field is possible , in this model, when

$$ t > 3.46\times10^{15} {\rm GeV}^{-1} = 7.2 \times 10^{-9} {\rm sec.}  \eqno(7.5.21)$$
as $a(t)$ remains almost constant upto this epoch. So, here, particle
production is expected, when the universe is around $2.3 \times 10^{-9} {\rm
  sec.}$ old. Also, these results show that particle production falls down
rapidly as time increases. Thus, like other models, here also particle
production is expected in the early stages of the universe. 

\smallskip

\noindent \underline{7.2 Primordial Nucleosynthesis}

\smallskip

Hydrogen ($H$) is the major component of baryonic matter in the universe. The
next main component is Helium-4 ($^4He$) . Occurrence of other light elements
and metals is very small. It is found  unlikely that abundance of $^4He$,
Deuterium ($D$) and other light elements, in the universe, being caused by burn out of $H$ in stars
\cite{sw, ek} So, it is argued that the required amount of these might have produced in
the early universe.

At the epoch of its formation, helium production depends upon neutron ($n$)
concentration, which is determined by weak interaction reactions given as

$$ n + \nu \leftrightarrows p + e^{-} , n + e^{+} \leftrightarrows p + {\bar
  \nu}  \eqno(7.5.22)$$
where $p$ stands for proton and $\nu$ for neutrino.
This chemical equillibrium is maintained till weak reaction rate $\Gamma_{\rm
  w} >> H $, where $H$ is the expansion rate of the universe given by
  eq.(7.4.9)   and $\Gamma_{\rm   w} \simeq 1.3 G_F^2 T^5 $ with Fermi
  constant $G_F = \pi  \alpha_{\rm   w}/{\sqrt{2} M^2_{\rm   w}} = 1.17 \times
  10^{-5} {\rm   GeV}^{-2}$. With the expansion of the universe, temperature
  decreases so weak interaction rate slows down . As a result, at the
  freeze-out temperature $T_*$,
$$\Gamma_{\rm   w} \simeq H.   \eqno(7.5.23a)$$ 

Connecting eqs.(7.4.7b), (7.4.9), (7.4.16) and (7.5.23a), $T_*$ is obtained that
$$ 1.3\times (1.17)^2 \times 10^{-10} T^5_* \simeq 2.89\times 10^{-16}
\Big(\frac{T_*}{T_{\rm ew}} \Big)^{1.9}, $$
which yields
$$ T_* = 0.9 {\rm MeV} . \eqno(7.5.23b)$$ 

Eqs.(7.4.10a,b) and (7.4.16) yield time dependence of temperature , for this
model, as 

$$ t_{\rm sec} = \frac{4.5}{ T^{1.9}_{\rm MeV}} . \eqno(7.5.24)$$ 

Using this result, the freeze-out time $t_*$, corresponding to $T_*$, is
obtained as
$$ t_* \simeq 5.55 sec. \eqno(7.5.25)$$

In SMU,$T_* \simeq 0.86 {\rm MeV}$ and $ t_* \simeq 1 sec$ \cite{muk}. It is
obtained that, in PM, $T_*$ is a bit higher and freeze-out takes place
later. The reason for these differences is the basic difference in developement
of these models. SMU is driven by elementary particles present in the early
universe and it expands adiabatically. The proposed model is driven by dark
energy, with very high density at the beginning and it expands with
acceleration.

Like \cite{re}, in this subsection, temperature $T_9$ is measured in units
$10^9 K$, so $1 {\rm MeV}$ becomes $11.6$ in $T_9$ temperature. Now, eq.(7.5.24)
looks as

$$ t = \frac{474}{ T_9^{1.9}}  .\eqno(7.5.26)$$

For large temperature $T_9 > 10,$ neutron abundance is given by

$$ X_n = \Big[ 1 + e^{Q_9/T_9} \Big]^{-1},  \eqno(7.5.27a)$$
where $Q_9 = m_n - m_p = 15.$ At freeze-out temperature $T_* \simeq 0.9 {\rm
  MeV},$ which is equivalent to $T_9 = 10.44$,

$$ X_n^* = 0.19. \eqno(7.5.27b)$$

Due to slightly higher freeze-out temperature, in the present model, neutron
concentration is obtained higher than $X_n^* = 0.16$ in SMU.

When $\Gamma_{\rm   w} << H $ i.e. when chemical equillibrium between
$n$ and $p$ freez-out, neutron concentration is determined through
neutron-decay
$$ n \to p + e^{-} + {\bar \nu}.   \eqno(7.5.28)$$

Thus, for $t > t_*,$ neutron abundance is given as 

$$ X_n(t) = 0.19 e^{-t/\tau}, \eqno(7.5.29)$$
where $\tau = (885.7 \pm 0.8) {\rm sec}$. is the neutron life-time.

First step, in the formation of complex nuclei, is 
$$ p + n \leftrightarrow D + \gamma$$
showing chemical equillibrium between deutron $(D)$ , nucleons and photon $(\gamma)$. This
equillibrium gives deutron abundance by weight \cite{ek} as

$$ X_D = 0.2 \times 10^{-12} (\Omega_B h^2) X_n X_p T_9^{3/2} exp(B_D/T_9), \eqno(7.5.30)$$
where $X_D = 2 n_D/n_B, X_n =  n_n/n_B, X_p = n_p/n_B, X_n + X_p = 1,$
deuterium binding energy $B_D = m_p + m_n - m_D = 2.23 {\rm MeV}$ which yields
$B_{D9} = 25.82 $ (in units of $T_9$) and $(\Omega_B h^2)$ is the baryon
number density.

Light heavy elements $^4He$ is formed through  reactions

$$ n + p \rightarrow D + \gamma,    \eqno(7.5.31a)$$

$$ (i)D + D \rightarrow ^3He + n, (ii) D + D \rightarrow ^3H + p
\eqno(7.5.31b,c)$$
and
$$ (ii) D + ^3H \rightarrow ^4He + n, (ii) ^3He + D \rightarrow ^4He + p,
\eqno(7.5.31d,e)$$
where $^3H$ stands for tritium.

These reactions show that sufficient deuterium abundance is required for
nucleosynthesis of $^4He$, which is supposed to be around $25\%$ of the
baryonic content of the universe. But after weak-interaction freeze out,
deuterium abundance is not sufficient unless temperature is very low. For
example, at $T_9 = 5.8 ( T = 0.5 {\rm MeV}), X_D = 2\times 10^{-12}$ is
obtained using eq.(7.5.30).

Rates for reactions (7.5.31b,c) are given as \cite[Appendix]{re}
\begin{eqnarray*}
\lambda_{DD(i)} &=& 3.97 \times 10^5 T_9^{-2/3} e^{-4.258/T_9^{1/3}}[1 + 0.098
T_9^{1/3} + 0.876 T_9^{2/3}
\\&& + 0.6 T_9 - 0.041 T_9^{4/3} - 0.071
T_9^{5/3}] (\Omega_B h^2) {\rm sec}^{-1}\\ \lambda_{DD(ii)} &=& 4.17 \times 10^5 T_9^{-2/3} e^{-4.258/T_9^{1/3}}[1 + 0.098 T_9^{1/3} + 0.518 T_9^{2/3} \\&& - 0.355 T_9 - 0.010 T_9^{4/3} -
0.018T_9^{5/3}] (\Omega_B h^2){\rm sec}^{-1}
\end{eqnarray*}
$$ \eqno(7.5.32a,b)$$
Making approximations in eqs.(7.5.31a,b), it is obtained that
\begin{eqnarray*}
\lambda_{DD} &=& \lambda_{DD(i)} + \lambda_{DD(ii)}
\\&& \sim 8.14\times 10^5 T_9^{-2/3} e^{-4.258/T_9^{1/3}}(\Omega_B h^2){\rm
  sec}^{-1} 
\end{eqnarray*}
$$ \eqno(7.5.33)$$

Mukhanov \cite{muk} has obtained the condition for conversion of sufficient
amount of $D$ to $^3He$ and $^3H$ as

$$ \frac{1}{2} \lambda_{DD} X_D t \sim 1  \eqno(7.5.34)$$
From eqs.(7.26), (7.29)-(7.32), it is obtained that
$$ 6.43 \times 10^{-6} T_{9i}^{-1.3} e^{\Big(\frac{25.82}{T_{9i}} -
  \frac{4.258}{(T_{9i})^{1/3}} \Big)}(\Omega_B h^2)^2  \sim 1  \eqno(7.5.35)$$
showing a relation between $\Omega_B h^2$ and $T_{9i}$. $T_{9i}$ is the
  temperature at which nucleosynthesis takes place.
For $\Omega_B h^2 = 0.05$, this equation yields
$$ T_{9i} \sim  1.16, \eqno(7.5.36)$$
which corresponds to $ T_i = 0.1 {\rm MeV}$. It shows that if baryon number
density $\Omega_B h^2$ is $\sim 0.05,$ nucleosynthesis of $^4He$ may take place
at temperature $ T_i = 0.1 {\rm MeV}$. In SMU, this temperature is $\sim
0.086{\rm MeV}$.

Eqs.(7.29) and (7.35) yield deuterium concencentration at $T_i$ as
$$X_{Di} \simeq 8.93 \times 10^{-6}    \eqno(7.5.37)$$

It happens at the cosmic time
$$ t_i = 357.5 {\rm sec} . \eqno(7.5.38)$$
Thus, it is obtained that, in the present model, nucleosynthesis of $^4He$
begins much later compared to SMU, where it happens at time $\sim 100{\rm
  sec}$ \cite{muk}. But, it is still less than life-time of a neutron, which
is required for the necleosynthesis.

The final $^4He$ abundance is determined by available free neutrons at cosmic
time $t_i$. As total weight of $^4He$ is due to neutron and proton, so its
final abundance by weight is given by \cite{re} as
$$ X^f_{\alpha} = 2 X_n^* exp(-t^f/\tau ) =  2 X_n^* exp(- 0.535/(T^f_{9})^{1.9} ) \eqno(7.5.39)$$
using eqs.(7.5.26) and $\tau = 886.5 {\rm sec}$. Here $\alpha$ stands for $^4He$.

The rate of production of $^4He$ is given by the equation
$$ {\dot X_{\alpha}} = 2 \lambda_{DD(ii)} X_D^2 . \eqno(7.5.40)$$
As free neutrons are captured in $^4He$, helium production dominates neutron
decay. It happens when \cite{re}

$$ 2{\dot X_{\alpha}} = \frac{X_n}{\tau}. \eqno(7.5.41)$$
Approximating $\lambda_{DD(ii)},$ given by eq.(7.5.32b), around $T_9 = 1.16,$ it
is obtained that

$$ \lambda_{DD(ii)} \simeq 6.98 \times 10^5 T_9^{-2/3} exp(- 4.258/T_9^{1/3})
(\Omega_B h^2). \eqno(7.5.42)$$

Eqs.(7.5.30), (7.5.39)-(7.5.41) yield

$$\frac{1}{2 \tau} \simeq 6.98 \times 10^5 T_9^{7/3} e^{(\frac{51.64}{T_9} -
  \frac{4.258}{T_9^{1/3}})} \times 0.19 e^{(- \frac{0.535}{T_9^{1.9}})} $$
$$ \times (0.2 \times 0.81 \times 10^{-12})^2 (\Omega_B h^2)^3,$$
which implies that

$$ T^f_9 = \frac{51.64}{37.96 - \frac{7}{3} ln T_9 - 3 ln(\Omega_B h^2) +
  \frac{4.258}{T_9^{1/3}} + \frac{0.535}{T_9^{1.9}} }. \eqno(7.5.43)$$

At $T = T_{9i} = 1.16,$

$$ T^f_9 = \frac{51.64}{42.08 - 3 ln(\Omega_B h^2)}. \eqno(7.5.44)$$

Using eq.(7.5.44) in eq.(7.5.39), helium abundance by weight, as function of baryon
number density $(\Omega_B h^2)$,is obtained as

$$ X^f_{\alpha} \simeq 0.38  exp \Big[ - 0.535\Big \{\frac{42.08 - 3
  ln(\Omega_B h^2)}{51.64} \Big\}^{1.9} \Big] . \eqno(7.5.45)$$

For $\Omega_B h^2 = 0.05,$ it is obtained that
$$ X^f_{\alpha} \simeq 0.23 . \eqno(7.5.46)$$

For $T < T^f_9,$ neutron abundance is governed by ${\dot X_n} = - 2 {\dot
  X_{\alpha}}$ with ${\dot   X_{\alpha}}$ given by eq.(7.5.39). So,

$$ {\dot X_n} = - 2 \lambda_{DD(ii)} X_D^2, $$ 
where $X_D$ and $\lambda_{DD(ii)}$ are given by eqs.(7.5.30) and (7.5.41)
respectively. Using eq.(7.5.26) and integrating, this equation yields neutron
abundance below $ T^f_9$ as

$$ X_n = \Big[(1/X_n^{\alpha}) + (\Omega_B h^2)^3 e^{-37.62} T_9^{0.43} \Big(
e^{51.64/T_9} - e^{51.64/T^f_9} \Big) \Big]^{-1}, \eqno(7.5.47a)$$
which shows that concentration of free neutrons drops very fast. It is due to
capture of free neutrons in $^4 He$. From eq.(7.5.30), it is obtained that
neutron and deuterium concentrations become equal at temperature $T^d_9$ given
as \cite{re}
$$ T^d_9 = \frac{25.82}{29.45 - ln(\Omega_B h^2) - \frac{3}{2} ln T^d_9 }.\eqno(7.5.47b)$$ 
Deutron concentration is given by \cite{re}

$$ {\dot X_D} \simeq - 2\lambda_{DD(ii)} X_D^2 = - 2\times 8.14\times 10^5
T^{-2/3} (\Omega_B h^2) e^{-4.258/T_9^{1/3}} X_D^2, $$
which integrates to
\begin{eqnarray*}
 X_D &=& \Big[\Big(1/X_D(T^d_9)\Big) + (\Omega_B h^2) e^{20.16}
\{(T^d_9)^{-2.8} e^{-4.258/(T^d_9)^{1/3}} \\&&- (T_9)^{-2.57}
e^{-4.258/(T_9)^{1/3}}\}\Big]^{-1},
\end{eqnarray*}
$$ \eqno(7.5.47c)$$
where $X_D(T^d_9) = X_n(T^d_9) .$

Eqs.(7.5.47a) and (7.5.47c) show that, below $T^f_9$, abundance of free neutron
and deuterium drop rapidly in PM also like SMU. 
 
\smallskip

\noindent \underline{(iii) Growth of inhomogeneities in the proposed model}
\smallskip

For small inhomogeneities, linear perturbation of Einstein equations is
enough. In this case, contrast density $\delta = \delta \rho_{\rm
  dm}/\rho_{\rm dm}$ (with $\rho_{\rm dm}$, being the cold dark matter energy
density and $\delta\rho_{\rm dm}$ is small fluctuation in $\rho_{\rm dm}$
presenting inhomogeneity), obeys a linear differential equation in the
homogeneous model of the universe \cite{pd}. Due to linearity, $\delta$ can be expanded
in modes $k$ as 

$$ \delta = {\sum_k} \delta_k e^{i {\vec k}.{\vec x}}.    \eqno(7.5.48)$$
For modes $k$ with proper wavelength $\digamma < H^{-1}(t) (H^{-1}(t) $ being
the Hubble's radius), the perturbation equation looks like 
$$ {\ddot \delta_k} + 2 \frac{\dot a}{a} {\dot \delta_k} + \Big( \frac{k^2
  v_s^2}{a^2} - 4 \pi G \rho_{\rm dm} \Big) \delta_k = 0,  \eqno(7.5.49)$$
which is free from gauge ambiguities \cite[eq.(4.158)]{pd}. So, $\delta_k$
  depends on $t$ only.

Eqs.(7.4.5c) and (7.4.13) give effective pressure for DM as 
$$ p_{\rm (dm, eff)} = (- 0.47 + {\rm w_{dm}}) \rho_{\rm dm}. \eqno(7.5.50)$$

It yields
$$ v_s^2 = \frac{d  p_{\rm (dm, eff)}}{d \rho_{\rm dm}} = - 0.47 + {\rm w}
.\eqno(7.5.51)$$

Connecting eqs.(7.4.6), (7.4.13), (7.5.49) and (7.5.51), the differential equation for
$\delta_k$ is obtained as
$${\ddot \delta_k} + 2 \frac{\dot a}{a} {\dot \delta_k} + \Big[(- 0.47 + {\rm
  w_{dm}}) \frac{k^2}{a^2} - \Big(0.345 H_0^2 a_0^{1.59}/a^{3(0.53 + {\rm
  w_{dm}} )} \Big)\Big] \delta_k = 0 \eqno(7.5.52)$$
with $\rho_{\rm cr,0} = 3 H^2_0/{8 \pi G}$.

Using eq.(7.4.10), eq.(7.5.52) looks like
$$ \tau \frac{d^2 \delta_k}{d \tau^2} + \frac{20}{9} \frac{d \delta_k}{d \tau} +  10^{31} \Big[(- 0.47 + {\rm
  w_{dm}}) \frac{k^2}{\tau^{11/9}}$$
$$ - \frac{0.345 H_0^2 a_0^{1.59}}{\tau^{10(0.23 +
  {\rm   w_{dm}} )/3}} \Big] \delta_k = 0 ,\eqno(7.5.53)$$
where $\tau$ is defined in eq.(7.5.4).

As mentioned above, for CDM $\rm {w_{dm}} = 0$, so eq.(7.5.51) is approximated as 
$$ \tau \frac{d^2 \delta_k}{d \tau^2} + \frac{20}{9} \frac{d \delta_k}{d \tau}  - \frac{3.45\times 10^{30} H_0^2 a_0^{1.59}}{\tau^{10(0.23 )/3}} \delta_k = 0 , \eqno(7.5.54)$$
which is integrated to

\begin{eqnarray*}
\delta_k &=& \tau^{-11/18} \Big[c_1 H^{(1)}_{110/21} (2 i b \tau^{7/60}) + c_2
H^{(2)}_{-110/21} (2 i b \tau^{7/60}) \Big] \\
& =&  \tau^{-7/6} \Big[(c_1+ c_2) J_{110/21} (2i b \tau^{7/60}) + i (c_1 - c_2)
Y_{110/21} (2 i b \tau^{7/60}) \Big], 
\end{eqnarray*}
 $$ \eqno(7.5.55)$$
where $i = \sqrt{-1}$ and  $b^2 = 3.45 \times 10^{30} H_0^2 a_0^{1.59}.$  Here $H^{(1)}_p(x)$ and
$H^{(2)}_p(x)$ are Hankel's functions as well as $J_p(x)$ and $Y_p(x)$ are
Bessel's functions. Moreover $c_1$ and $c_2$ are integration constants.

As discussed in section 7.4, CDM dominates HDM in the late universe, so
structure formation is expected for large $\tau$. Using approximations
(7.7c,d)for Bessel functions, for large $\tau$ in eq.(7.5.55)and doing some manipulations,
it is obtained that

$$\delta_k \approx \tau^{-0.67} e^{b \tau^{7/60}} , \eqno(7.5.56)$$
which shows growth of structure formation as cosmic time $t$ increases.

\bigskip

\centerline{\bf 8 Summary of Results  }

\smallskip

Thus it is found above that dual role of the Ricci scalar as a geometrical
field as well as a scalar physical field with its particle concept termed as
$riccion$ (which is different from `scalaron' or `graviton') reflects many
important features of gravity and cosmology. Interestingly, these features are inspired by
geometry of the space-time. In section 3, it is obtained that $riccion$
behaves like an $instanton$. The instanton solution, so obtained (using dual
role of the Ricci scalar), causes inflation prior to GUT phase
transition. This inflationary scenario is called `primordial
inflation'. Riccions, emerging from higher dimensional space-time, decompose
into fermionic and anti-fermionic counterparts, called $riccions$ and
$anti-riccions$ when parity is violated. As a riccion and a riccino have the
same mass, these are supers-symmetric partners too. Mass of riccions and
riccinos, being inversely proportional to gravitational constant, is very
high. So,during the course of expansion of the universe, these might have
decayed due to emission of observed elementary particles. If extra-dimension
$D$ of the $(4+D)-$dimensional space-time is energy mass scale dependent,
results of one-loop renormalization yield $D$ as positive real number above
the energy scale $1.76 \times 10^{16} {\rm GeV}$, not as
positve integer. It supports $fractal$ dimension, not topological dimension
above this scale. But when $D$ is taken independent of energy mass scale,
renormalization group equations for riccion yield $D=6$, making dimension of
the space-time $D + 4 = 10$ as required in super-string theory.

Riccion yields many interseting cosmological consequences too. Homogeneous and
inhomogeneous models are dervied here using dual role of the Ricci scalar. It
is shown that the universe bounces at the critical temperature $\sim 10^{18}
{\rm GeV}$ supporting a non-singular biginning of cosmology. Creation of
spinless and spin-1/2 particles in these models are discussed here. Lastly, a
new cosmological scenario, inspired by dual nature of the Ricci scalar,is
proposed here. It is discussed at length in the preceding section and briefly
it is as follows.

In contrast to the original $big-bang$ theory, the proposed
cosmology answers many basic questions as (i)``What is the $fireball$?'', (ii)
``How does it burst out?'', (iii) ``What is the background radiation?'' and
``What are initial values of temperature and energy density?'' Moreover, the
proposed cosmological model is free from initial singularity. It is found that
the $dark$ $energy$ violates the $strong$ $energy$ $condition$ showing
``bounce'' of the universe, which is consistent with $singularity - free$
model of cosmology \cite{skp93,  sk98, sk99}. The initial scale factor is computed to be $a_{\rm ew} = 0.25$. It is in contrast to the standard model of cosmology SMU, which encounters with singularity having zero scale factor, infinite energy density and infinite temperature.

The present value of $dark$ $energy$ density is supposed to be $\sim 7.3
\times 10^{-48} {\rm Gev}^4$, which is 53 orders below its initial
values $10^6 {\rm Gev}^4$ . The question ``How does $dark$ $energy$ falls by 53 orders in the
current universe?'' is answered by the result, in section 6(a), showing that
$dark$ $energy$ decays to $dark$ $matter$, obeying the rule, given by
eq.(6.7). Upto the decoupling time $t_d \simeq 386 kyr$, matter remains in
thermal equllibrium with the background radiation, so produced $dark$ $matter$
upto $t_d$ is supposed to be HDM. But when $t >t_d$, production of CDM is more
than HDM. The current ratio of HDM density and CDM density is found to be
$5 \times 10^{-30}$. In section 6, it is found that creation of HDM raises
entropy of the universe upto $\sim 10^{87}$.   During the phase of
accelerated expansion, temperature falls as $a(t)^{- 50/103}$.

It is found that one of the two types of  dynamical changes of the universe
are possible, beyond $3.7 t_0  = 50.69{\rm Gyrs}$, (i) decelerated expansion and (ii) contraction. If future universe expands with deceleration , it will expand for ever. But in the case of contraction, it will collapse by $247.73{\rm Gyrs}$.

As $dark$ $energy$ dominates over $dark$ $matter$, from the beginning upto the time $3.03 t_0$, no $cosmic$ $coincidence$ $problem$ arises in the proposed  scenario. 
Moreover, in the preceding section, some  other basic problems such as
creation of particles in the early universe, primordial nucleosynthesis and
structure formation in the late universe. In the proposed model, it is shown
that particles are created due to topological changes which is unlike the SMU,
where it is assumed that elementary particles are created at the time of
big-bang. It is also shown, in section 7.2, that process of primordial
nucleosynthesis goes very well in the proposed model predicting $\sim 23\%$
Helium-4 abundance by weight. In section 7.3, it is shown that ,in the late
universe, inhomogeneities in CDM grow exponentially with time. Thus  the proposed model is able to provide possible solutions to many
cosmological problems with prediction for the future universe.

Thus, it is found that dual nature of the Ricci scalar has enough potential to
reveal many interesting aspects of gravitation and cosmology. It many yield
some interesting results, if riccion is obtained from brane-gravity and
Einstein - Gauss-Bonnet theory.

\bigskip

\centerline{\bf Appendix A}

\centerline{\underline{\bf Riccion and Graviton}}

\smallskip

From the action

$$ S = \int {d^4 x} {d^D y}  \sqrt{- g_{(4+D)}} \quad
\Big[ \frac{M^{(2+D)}R_{(4+D)}}{16 \pi
} + {\alpha_{(4+D)}} R_{(4+D)}^2 + $$
$$\gamma_{(4+D)} ( R_{(4+D)}^3   - \frac{6(D+3)}{D-2)}{\Box}_{(D+4)}R^2_{(D+4)})\Big], \eqno(A.1)$$
the gravitational equations are obtained as

$$ \frac{M^{(2+D)}}{16 \pi} (R_{MN} - {1 \over 2} g_{MN} R_{(4+D)} ) + {\alpha_{(4+D)}} H^{(1)}_{MN} + {\gamma_{(4+D)}} H^{(2)}_{MN} = 0,\eqno(A.2a)$$ 
where

$$ H^{(1)}_{MN} = 2 R_{; MN} - 2 g_{MN} {\Box}_{(4+D)} R_{(4+D)} - {1\over 2} g_{MN} R^2_{(4+D)} + 2 R_{(4+D)} R_{MN}, \eqno(A.2b)$$  
and
$$ H^{(2)}_{MN} = 3 R^2_{; MN} - 3 g_{MN} {\Box}_{(4+D)} R^2_{(4+D)} - \frac{6(D+3)}{(D-2)}\{- { 1 
\over 2} g_{MN}{\Box}_{(4+D)} R^2_{(4+D)}$$  $$+ 2{\Box}_{(4+D)}R_{(D+4)}R_{MN} + R^2_{; MN}\} - { 1 \over 2} g_{MN} R^3_{(4+D)} + 3 R^2_{(4+D)} R_{MN} . \eqno(A.2c) $$

Taking $g_{MN} = \eta_{MN} + h_{MN}$ with $\eta_{MN}$ being $(4+D)$-dimensional Minkowskian metric tensor components and $h_{MN}$ as small fluctuations, the equation for $graviton$ are obtained as

$${\Box}_{(4+D)} h_{MN} = 0   \eqno(A.3) $$
neglecting  higher-orders of $h$.

On compactification of $M^4 \otimes S^D$ to $M^4$, eq.(A.3) reduces to the equation for 4-dimensional $graviton$ as
$${\Box} h_{\mu\nu} + \frac{l(l+D-1)}{\rho^2}h_{\mu\nu} = 0   \eqno(A.4) $$
for the space time 

$$ d S^2 = g_{\mu\nu} d x^{\mu} d x^{\nu} - \rho^2 [ d\theta_1^2 + sin^2\theta_1 d\theta^2_2 + \cdots +
sin^2\theta_1 \cdots sin^2\theta_{(D-1)} d\theta^2_D.  \eqno(A.5)$$

The 4-dimensional graviton equation (A.4) is like usual 4-dimensional graviton equation ( the equation derived from 4-dimensional action) only for $l = 0$. Thus the massless graviton is obtained for $l = 0$ only.

As explained, in section 2, the trace of equations (A.2) leads to the $riccion$ equation

$$[{\Box} + {1 \over2} \xi R + m^2_{\tilde R} + \frac{\lambda}{3!} {\tilde R}^2]{\tilde R} + \vartheta  = 0,  \eqno(A.6a)$$
where

\begin{eqnarray*}
 \xi & = & \frac{D}{2(D+3)} +  \eta^2 \lambda R_D \\ m^2_{\tilde R} & = & - \frac{(D + 2) \lambda V_D}{16 \pi G_{(4+D)}} + \frac{D R_D}{2(D+3)} + \frac{1}{2} \eta^2 \lambda R_D^2 \\ \lambda  & = & \frac{1}{4(D+3)\alpha}, \\ \vartheta & = & - \eta \Big[- \frac{(D + 
2) \lambda M^{(2+D)}V_D}{16 \pi} + \frac{D  R_D^2}{4 (D+3)} + \frac{1}{6} \eta^2 \lambda R_D^3 \Big].
 \end{eqnarray*} 
$$ \eqno(A.6 b,c,d,e)$$

The graviton $h_{\mu\nu}$ has 5 degrees of freedom ( 2 spin-2 graviton, 2 spin-1 gravi-vector (gravi-photon) and 1 scalar). The scalar mode $f$ satisfies the equation

$${\Box} f + \frac{l(l+D-1)}{\rho^2}f= 0   \eqno(A.7) $$
Comparison of eqs.(A.6) and (A.7) show many differences between the $scalar$ $mode$ $f$ of $graviton$ and the $riccion$ ($\tilde R$) e.g. $\xi, \lambda$ and $ \vartheta$ are vanishing for $f$, but these are non-vanishing for $\tilde R$. Eq.(A.7) shows $(mass)^2$ for $f$ as

$$m_f^2 =  \frac{l(l+D-1)}{\rho^2} ,\eqno(A.8) $$ 
whereas $(mass)^2$ for $\tilde R$ ($m_{\tilde R}^2$), given by eq.(A.6c), depends on $G_{(4+D)}, V_D$ and $R_D$ (given in section 2). $m_f^2 = 0$ for $l = 0$, but $m^2_{\tilde R}$ can vanish only when gravity is probed upto $\sim 10^{-33} cm$. As mentioned above, gravity could be probed upto minimum distance 1cm only. Thus, $(mass)^2$ of $riccion$ can not vanish.  

Thus, even though $f$ and $\tilde R$ are scalars arising from gravity, both are different. $Riccion$ can $not$ arise without higher-derivative terms in the action,  but $graviton$ is also obtained from Einstein-Hilbert action with lagrangian density $R/{8 \pi G}$.

In a locally inertial co-ordinate system
 $$ {\Box} = {1\over \sqrt{-g}}{\frac{\partial}{\partial x^{\mu}}}\Big( \sqrt{-g}\quad g^{\mu\nu} \frac{\partial}{\partial x^{\nu}}\Big)= \frac{\partial^2}{\partial X^i \partial X^j} $$
showing that in the ${\Box}{\tilde R}$, $\tilde R$ and $\Box$ are like other scalar fields $\phi$ and $\Box$ in Klein -Gordon equation for $\phi$, according to the principle of covariance. So, there is no harm in making loop quantum corrections and renormalization for $\tilde R$ like $\phi$.

\centerline{\bf Appendix B}

\bigskip
In eq.(6.4.6), using the convolution theorem the function $e^{i k.r} \Big[1
- \frac{r_0}{r} \Big]^{-4/7}$ can be written as
$$e^{i k.r} \Big[1 - \frac{r_0}{r} \Big]^{-4/7} = \eta^{-1} e^{i k.r}
{\int_{r_0 + \epsilon}^{\infty}}{d y} e^{-i k.y} \Big[1 - \frac{r_0}{y}
\Big]^{-4/7} , \eqno(B.1)$$ 
where $\eta$ has dimension of length, as used in section 2,for dimensional
correction. Here in the integral $d y$ introduces a quantity of length
dimension which is compensated by $\eta.$

Evaluation of the integral in eq.(B.1), is done as follows.
\begin{eqnarray*}
{\int_{r_0 + \epsilon}^{\infty}}{d y} \frac{e^{-i k.y}}{ \Big[1 - \frac{r_0}{y}
\Big]^{4/7}} &=& {\int_{r_0 + \epsilon}^{\infty}}{d y} \frac{y^{4/7} (y -
r_0)^{\omega}}{(y - r_0)^{\omega + 4/7} } {\sum_{n =0}^{\infty}} \frac{( -
i k y)^n}{n!}\\ &=&{\sum_{n =0}^{\infty}} \frac{( -i k )^n}{n!}
\frac{1}{\Gamma(\omega + 4/7)} {\int_{r_0 + \epsilon}^{\infty}} {d y} y^{n +
4/7} (y - r_0)^{\omega}\times \\&& \Big[ 
{\int_0^{\infty}} x^{\omega - 3/7} e^{- x (y - r_0)} {d x} \Big] 
\\ &=& {\sum_{n =0}^{\infty}} \frac{( -i k )^n}{n!} \frac{1}{\Gamma(\omega
+ 4/7)} 
{\int_0^{\infty}} {d x} x^{\omega - 3/7} e^{ x r_0)} \times \\&& \Big[{\int_{r_0 +
\epsilon}^{\infty}}  y^{n + 4/7} (y - r_0)^{\omega} e^{- x y} {d y} \Big]
 \\& \simeq& {\sum_{n =0}^{\infty}} \frac{( -i k )^n}{n!}
\frac{1}{\Gamma(\omega + 4/7)} 
{\int_0^{\infty}} {d x}  x^{\omega - 3/7} e^{- x \epsilon} \times \\&& (r_0 +
\epsilon)^{(n + 4/7)} \epsilon^{\omega} \Big[\frac{1}{x} +
\frac{\omega}{\epsilon x^2}  + \frac{\omega (\omega - 1)}{\epsilon^2  x^3}
+ \cdots \Big] 
\end{eqnarray*}
$$ \eqno(B.2)$$
neglecting terms containing $(r_0 + \epsilon)^{-1}$ compared to terms
containing $ \epsilon^{-1}.$

Performing further integration in eq.(B.2), one obtains that
\begin{eqnarray*}
{\int_{r_0 + \epsilon}^{\infty}}{d y} \frac{e^{-i k.y}}{ \Big[1 - \frac{r_0}{y}
\Big]^{4/7}} & \simeq& {\sum_{n =0}^{\infty}}\frac{( -i k )^n}{n!}
\frac{1}{\Gamma(\omega + 4/7)} (r_0 + \epsilon)^{(n + 4/7)}\times \\&&
\epsilon^{3/7} 
\Big[\Gamma(\omega - 3/7) + \omega \Gamma(\omega - 10/7) + \\&& 
 \omega (\omega - 1) \Gamma(\omega - 17/7) + \cdots \Big]
 \\&=& {\sum_{n =0}^{\infty}}\frac{( -i k )^n}{n!}
\frac{1}{\Gamma(\omega + 4/7)} (r_0 + \epsilon)^{(n + 4/7)} \epsilon^{3/7}
\\&=& (r_0 + \epsilon)^{ 4/7} \epsilon^{3/7} cos k(r_0 +
\epsilon),
\end{eqnarray*}
$$ \eqno(B.3)$$  
when $\omega \to \infty$. Here, symmetry under $k \to -k$ caused by
symmetry of the model under $r \to -r$ is used.

The same procedure yields
$${\int_{r_0 + \epsilon}^{\infty}}{d y} e^{-i k.y} y^{-2} \Big[1 -
\frac{r_0}{y} \Big]^{-4/7} = (r_0 + \epsilon)^{-10/7} \epsilon^{3/7} cos
k(r_0 + \epsilon) ,  \eqno(B.4)$$

$${\int_{r_0 + \epsilon}^{\infty}}{d y} e^{-i k.y} y^{-3/7} [y - r_0]^{-18/7}
 = (r_0 + \epsilon)^{-3/7} \epsilon^{-11/7} cos
k(r_0 + \epsilon) ,  \eqno(B.5)$$

$${\int_{r_0 + \epsilon}^{\infty}}{d y} e^{i k.y} y^{-4/7} [y - r_0]^{4/7}
 = (r_0 + \epsilon)^{-4/7} \epsilon^{11/7} cosk(r_0 + \epsilon) ,  \eqno(B.6)$$
 
$${\int_{r_0 + \epsilon}^{\infty}}{d y} e^{-i k.y} y^{-10/7} [y - r_0]^{-18/7}
 = (r_0 + \epsilon)^{-10/7} \epsilon^{-11/7}cosk(r_0 + \epsilon) .
\eqno(B.7)$$

\centerline{\bf Acknowledgements}

I feel indebted to Prof.K.P.Sinha for useful discussion on riccion, when
this idea came to my mind. Actually, particle concept of the Ricci scalar was
suggested by him. Moreover,I am grateful to Prof.A.A.Starobinsky for
his discussion and useful suggestions on the physical aspect of the Ricci
scalar, when I met him at IUCAA, Pune during GR97 conference. During this
discussion, he mentioned about his work on scalaron.

\end{document}